\documentclass[useAMS,usenatbib]{mn2e}

\usepackage{lscape}
\usepackage{epsfig}
\usepackage{times}

\def\gtsim{\mathrel{\spose{\lower.5ex \hbox{$\mathchar"218$}}
     \raise.4ex\hbox{$\mathchar"13E$}}}
\def\ltsim{\mathrel{\spose{\lower.5ex\hbox{$\mathchar"218$}}
     \raise.4ex\hbox{$\mathchar"13C$}}}
\def\aFe{[$\alpha/{\rm Fe}$]}

\def\Hb{${\rm H}{\beta}$}
\def\hb{${\rm H}{\beta}$}

\def\Mgb{{\rm Mg}\,$_b$}
\def\Fe{$\langle {\rm Fe}\rangle$}

\def\ZH{[$Z/{\rm H}$]}
\def\MgFe{[${\rm MgFe}$]$'$}

\def\Mgd{{\rm Mg}\,$_2$}

\def\rbd{$r_{\rm{ bd}}$}

\def\reff{$r_{\rm{ e}}$}

\def\kms{$\rm km\;s^{-1}$}

\def\apj{ApJ}
\def\aj{AJ}
\def\apjl{ApJL}
\def\apjs{ApJS}
\def\mnras{MNRAS}

\def\aap{A\&A}

\def\araa{ARA\&A}
\def\pasp{PASP}
\def\spose#1{\hbox to 0pt{#1\hss}}

\def\aj{AJ}                   
\def\araa{ARA\&A}             
\def\apj{ApJ}                 
\def\apjl{ApJ}                
\def\apjs{ApJS}               
\def\aap{A\&A}                
\def\mnras{MNRAS}             
\def\pasp{PASP}               

\def\kms{$\rm km\;s^{-1}$}

\def\Hb{${\rm H}{\small{\beta}}$}

\begin{document}

\title[Stellar populations in the bulges of isolated galaxies]{Stellar
  populations in the bulges of isolated galaxies}
\author[L. Morelli et al.]{L.~Morelli$^{1,2}$\thanks{E-mail: 
  lorenzo.morelli@unipd.it}, M. Parmiggiani$^{1,2}$, E.~M.~Corsini$^{1,2}$, 
  L.~Costantin$^{1}$, E. Dalla Bont\`a$^{1,2}$,
  \and J.~M\'endez-Abreu$^{3}$, and A.~Pizzella$^{1,2}$\\ 
$^1$ Dipartimento di Fisica e Astronomia ``G. Galilei'', Universit\`a
     di Padova, vicolo dell'Osservatorio 3, I-35122 Padova, Italy\\ 
$^2$ INAF--Osservatorio Astronomico di Padova,
     vicolo dell'Osservatorio 5, I-35122 Padova, Italy\\
$^3$ School of Physics and Astronomy, University of St. Andrews, 
    SUPA, North Haugh, KY16 9SS St. Andrews, UK}
\date{{\it Draft version on \today}}

\maketitle


\begin{abstract}
We present photometry and long-slit spectroscopy for 12 S0 and spiral
galaxies selected from the Catalogue of Isolated Galaxies. The
structural parameters of the sample galaxies are derived from the
Sloan Digital Sky Survey $i$-band images by performing a
two-dimensional photometric decomposition of the surface brightness
distribution. This is assumed to be the sum of the contribution of a
S\'ersic bulge, an exponential disc, and a Ferrers bar characterized
by elliptical and concentric isophotes with constant ellipticity and
position angles. The rotation curves and velocity dispersion profiles
of the stellar component are measured from the spectra obtained along
the major axis of galaxies. The radial profiles of the \Hb, Mg and Fe
line-strength indices are derived too. Correlations between the
central values of the \Mgd\ and \Fe\ line-strength indices and the
velocity dispersion are found. The mean age, total metallicity and
total $\alpha$/Fe enhancement of the stellar population in the centre
and at the radius where the bulge gives the same contribution to the
total surface brightness as the remaining components are obtained
using stellar population models with variable element abundance
ratios. We identify intermediate-age bulges with solar metallicity and
old bulges with a large spread in metallicity.  Most of the sample
bulges display super-solar $\alpha$/Fe enhancement, no gradient in age
and negative gradients of metallicity and $\alpha$/Fe
enhancement. These findings support a formation scenario via
dissipative collapse where environmental effects are remarkably less
important than in the assembly of bulges of galaxies in groups and
clusters.
\end{abstract}


\section{Introduction}
\label{sec:introduction}

Stellar populations are a powerful diagnostics to constrain the
assembly history of galaxy bulges. In the current picture, dissipative
collapse \citep[e.g.,][]{gilwys98}, merging and acquisition events
\citep[e.g.,][]{coletal00}, and secular evolution
\citep[e.g.,][]{korken04} are considered as possible processes driving
the formation of bulges. According to theoretical models, these
processes give rise to different properties of the stellar populations
in galaxy centres and to different trends of age, metallicity, and
star-formation timescale as a function of the galactocentric distance.

For example, from the metallicity gradient it is possible to
  extract information about the gas dissipation processes and the
  importance of secular processes and merging history. Stars form at
all galactocentric distances during the dissipative collapse of a
protogalactic cloud and they remain on their orbits with little
migration towards the center. On the contrary, the gas dissipates
inward and it is continuously enriched by the evolving stars. In
  consequence of this, the stars formed in the outskirts of a galaxy
  are expected to have a lower metal contents with respect to those in
  the central regions. Also galactic winds induced by the supernovae
  \citep{aryo87,Creasey2013} have a relevant role in the evolution
  history of the galaxy. High-resolution simulations
\citep{Hirschmann2013} demonstrated that the stellar accretion in
galaxies with galactic winds is steepening the galactic gradient of
about 0.2 dex \citep{Hirschmann2015}. These winds, indeed, eliminate
the gas suppressing the fuel needed for star formation. The outer
regions develop the winds before the central ones, where the star
formation and chemical enrichment continue for a longer time. Strong
negative gradients are expected in dissipative collapse models as both
star formation and galactic winds act in steepening any incipient
gradient. In hierarchical formation models, the situation is somewhat
contradictory. Some authors suggest that clustering and wet or dry
merging erase the metallicity gradient
\citep[e.g.,][]{besh99,dimatteo2009}, while others argue that the
metallicity gradient is moderately affected by interactions since the
violent relaxation preserves the position of the stars in the local
potential \citep[e.g.,][]{vanAlbada1982}. Such a dichotomy possibly
depends on how the properties of the resulting galaxy are related to
the gas-to-stellar mass ratio of the progenitors. If they are
characterized by a large gas fraction, the resulting metallicity
gradient is indeed steeper. In the secular evolution scenario,
  the bulge is the result of a redistribution of the disc stars due to
  the instabilities triggered by bars, ovals, and spiral arms. The
  theoretical model predictions for the metallicity gradient in these
  bulges are ambiguous. It could be erased as consequence of disc
  heating or amplified from the reduction of the scalelength of the
  final resulting spheroid \citep{mooretal06}.

In the last decade, a major observational effort was performed to
derive the stellar population properties in large number of bulges
\citep[e.g.,][]{jabletal07, moreetal08, morelli2012,
  gonzalezdelgado2014, seidel2015, Wilkinson2015} to be compared to
those of elliptical galaxies \citep[e.g.,][]{sancetal06p, annietal07,
  kuntschner2010, mcdermid2015} and galaxy disks \citep{sancetal14,
  morelli2015b}.
Stellar populations of bulges show a complex variety of
properties. The ages of bulges are spread between 1 and 15 Gyr. Such a
large difference seems to be driven by the morphological type of the
host galaxy with the late type younger than the early type
\citep{gandetal07}. The timescale of the last major star-formation
burst spans between 1 to 5 Gyr, as derived from the central values of
$\alpha/$Fe abundance ratio \citep{thda06}. In general, $\alpha$/Fe is
constant over the observed radial ranges and many bulges have a solar
abundance ratio \citep{jabletal07, moreetal08,
  morelli2012}. Independently of their structural properties and
whether they reside in low or high surface-brightness discs, most
bulges are characterised by a negative metallicity gradient, which is
one of the tighter predictions made by theoretical models for the
dissipative collapse \citep{gilwys98,pipietal10}. On the other hand,
the absence of stellar population gradients measured in some bulges is
an clear indication that bulge stars were redistributed as a
consequence of external and internal processes, like minor mergers and
slow rearrangement of the disk material, respectively \citep{besh99,
  cobari99}.

In many cases the difficulty in determining the mechanism driving the
assembly history of the bulge is probably due to the fact that the
dissipative collapse, minor and major mergers, and secular evolution
are all having an effect in reshaping the structure of disk
galaxies. Furthermore, phenomena driven by the environment like gas
stripping, harassment, and strangulation are likely to play a role in
mixing up the properties of the stellar populations
\citep{labarbera2014}. However, to date the observational evidences on
how the environment influences the stellar populations of bulges are
sparse and the analysis of both the central values of age,
metallicity, and star formation timescale
\citep{denietal05,redaetal07} and their radial gradients
\citep{katkov2015} does not lead to any firm conclusion. In addition,
the comparison of the results obtained for galaxies in different
environments is not straightforward. Part of the difficulty lies in
addressing the relative importance of one-to-one interactions and the
local galaxy density, and this reflects the lack of suitable control
samples to which the properties of bulges of interacting and/or
cluster galaxies can be compared. As a matter of fact, the samples of
field galaxies studied so far include also galaxies in pairs and loose
groups.

A way to make simpler the observational picture is studying the bulges
of isolated galaxies, for which the interactions with the surrounding
environment or with other galaxies are likely to be negligible
\citep{Hirschmann2013_is}. Therefore, it could be possible to use the
stellar population diagnostics to disentangle between bulges formed
from dissipative collapse and those assembled via secular
evolution. To this aim, here we analyse the stellar populations of the
bulges of a carefully selected sample of high surface-brightness
isolated disc galaxies to be compared with the complementary samples
of bulges in high surface-brightness cluster galaxies and giant low
surface-brightness galaxies which we studied in in the past several
years \citep{pizzetal08, moreetal08, morelli2012, morelli2015a}.

The paper is organized as follows. We present the selection of the
sample of isolated disc galaxies in Section~\ref{sec:sample}, and we
describe the analysis of the photometric and spectroscopic data in
Sections \ref{sec:photometry} and \ref{sec:spectroscopy},
respectively. We analysed the stellar population properties in
Section~\ref{sec:populations}. We discuss conclusions and summarise
results in Section~\ref{sec:conclusions}.

\section{Sample selection}
\label{sec:sample}

\begin{figure}
  \centering
  \includegraphics[width=0.49\textwidth]{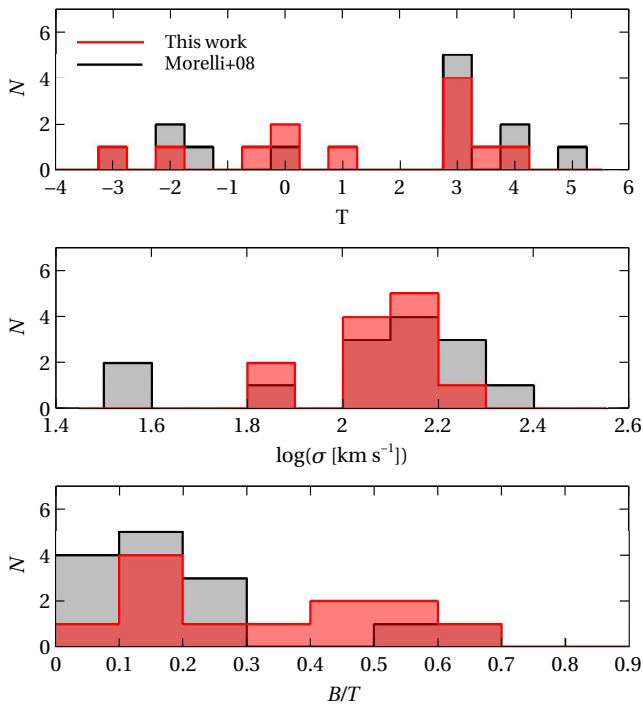}
  \caption{Distribution of the morphological type (upper panel),
    central velocity dispersion (middle panel), and $B/D$ ratio (lower
    panel) for the sample galaxies (red histograms). The distribution
    of the same quantities for the group and cluster galaxies studied
    by \citet{moreetal08} is plotted for a comparison (grey
    histograms).}
  \label{fig:histo_sample}
\end{figure}

We selected the sample galaxies from the Catalogue of Isolated
Galaxies \citep[CIG, ][]{Karachentseva1973}. The CIG comprises 1051
galaxies with apparent magnitude $M_{ZW} < 15.7$ and
$\delta<-3^\circ$). In the CIG are included galaxies with radius $R$
and with no other galaxy with radius $1/4R<r<4R$ within a projected
distance of $40r$. Assuming an average radius of $R=10$ kpc and an
average field velocity of $V=150$ \kms, the CIG galaxies should not
have passed close to a mass perturber in the last $3\times10^9$
years. We mined the CIG to find nearby ($D<150$ Mpc)
early-to-intermediate disc galaxies (from S0 to Sbc) in order to
spatially resolve the bulge component. They were chosen to have a
low-to-intermediate inclination ($i < 70^\circ$) to allow a reliable
photometric decomposition, available images in the Sloan Digital Sky
Survey \citep[SDSS,][]{2000AJsdss}, and to be spectrocopically
observed at the Telescopio Nazionale Galileo (TNG, $\delta>0^\circ$)
Our final sample consist of 16 spiral galaxies, with multiband
photometry already available and which are fully representative of the
whole CIG sample.  We actually observed 12 galaxies at the TNG
telescope and they constitute the sample of isolated galaxies we
studied in this paper.  We report their basic properties in
Table~\ref{tab:sample}.
%
\begin{table*}
\caption{Parameters of the sample galaxies. The columns show the
  following. (1): galaxy name; (2): morphological classification from
  Lyon Extragalactic Database (LEDA); (3): numerical morphological
  type from LEDA; (4): apparent isophotal diameter measured at a
  surface-brightness level of $\mu_B = 25$ mag arcsec$^{-2}$ from
  LEDA; (5): total observed blue magnitude from LEDA; (6): radial
  velocity with respect to the cosmic microwave background (CMB)
  reference frame from LEDA; (7): distance obtained as $V_{\rm
    CMB}/H_0$ with $H_0= 75$ km s$^{-1}$ Mpc$^{-1}$; (8): absolute
  total blue magnitude from $B_T$ corrected for extinction as in LEDA
  and adopting $D$;}
\begin{center}
\begin{small}
\begin{tabular}{llc cr ccc c}
\hline
\noalign{\smallskip}
\multicolumn{1}{c}{Galaxy} &
\multicolumn{1}{c}{Type} &
\multicolumn{1}{c}{$T$} &
\multicolumn{1}{c}{$D_{25}\,\times\,d_{25}$} &
\multicolumn{1}{c}{$B_{\rm T}$} &
\multicolumn{1}{c}{$V_{\rm CMB}$} &
\multicolumn{1}{c}{$D$} &
\multicolumn{1}{c}{$M_{B_{\rm T}}$} \\ 
\noalign{\smallskip}
\multicolumn{1}{c}{} &
\multicolumn{1}{c}{} &
\multicolumn{1}{c}{} &
\multicolumn{1}{c}{(arcmin)} &
\multicolumn{1}{c}{(mag)} &
\multicolumn{1}{c}{(\kms)} &
\multicolumn{1}{c}{(Mpc)} &
\multicolumn{1}{c}{(mag)} \\
\noalign{\smallskip}
\multicolumn{1}{c}{(1)} &
\multicolumn{1}{c}{(2)} &
\multicolumn{1}{c}{(3)} &
\multicolumn{1}{c}{(4)} &
\multicolumn{1}{c}{(5)} &
\multicolumn{1}{c}{(6)} &
\multicolumn{1}{c}{(7)} &
\multicolumn{1}{c}{(8)} \\
\noalign{\smallskip}
\hline
\noalign{\smallskip}  
CGCG 034-050 & Sb   & $ 1.0$ & $0.95\times0.52$ & 14.53 & 3808 &  50.7 & $-18.99$ \\
CGCG 088-060 & S0a  & $ 0.0$ & $0.83\times0.67$ & 15.21 & 4725 &  63.0 & $-18.79$ \\
CGCG 152-078 & E/S0 & $-2.8$ & $0.48\times0.47$ & 15.81 & 6227 &  83.0 & $-18.78$ \\
CGCG 206-038 & S0a  & $-0.5$ & $0.81\times0.69$ & 14.86 & 6103 &  81.3 & $-19.69$ \\
IC 2473      & Sbc  & $ 3.5$ & $1.07\times0.74$ & 14.72 & 8325 & 111.0 & $-20.50$ \\
NGC 2503     & Sbc  & $ 4.0$ & $0.95\times0.79$ & 14.80 & 5711 &  76.1 & $-19.61$ \\
NGC 2712     & SBb  & $ 3.1$ & $2.95\times1.58$ & 12.78 & 2001 &  26.7 & $-19.35$ \\
NGC 2955     & SABb & $ 3.2$ & $1.50\times0.82$ & 13.58 & 7254 &  96.7 & $-21.34$ \\
UGC 4000     & SABb & $ 2.9$ & $1.44\times0.46$ & 14.88 & 9491 & 126.5 & $-20.63$ \\
UGC 4341     & S0a  & $-0.1$ & $1.17\times0.59$ & 14.66 & 6081 &  81.1 & $-19.88$ \\
UGC 5026     & S0   & $-2.0$ & $0.91\times0.62$ & 14.30 & 4480 &  59.7 & $-19.57$ \\
UGC 5184     & Sb   & $ 3.0$ & $1.02\times0.65$ & 14.73 & 6806 &  90.7 & $-20.05$ \\
\noalign{\smallskip}
\hline
\noalign{\medskip}
\end{tabular}
\end{small}
\label{tab:sample}
\end{center}
\end{table*}
%
In Fig.~\ref{fig:histo_sample} we compare the properties the
morphological type, bulge-to-disc ratio ($B/D$) and central velocity
dispersion of the sample galaxies with those of the group and cluster
galaxies analysed by \citet{moreetal08}.

\section{Broad-band imaging}
\label{sec:photometry}

\subsection{Data reduction}
\label{sec:sky}

We retrieved the $i$-band images of the sample galaxies from the Data
Archive Server (DAS) of the SDSS Data Release 9
\citep[DR9,][]{Ahn2012}. The images were reduced using {\sc
    iraf}\footnote{Image reduction and Analysis Facility is
    distributed by the National Optical Astronomy Observatories, which
    are operated by the Association of Universities for Research in
    Astronomy under cooperative agreement with the National Science
    Foundation.} routines and trimmed selecting a field of view (FOV)
  of at least $400\times400$ pixels (corresponding to $2.6\times2.6$
  arcmin$^2$) centred on the galaxies.
 
To estimate the goodness of the SDSS sky subtraction, we measured the
surface-brightness radial profile of the galaxies at large radii by
fitting their isophotes with ellipses using the {\sc ellipse}
algorithm described by \citet{jed87}.  We masked foreground stars,
nearby and background galaxies, residual cosmic rays, and bad pixels
in each galaxy image before fitting the galaxy isophotes. As a first
step, we allowed to vary the centres, ellipticities, and position
angles of fitting ellipses. Then, we adopted the centre of the inner
ellipses in the galaxy nucleus ($a<5$ arcsec) and the ellipticity and
position angle of the outer ones in the galaxy outskirts ($a>10$
arcsec). We assumed the constant value of the surface brightness
measured at large radii ($a\sim50$ arcsec where there was no light
contribution from the galaxy) as the residual sky level to be
subtracted from the image. We measured the standard deviation of the
image background after sky subtraction, $\sigma_{\rm sky}$, in regions
free of sources at the edges of the FOV.

Finally, we ran {\sc ellipse} on the sky-subtracted images to derive
the radial profiles of the azimuthally averaged surface brightness,
$\mu$, ellipticity, $\epsilon$, and position angle, PA, of the
galaxy isophotes. These profiles and the mask images were adopted for
the photometric decomposition.

\subsection{Photometric decomposition}
\label{sec:decomposition}

The structural parameters of the bulge, disc, and bar components of
the sample galaxies were derived from the sky-subtracted and masked
images by applying the Galaxy Surface Photometry Two-Dimensional
Decomposition ({\sc gasp2d}) algorithm \citep{mendetal08,
  mendetal14}. 

We modelled the surface brightness of the bulge, disc , and bar using
a \citet{sersic68}, \citet{free70}, and \citet[][see also
  \citealt{agueetal09} for the choice of the shape
  parameter]{ferrers1877} function, respectively. More details of the
procedures and algorithm used to retrieve the structural parameters of
the sample galaxies can be found in \citet{morelli2012}.

In our analysis we did not take into account for any other galaxy
component, such as spiral arms or inner and outer rings. Despite their
inclusion in the galaxy model slightly modifies the best-fitting
values of the structural parameters of our bulges, they result in a
less constrained solution due to the increase of free parameters. In
order to reduce the influence of these components on the fit, we
masked their corresponding regions in the galaxy images and excluded
them from the fitting process.

A bar was adopted for CGCG 088-060, CGCG 206-038, and IC 2473, whereas
we do not include the bar to build the surface-brightness models of the other
sample galaxies.
The best-fitting parameters were used to determine the radius \rbd,
where half of the total surface brightness is due to the bulge
only. Since we measured the bulge kinematics and stellar population
from long-slit spectra, the \rbd\ value has to be determined along the
slit direction which was always chosen to be aligned with the galaxy
major axis (Table \ref{tab:log}). To this aim, we built a
PSF-convolved image for each galaxy component using the structural
parameters from the photometric decomposition. We computed \rbd\ by
comparing the surface-brightness profiles of the components we
extracted along a 1-pixel wide strip crossing the galaxy centre and
aligned with the galaxy major axis.
We list the structural parameters of the bulges together with the
radius \rbd\ in Table \ref{tab:parameters_bulges}.
\begin{table*}
\caption{Photometric parameters of the bulges of the sample
  galaxies. The columns show the following. (2): effective surface
  brightness; column (3): effective radius; column (4): shape
  parameter; column (5): axial ratio; column (6): position angle of
  the major axis; (7): bulge-to-total luminosity ratio; (8): radius
  where the bulge contributes half of the galaxy surface brightness. }
\label{tab:parameters_bulges}
\begin{tabular}{lrrrrrrc}
\hline
\multicolumn{1}{c}{Galaxy} &
\multicolumn{1}{c}{$\mu_{\rm e}$} &
\multicolumn{1}{c}{$r_{\rm e}$} &
\multicolumn{1}{c}{$n$} &
\multicolumn{1}{c}{$q_{\rm b}$} &
\multicolumn{1}{c}{PA$_{\rm b}$} &
\multicolumn{1}{c}{$B/T$}&
\multicolumn{1}{c}{\rbd} \\
\multicolumn{1}{c}{} &
\multicolumn{1}{c}{(mag arcsec$^{-2}$)} &
\multicolumn{1}{c}{(arcsec)} &
\multicolumn{1}{c}{} &
\multicolumn{1}{c}{} &
\multicolumn{1}{c}{($^{\circ}$)} &
\multicolumn{1}{c}{} &
\multicolumn{1}{c}{(arcsec)} \\
\multicolumn{1}{c}{(1)} &
\multicolumn{1}{c}{(2)} &
\multicolumn{1}{c}{(3)} &
\multicolumn{1}{c}{(4)} &
\multicolumn{1}{c}{(5)} &
\multicolumn{1}{c}{(6)} &
\multicolumn{1}{c}{(7)}&
\multicolumn{1}{c}{(8)} \\
\noalign{\smallskip}
\hline
\noalign{\smallskip}
CGCG 034-050 & $18.38 \pm 0.12$ & $ 2.93 \pm 0.10$ & $2.57 \pm 0.15$ & $0.55 \pm 0.09$ & $ 37.54 \pm 0.10$ & 0.56& $5.7$ \\
CGCG 088-060 & $18.70 \pm 0.12$ & $ 1.50 \pm 0.10$ & $1.85 \pm 0.15$ & $0.95 \pm 0.09$ & $  1.46 \pm 0.10$ & 0.26& $2.4$ \\
CGCG 152-078 & $19.09 \pm 0.12$ & $ 1.68 \pm 0.10$ & $3.48 \pm 0.15$ & $0.92 \pm 0.09$ & $102.64 \pm 0.10$ & 0.49& $3.8$ \\
CGCG 206-038 & $18.44 \pm 0.12$ & $ 2.09 \pm 0.10$ & $1.27 \pm 0.15$ & $0.85 \pm 0.09$ & $ 54.94 \pm 0.10$ & 0.41& $2.8$ \\
IC 2473      & $18.87 \pm 0.12$ & $ 1.98 \pm 0.10$ & $1.31 \pm 0.15$ & $0.81 \pm 0.09$ & $ 30.75 \pm 0.10$ & 0.18& $3.7$ \\
NGC 2503     & $22.89 \pm 0.12$ & $ 8.84 \pm 0.10$ & $4.26 \pm 0.15$ & $0.54 \pm 0.09$ & $153.67 \pm 0.10$ & 0.11& $1.8$ \\
NGC 2712     & $18.12 \pm 0.06$ & $ 2.22 \pm 0.06$ & $0.99 \pm 0.10$ & $0.66 \pm 0.05$ & $  1.72 \pm 0.05$ & 0.09& $3.8$ \\
NGC 2955     & $23.10 \pm 0.12$ & $26.60 \pm 0.10$ & $5.56 \pm 0.10$ & $0.72 \pm 0.09$ & $  0.00 \pm 0.10$ & 0.59& $4.1$ \\
UGC 4000     & $20.64 \pm 0.12$ & $ 4.12 \pm 0.10$ & $2.69 \pm 0.10$ & $0.83 \pm 0.09$ & $ 27.60 \pm 0.10$ & 0.33& $4.2$ \\
UGC 4341     & $18.61 \pm 0.12$ & $ 2.06 \pm 0.10$ & $1.30 \pm 0.10$ & $0.73 \pm 0.09$ & $ 39.62 \pm 0.10$ & 0.18& $2.3$ \\
UGC 5026     & $20.64 \pm 0.12$ & $ 9.49 \pm 0.10$ & $3.85 \pm 0.10$ & $0.48 \pm 0.09$ & $ 12.62 \pm 0.10$ & 0.69& $6.1$ \\
UGC 5184     & $19.25 \pm 0.12$ & $ 1.53 \pm 0.10$ & $1.66 \pm 0.10$ & $0.71 \pm 0.09$ & $ 98.78 \pm 0.10$ & 0.10& $1.2$ \\
\noalign{\smallskip}
\hline
\noalign{\smallskip}
\end{tabular}
\end{table*}
The structural parameters of the discs and bars are given in
Tables~\ref{tab:parameters_discs} and \ref{tab:parameters_bars},
respectively.
\begin{table*}
\caption{Photometric parameters of the discs of the sample
  galaxies. The columns show the following. (2): central surface
  brightness; (3): scale length; (4): axial ratio; (5): position angle
  of the major axis; (6): disc-to-total luminosity ratio.}
\label{tab:parameters_discs}
\begin{tabular}{lcrrrr}
\hline
\noalign{\smallskip}
\multicolumn{1}{c}{Galaxy} &
\multicolumn{1}{c}{$\mu_0$} &
\multicolumn{1}{c}{$h$} &
\multicolumn{1}{c}{$q_{\rm disc}$} &
\multicolumn{1}{c}{PA$_{\rm disc}$}&
\multicolumn{1}{c}{$D/T$} \\
\multicolumn{1}{c}{ } &
\multicolumn{1}{c}{(mag arcsec$^{-2}$)} &
\multicolumn{1}{c}{(arcsec)} &
\multicolumn{1}{c}{} &
\multicolumn{1}{c}{($^{\circ}$)}&
\multicolumn{1}{c}{}\\
\multicolumn{1}{c}{(1)} &
\multicolumn{1}{c}{(2)} &
\multicolumn{1}{c}{(3)} &
\multicolumn{1}{c}{(4)} &
\multicolumn{1}{c}{(5)} &
\multicolumn{1}{c}{(6)} \\
\hline
\noalign{\smallskip}
CGCG 034-050 & $19.42 \pm 0.11$ & $ 7.48 \pm 0.10$ & $0.51 \pm 0.10$ & $ 35.39 \pm 0.10$&$0.44 $\\
CGCG 088-060 & $20.32 \pm 0.11$ & $ 8.87 \pm 0.10$ & $0.76 \pm 0.10$ & $ 36.65 \pm 0.10$&$0.65 $\\
CGCG 152-078 & $20.57 \pm 0.11$ & $ 6.32 \pm 0.10$ & $0.89 \pm 0.10$ & $122.65 \pm 0.10$&$0.51 $\\
CGCG 206-038 & $20.88 \pm 0.11$ & $10.69 \pm 0.10$ & $0.29 \pm 0.10$ & $157.33 \pm 0.10$&$0.18 $\\
IC 2473      & $20.52 \pm 0.11$ & $13.28 \pm 0.10$ & $0.66 \pm 0.10$ & $ 88.42 \pm 0.10$&$0.69 $\\
NGC 2503     & $20.34 \pm 0.11$ & $11.98 \pm 0.10$ & $0.87 \pm 0.10$ & $  2.20 \pm 0.10$&$0.89 $\\
NGC 2712     & $19.35 \pm 0.05$ & $19.43 \pm 0.06$ & $0.55 \pm 0.04$ & $ 13.49 \pm 0.06$&$0.91 $\\
NGC 2955     & $19.59 \pm 0.11$ & $12.47 \pm 0.10$ & $0.38 \pm 0.10$ & $154.95 \pm 0.10$&$0.41 $\\
UGC 4000     & $20.21 \pm 0.11$ & $12.94 \pm 0.10$ & $0.34 \pm 0.10$ & $ 35.76 \pm 0.10$&$0.67 $\\
UGC 4341     & $18.96 \pm 0.11$ & $ 9.95 \pm 0.10$ & $0.42 \pm 0.10$ & $ 47.81 \pm 0.10$&$0.82 $\\
UGC 5026     & $19.22 \pm 0.11$ & $ 6.06 \pm 0.10$ & $0.50 \pm 0.10$ & $ 26.41 \pm 0.10$&$0.30 $\\
UGC 5184     & $19.12 \pm 0.11$ & $ 8.63 \pm 0.10$ & $0.44 \pm 0.10$ & $107.24 \pm 0.10$&$0.90 $\\
\noalign{\smallskip} 
\hline
\noalign{\smallskip}
\end{tabular}
\end{table*}
\begin{table*}
\caption{Photometric parameters of the bars of the sample
  galaxies. The columns show the following. (2): central surface
  brightness; (3): length; (4): axial ratio; (5): position angle of
  the major axis; (6): bar-to-total luminosity ratio.}
\label{tab:parameters_bars}
\begin{tabular}{lccccr}
\hline
\noalign{\smallskip}
\multicolumn{1}{c}{Galaxy} &
\multicolumn{1}{c}{$\mu_{\rm bar}$} &
\multicolumn{1}{c}{$r_{\rm bar}$} &
\multicolumn{1}{c}{$q_{\rm bar}$}&
\multicolumn{1}{c}{PA$_{\rm bar}$}&
\multicolumn{1}{c}{$Bar/T$} \\
\multicolumn{1}{c}{ } &
\multicolumn{1}{c}{(mag arcsec$^{-2}$)} &
\multicolumn{1}{c}{(arcsec)} &
\multicolumn{1}{c}{} &
\multicolumn{1}{c}{($^{\circ}$)}&
\multicolumn{1}{c}{ }\\
\multicolumn{1}{c}{(1)} &
\multicolumn{1}{c}{(2)} &
\multicolumn{1}{c}{(3)} &
\multicolumn{1}{c}{(4)} &
\multicolumn{1}{c}{(5)} &
\multicolumn{1}{c}{(6)} \\
\hline
\noalign{\smallskip}
CGCG 088-060 & $20.52 \pm 0.14$ & $12.59 \pm 0.15$ & $0.39 \pm 0.10$&$  39.30\pm 0.10$ & $0.08$ \\
CGCG 206-038 & $20.93 \pm 0.14$ & $26.36 \pm 0.15$ & $0.78 \pm 0.10$&$ 143.76\pm 0.10$ & $0.41$ \\
IC 2473      & $21.24 \pm 0.14$ & $37.04 \pm 0.15$ & $0.21 \pm 0.10$&$  83.47\pm 0.10$ & $0.13$ \\
\noalign{\smallskip} 
\hline
\noalign{\smallskip}
\end{tabular}
\end{table*}
We show the {\sc gasp2d} fits and results of the photometric
decomposition of the SDSS images of the sample galaxies in
Fig.~\ref{fig:decomposition}.

To derive the uncertainties on the structural parameters we adopted a
  series of Monte Carlo simulations as done in
  \citet{morelli2012,morelli2015a}. We generated a set of barred and
  unbarred galaxies with a total $i$-band magnitude within $11.5 \leq
  i_{\rm T} \leq 14.5$ mag, mimicking the instrumental setup of the
  SDSS images.  We randomly chose the structural parameters of the
  artificial galaxies to cover the ranges obtained for our galaxies
  with
\begin{equation}
0.1 \leq r_{\rm e} \leq 13~{\rm kpc},\ 0.4 \leq q_{\rm bulge} \leq 1.0,{\rm and}\ 
0.5 \leq n \leq 6.5
\end{equation}
for the S\'ersic bulges, 
\begin{equation}
1 \leq h\leq 8~{\rm kpc}\ {\rm and}\ 0.2 \leq q_{\rm disc} \leq 0.9
\end{equation}
for Freeman disc, 
\begin{equation}
3 \leq r_{\rm bar} \leq 21~{\rm kpc}\ {\rm and}\ 0.2 \leq q_{\rm bar} \leq 0.8 
\end{equation}
for the Ferrers bars, and
\begin{equation}
q_{\rm bar} \leq q_{\rm disc} \leq q_{\rm bulge}.
\end{equation}
We assumed the artificial galaxies to be at a distance of 80 Mpc
corresponding to a spatial scale of 388 pc arcsec$^{-1}$. We added to
the simulated images a background level (120 ADU) and photon noise to
yield a signal-to-noise ratio ($S/N$) similar to that of the SDSS
ones. Finally, the simulated images were convolved with a Moffat PSF
with $FWHM = 2.77$ pixels and $\beta = 3.05$. We analysed the images
of the artificial galaxies with {\sc gasp2d} as if they were real and
the systematic and typical errors for the best-fitting parameters were
derived as in \citet{morelli2015a}. The computed errors account for
the systematic as well as the parameter covariance errors. However,
they do not include possible deviations from the actual solution due
to the presence of additional galaxy components, such as spiral arms
or rings.

\section{Long-slit spectroscopy}
\label{sec:spectroscopy}

\subsection{Observations and data reduction}
\label{sec:observations}

We carried out the spectroscopic observations of the sample galaxies
with 3.6-m TNG telescope at the Observatorio del Roque de los
Muchachos in La Palma (Spain) on 2012 January 20-23. The low
resolution spectrograph (DOLORES) mounted the volume phase holographic
VHR-V grism with 566 grooves mm$^{-1}$ in combination with the $0.7$
arcsec $\times$ 8.1 arcmin slit and the thinned back-illuminated E2V
CCD with $2048\times2048$ pixels of $13.5\times13.5$ $\mu m^2$. The
spectral range covered from 4700 to 6750 \AA. The reciprocal
dispersion was 0.95 \AA\ pixel$^{-1}$ and the measured instrumental
dispersion after the wavelength calibration was $2.7$
\AA\ (FWHM). This corresponds to $\sigma_{\rm inst}\sim 60$ \kms\ at
5725 \AA . The angular sampling was 0.252 arcsec pixel$^{-1}$.

We observed the sample galaxies along the major axis. The value of the
seeing FWHM measured on the guide star during the observation ranged
between 0.7 and 1.5 arcsec. The integration time of each exposure,
total integration time, and slit position angle of the galaxy spectra
are given in Table~\ref{tab:log}.
We also observed a number giant stars selected from the sample by
\citet{wortetal94} to use their spectra as templates for
calibrating the line-strength indices. In addition, we obtained
different spectra of a least one spectrophotometric standard star per
night to calibrate the flux of the galaxy and template star. 

\renewcommand{\tabcolsep}{4pt}
\begin{table}
\caption{Log of the spectroscopic observations. The columns show the
  following. (2): date; (3): number and exposure time of the single
  exposures; (4): total exposure time; (5): slit position angle.}
\label{tab:log}
\begin{tabular}{lcccc}
\hline
\noalign{\smallskip}
\multicolumn{1}{c}{Galaxy} &
\multicolumn{1}{c}{Date} &
\multicolumn{1}{c}{Single Exp. T.} &
\multicolumn{1}{c}{Total. Exp. T.} &
\multicolumn{1}{c}{PA} \\
\multicolumn{1}{c}{ } &
\multicolumn{1}{c}{ } &
\multicolumn{1}{c}{(s)} &
\multicolumn{1}{c}{(h)} &
\multicolumn{1}{c}{($^\circ$)} \\
\multicolumn{1}{c}{(1)} &
\multicolumn{1}{c}{(2)} &
\multicolumn{1}{c}{(3)} &
\multicolumn{1}{c}{(4)} &
\multicolumn{1}{c}{(5)} \\
\hline
\noalign{\smallskip}
CGCG 034-050 &$ 2012-01-21$ & $3\times2400$ &$2.0$ &$   37.1$  \\
CGCG 088-060 &$ 2012-01-23$ & $4\times2400$ &$2.6$ &$   32.5$ \\
CGCG 152-078 &$ 2012-01-23$ & $3\times2400$ &$2.0$ &$  120.0$ \\
CGCG 206-038 &$ 2012-01-22$ & $3\times2400$ &$2.0$ &$  150.0$ \\
IC 2473      &$ 2012-01-21$ & $3\times2400$ &$2.0$ &$   95.9$ \\
NGC 2503     &$ 2012-01-23$ & $4\times2400$ &$2.6$ &$    4.5$ \\
NGC 2712     &$ 2012-01-20$ & $3\times2400$ &$2.0$ &$    4.3$ \\
NGC 2955     &$ 2012-01-20$ & $5\times1800$ &$2.5$ &$  159.4$ \\
UGC 4000     &$ 2012-01-20$ & $3\times2400$ &$2.0$ &$   37.4$ \\
UGC 4341     &$ 2012-01-22$ & $3\times2400$ &$2.0$ &$   45.3$ \\
UGC 5026     &$ 2012-01-22$ & $3\times2400$ &$2.0$ &$   22.8$ \\
UGC 5184     &$ 2012-01-22$ & $3\times2400$ &$2.0$ &$   84.8$ \\
\noalign{\smallskip} 
\hline
\noalign{\smallskip}
\end{tabular}
\end{table}

The spectroscopic data reduction was performed using standard {\sc
  iraf} routines as done in \citet{morelli2012}.  In this case we
corrected the rebinned spectra for the wavelength shift arising from
DOLORES flexures, which produce shifts of the arc-lamp emission lines
as a function of derotator angle \citep{marinoni2013}. We finally
measured an error of 0.3 \AA\ in the wavelength calibration resulting
in an accuracy of $\sim2$ \kms\ at 5725 \AA.

\subsection{Stellar kinematics}
\label{sec:kinematics}

We measured the line-of-sight velocity distribution (LOSVD) of the
stellar component of the sample galaxies from the absorption lines in
the observed wavelength range using the Penalized Pixel Fitting
\citep[{\sc ppxf},][]{capems04} and Gas and Absorption Line
Fitting \citep[{\sc gandalf},][]{sarzetal06} {\sc idl} codes which we
adapted to deal with TNG spectra. The LOSVD was assumed to be a
Gaussian plus third- and fourth-order Gauss-Hermite polynomials
\citep{vdmarfran93, Gerhard93}.
 
We rebinned each galaxy spectrum along the dispersion direction to a
logarithmic scale, and along the spatial direction to obtain a
$S/N\geq20$ per resolution element. For each radial bin we built an
optimal template spectrum by convolving a linear combination of simple
stellar population (SSP) spectra \citep{vazdekisetal2010} based on the
Medium Resolution Isaac Netwon Telescope Library of Empirical Spectra
\citep[MILES,][]{sancetal06lib} with the LOSVD in order to fit the
galaxy spectrum. The optimal template spectrum and LOSVD moments were
obtained by $\chi^2$ minimisation in pixel space. Before fitting the
SSP spectra were logarithmically rebinned and dereshifted to rest
frame. Moreover, we degraded the spectral resolution of the galaxy
spectrum by convolving it with a Gaussian function in order to match
the MILES spectral resolution \citep[$\rm FWHM=2.5$
  \AA,][]{falcon_miles2011, beifiori11,}. In addition, we
simultaneously fitted the ionized-gas emission lines detected with a
$S/N>3$.  We masked the bad pixels coming from imperfect subtraction
of sky emission lines and excluded them from the fitting procedure. We
added a low-order multiplicative Legendre polynomial to correct for
the different shape of the continuum in the spectra of the galaxy and
optimal template due to reddening and large-scale residuals of
flatfielding and sky subtraction.

By measuring the LOSVD moments in all the available radial bins along
the spatial direction we derived the radial profiles of the
LOS velocity $v$, velocity dispersion $\sigma$, third- and
fourth-order Gauss-Hermite moments $h_3$ and $h_4$ of the stars. We
estimated the uncertainties on the LOSVD moments running Monte Carlo
simulations. For each radial bin we built a set of simulated galaxy
spectra by randomly perturbing the best-fitting galaxy spectrum. We
added to the counts of each pixel of the best-fitting galaxy spectrum
a random value chosen from a Gaussian distribution with a mean of zero
and the same standard deviation of the difference between the observed
and best-fitting galaxy spectra in the wavelength range used in the
fit and excluding the emission lines. We measured the simulated
spectra as if they were real. For each LOSVD moment we adopted as
error the standard deviation of the distribution of the values derived
for the simulated galaxy spectra. We found no bias of the {\sc ppxf}
method with the adopted instrumental setup and spectral sampling in
the ranges of $S/N$ and $\sigma$ which characterise the spectra of the
sample galaxies. Indeed, the values of $h_3$ and $h_4$ measured in a
set of simulated galaxy spectra obtained by convolving the
best-fitting galaxy spectrum with a Gauss-Hermite LOSVD, and adding
photon, readout, and sky noise to mimic actual observations differ
from the intrinsic ones only within the estimated errors.
We give the measured stellar kinematics in
Table~\ref{tab:kinematics_indices} and plot the folded profiles of
$v$, $\sigma$, $h_3$, and $h_4$ as a function of radius in
Fig.~\ref{fig:kinematics}.

The regularity and symmetry of the kinematic profiles is further
indication that the sample galaxies are isolated and have not recently
experienced merging or interaction with other galaxies \citep[e.g.][]{barton1999}.

\subsection{Line-strength indices}
\label{sec:linestrength}

We measured the Mg, Fe, and \Hb\ line-strength indices of the Lick/IDS
system \citep{faberetal85, wortetal94}, the iron index
$\rm{\left<Fe\right> = (Fe5270 + Fe5335)/2}$ \citep{gorgetal90},
the combined magnesium-iron index $[{\rm MgFe}]^{\prime}=\sqrt{{\rm
    Mg}\,b\,(0.72\times {\rm Fe5270} + 0.28\times{\rm Fe5335})}$
\citep{thmabe03} and their errors by following the same procedure and
method adopted by \citet{moreetal04, morelli2015a}.
The offsets between our line-strength measurements and Lick/IDS
line-strength values \citep{wortetal94} were smaller than the mean
error of the differences and therefore we did not apply any offset
correction to our line-strength measurements.
We list in Table \ref{tab:kinematics_indices} and plot in
Fig.~\ref{fig:indices} the measured values of \Hb , \MgFe, \Fe, \Mgb,
and \Mgd\ as a function of radius for all the sample galaxies.

\begin{figure}
\centering
\includegraphics[angle=0,width=0.44\textwidth]{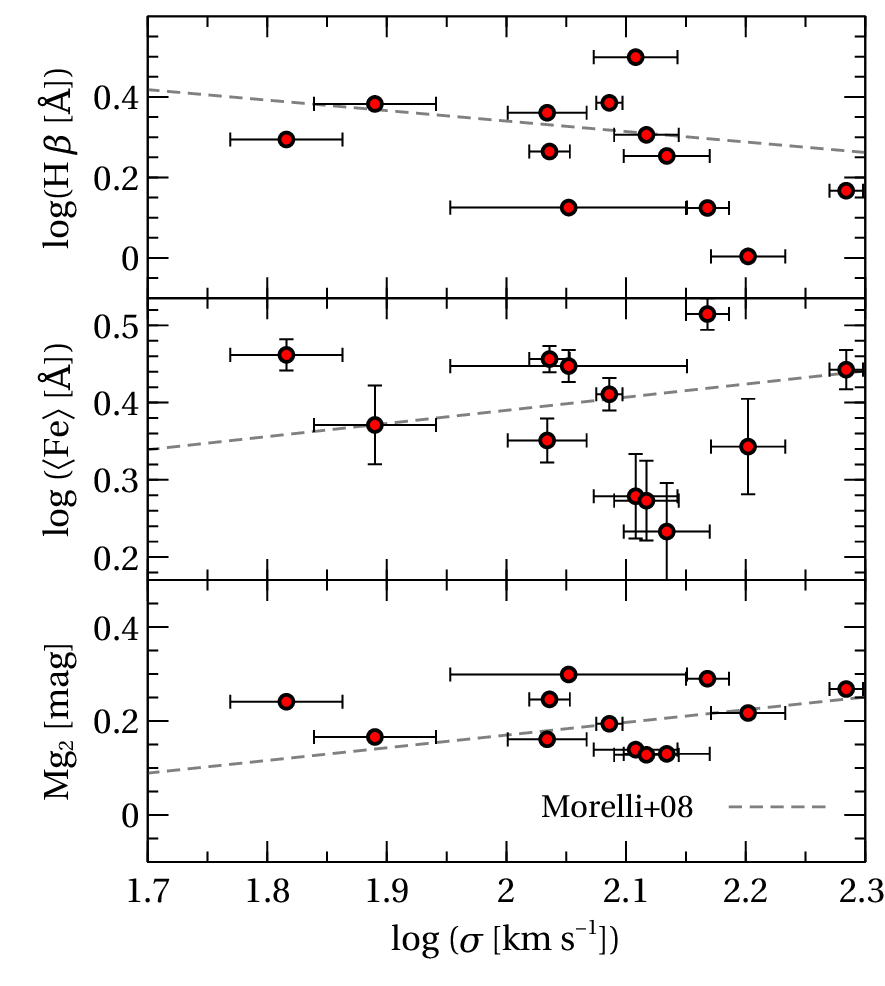}
\caption{{ Central values of the \hb\ (upper panel), \Fe\ (middle
    panel), and \Mgd\ (lower panel) line-strength indices measured
    over an aperture of $0.3 r_{\rm e}$ for the sample galaxies as a
    function of their central velocity dispersion. In each panel the
    blue dashed line represents the correlation found by
    \citet{moreetal08} for the bulges of galaxies in groups and
    clusters.}
\label{fig:censmg2fe}}
\end{figure}

We derived the central values of \Mgb, \Mgd, \Hb, \Fe, and \MgFe\/ as
a $S/N$ weighted mean of the values measured within an aperture of
radius $0.3\,r_{\rm e}$ along the major axis of the
galaxies. Similarly, we derived the central value of the velocity
dispersion. The central values of $\sigma$, \Mgb, \Mgd, \Hb, \Fe, and
\MgFe\/ are reported in Tab.~\ref{tab:central_values}.
%
\begin{table*}
\caption{Central values of the velocity dispersion and line-strength
  indices of the sample galaxies measured within an aperture of radius
  $0.3\,r_{\rm e}$.}
\begin{tabular}{lrcccccc}
\hline
\noalign{\smallskip}
\multicolumn{1}{c}{Galaxy} &
\multicolumn{1}{c}{$\sigma$} &
\multicolumn{1}{c}{\Fe} &
\multicolumn{1}{c}{\MgFe} &
\multicolumn{1}{c}{\Mgd} &
\multicolumn{1}{c}{\Mgb} &
\multicolumn{1}{c}{\Hb}  \\
\multicolumn{1}{c}{ } &
\multicolumn{1}{c}{(\kms)} &
\multicolumn{1}{c}{(\AA)} &
\multicolumn{1}{c}{(\AA)} &
\multicolumn{1}{c}{(mag)} &
\multicolumn{1}{c}{(\AA)} &
\multicolumn{1}{c}{(\AA)}  \\
\multicolumn{1}{c}{(1)} &
\multicolumn{1}{c}{(2)} &
\multicolumn{1}{c}{(3)} &
\multicolumn{1}{c}{(4)} &
\multicolumn{1}{c}{(5)} &
\multicolumn{1}{c}{(6)} &
\multicolumn{1}{c}{(7)}  \\
\noalign{\smallskip}
\hline
\noalign{\smallskip}
CGCG 034-050 & $112.6 \pm 28.8$ & $2.801 \pm 0.137$ & $3.630 \pm 0.036$ & $0.299 \pm 0.004$ & $4.642 \pm 0.144$ & $1.334 \pm  0.117$ \\
CGCG 088-060 & $ 65.4 \pm  7.5$ & $2.896 \pm 0.131$ & $3.400 \pm 0.032$ & $0.241 \pm 0.004$ & $3.932 \pm 0.144$ & $1.969 \pm  0.114$ \\
CGCG 152-078 & $108.6 \pm  4.4$ & $2.860 \pm 0.132$ & $3.465 \pm 0.023$ & $0.246 \pm 0.005$ & $4.081 \pm 0.102$ & $1.838 \pm  0.092$ \\
CGCG 206-038 & $147.2 \pm  6.1$ & $3.271 \pm 0.142$ & $3.960 \pm 0.041$ & $0.290 \pm 0.004$ & $4.775 \pm 0.147$ & $1.331 \pm  0.116$ \\
IC 2473      & $136.2 \pm 11.9$ & $1.710 \pm 0.304$ & $2.036 \pm 0.076$ & $0.130 \pm 0.006$ & $2.505 \pm 0.244$ & $1.792 \pm  0.220$ \\
NGC 2503     & $ 77.6 \pm  9.7$ & $2.350 \pm 0.299$ & $2.623 \pm 0.097$ & $0.166 \pm 0.007$ & $2.857 \pm 0.247$ & $2.413 \pm  0.236$ \\
NGC 2712     & $108.2 \pm  8.6$ & $2.244 \pm 0.155$ & $2.653 \pm 0.034$ & $0.161 \pm 0.004$ & $3.052 \pm 0.166$ & $2.294 \pm  0.116$ \\
NGC 2955     & $128.2 \pm 10.7$ & $1.900 \pm 0.265$ & $2.236 \pm 0.061$ & $0.139 \pm 0.007$ & $2.616 \pm 0.207$ & $3.153 \pm  0.168$ \\
UGC 4000     & $147.4 \pm 10.6$ & $2.203 \pm 0.337$ & $2.982 \pm 0.182$ & $0.217 \pm 0.009$ & $3.981 \pm 0.362$ & $1.009 \pm  0.265$ \\
UGC 4341     & $192.5 \pm  6.2$ & $2.771 \pm 0.167$ & $3.610 \pm 0.060$ & $0.268 \pm 0.005$ & $4.595 \pm 0.200$ & $1.469 \pm  0.173$ \\
UGC 5026     & $121.9 \pm  3.1$ & $2.575 \pm 0.145$ & $2.975 \pm 0.029$ & $0.194 \pm 0.004$ & $3.371 \pm 0.134$ & $2.428 \pm  0.116$ \\
UGC 5184     & $130.9 \pm  8.4$ & $1.875 \pm 0.237$ & $2.405 \pm 0.066$ & $0.128 \pm 0.007$ & $2.803 \pm 0.232$ & $2.022 \pm  0.201$ \\
\hline
\noalign{\bigskip}
\label{tab:central_values}
\end{tabular}
\end{table*}
{Fig.~\ref{fig:censmg2fe} shows the correlations between the
  central values of the \hb , \Fe , and \Mgd\/ line-strength indices
  with the central velocity dispersion for the sample galaxies. They
  are compared with the results obtained by \citet{moreetal08} for the
  bulges of group and cluster galaxies.}

Tight \Mgd$-\sigma$ and \Fe$-\sigma$ correlation have been predicted
from theoretical models \citep[e.g.,][]{kodaetal98} and observed in
spheroids of galaxies being them early-type galaxies \citep[see][]{
  fishetal96, jorgen99, tragetal98, mehletal03} or bulges of spiral
galaxies \citep{idiaetal96, prugetal01, procetal02, gandetal07,
  morelli2012}. The correlations, in general, show that more massive
systems host a more metal-rich stellar population
\citep{Thomasetal2010}. The values of the indices we obtained
(Fig.~\ref{fig:censmg2fe}) are consistent with the relations found for
a similar sample of cluster galaxies by \citet{moreetal08}. We
  also found an anticorrelation between \hb\ and $\sigma$ in our
  sample of isolated galaxies as we did in \citet{moreetal08} and
  \citet{morelli2012}. Less massive galaxies are on average younger
  than more massive galaxies and, in spite of the large scatter, this
  result is consistent with our previous works and with the findings
  of \citet{gandetal07}.

\begin{figure}
  \centering
  \includegraphics[width=0.51\textwidth]{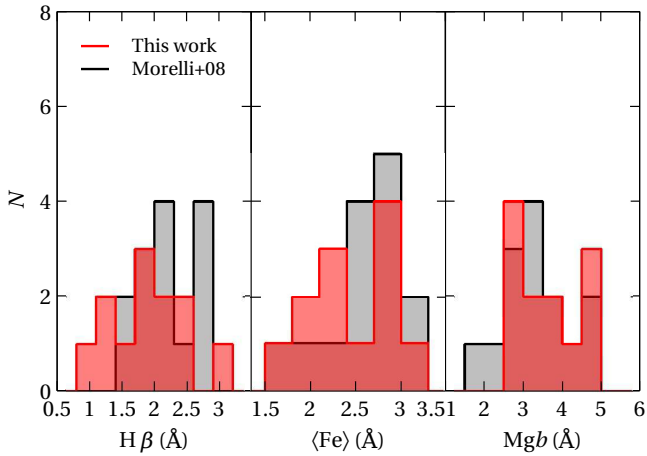}
  \caption{ Distribution of the central values of the
    \hb\ (left-hand panel), \Fe\/ (middle panel), and \Mgb\/
    (right-hand panel) for the sample bulges (red histograms).  The
    distribution of the same quantities for the bulges of group and
    cluster galaxies studied by \citet{moreetal08} is plotted for a
    comparison (grey histograms). }
  \label{fig:histo_compare_centre_index}
\end{figure}

In Fig.~\ref{fig:histo_compare_centre_index}, the distribution of the
central values of \hb\/, \Fe\/, and \Mgb\/ line-strength indices for
the bulges of our sample of isolated galaxies is compared with the
values obtained for the bulges of group and cluster galaxies of
\citet{moreetal08}. The distributions are similar, although the \Mgb\/
values in the isolated galaxies seems to be slightly shifted to the
higher end.

\section{Properties of the stellar populations}
\label{sec:populations}

\subsection{Central values of age, metallicity, and $\alpha/$Fe enhancement}
\label{sec:central}

\begin{figure}
\centering
\includegraphics[width=0.51\textwidth]{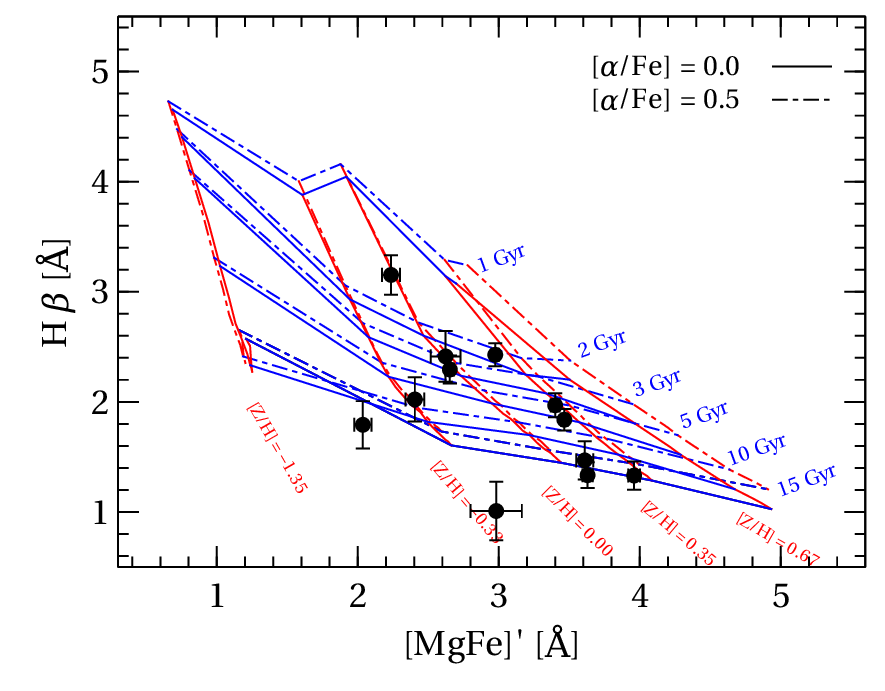} 
\includegraphics[width=0.51\textwidth]{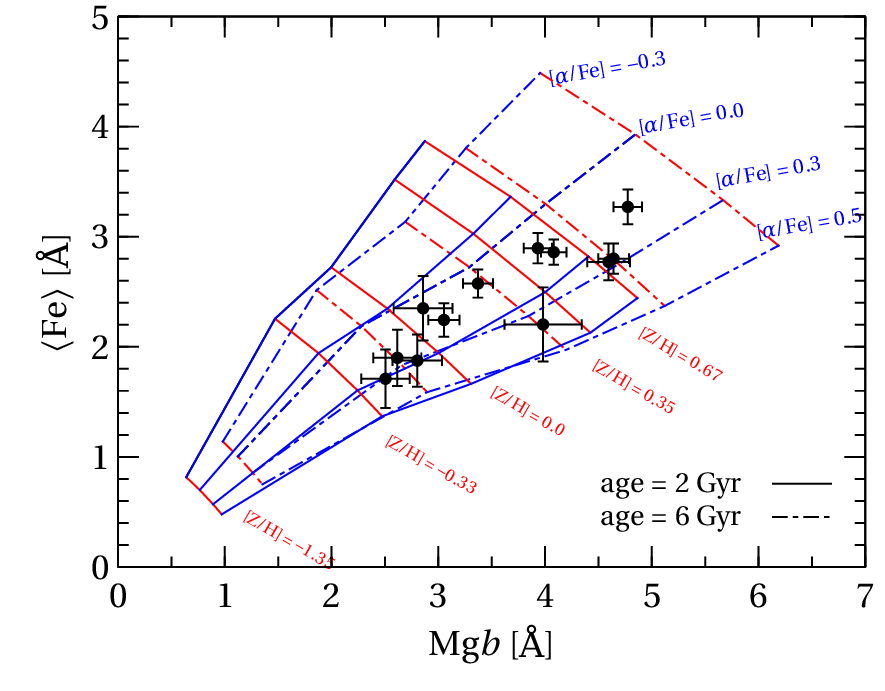}
\caption{Distribution of the central values of \Hb\ and \MgFe\ indices
  (top panel) and \Fe\ and \Mgb\ indices (bottom panel) for the sample
  galaxies. The lines indicate the models by \citet{thmabe03}. }
  \label{fig:idb_hbmfdiag}
\end{figure}

\begin{figure}
  \centering
  \includegraphics[width=0.51\textwidth]{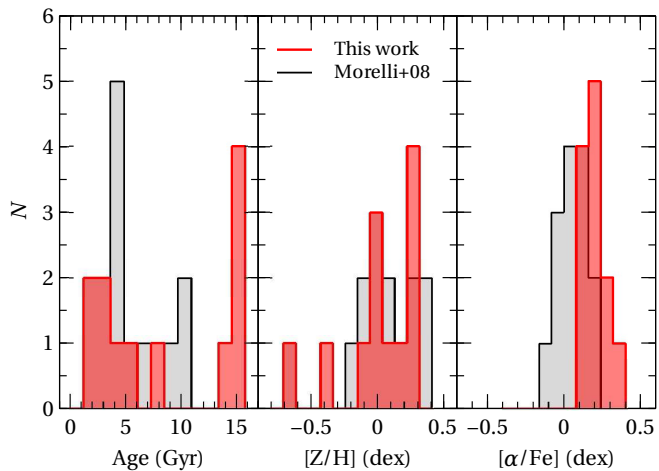}
  \caption{Distribution of the mean age (left-hand panel), total
    metallicity (central panel), and total $\alpha$/Fe enhancement
    (right-hand panel) for the stellar population of the bulges of the
    sample galaxies (red histograms). The distribution of the same
    quantities for the bulges of group and cluster galaxies studied by
    \citet{moreetal08} is plotted for a comparison (grey
    histograms).}
  \label{fig:idb_histo_stpop}
\end{figure}

We derived the stellar population properties in the centre of the
sample galaxies by comparing the measurements of the line-strength
indices with the model predictions by \citet{thmabe03} for the single
stellar population as a function of age, metallicity, and $\alpha/$Fe
enhancement. We plot the central values of \Hb, \MgFe, \Fe, and
\Mgb\ and model predictions in Fig.~\ref{fig:idb_hbmfdiag}.
%
%
We calculated the age, metallicity, and $\alpha$/Fe enhancement in the
centre of the sample galaxies from the central values of line-strength
indices given in Table \ref{tab:central_values} by following
\citet{morelli2012}.
%
\begin{table}
\caption{Mean age, total metallicity, and total $\alpha/$Fe
  enhancement of the stellar populations of the bulges of the sample
  galaxies.}
\begin{center}
\begin{small}
\begin{tabular}{lrrr}
\hline
\noalign{\smallskip}
\multicolumn{1}{c}{Galaxy} &
\multicolumn{1}{c}{Age} &
\multicolumn{1}{c}{\ZH} &
\multicolumn{1}{c}{\aFe} \\
\noalign{\smallskip}
\multicolumn{1}{c}{} &
\multicolumn{1}{c}{(Gyr)} &
\multicolumn{1}{c}{(dex)} &
\multicolumn{1}{c}{(dex)} \\
\noalign{\smallskip}
\multicolumn{1}{c}{(1)} &
\multicolumn{1}{c}{(2)} &
\multicolumn{1}{c}{(3)} &
\multicolumn{1}{c}{(4)} \\
\noalign{\smallskip}
\hline
\noalign{\smallskip}
CGCG~034-050 & $15.0\pm3.2$ & $ 0.05\pm0.07$ & $0.22\pm0.05$ \\
CGCG~088-060 & $ 4.0\pm1.4$ & $ 0.30\pm0.09$ & $0.12\pm0.06$ \\
CGCG~152-078 & $ 5.4\pm1.7$ & $ 0.27\pm0.07$ & $0.15\pm0.06$ \\
CGCG~206-038 & $14.9\pm3.8$ & $ 0.28\pm0.09$ & $0.11\pm0.05$ \\
IC~2473      & $ >15.0    $ & $-0.65\pm0.10$ & $0.21\pm0.15$ \\
NGC~2503     & $ 2.6\pm1.3$ & $ 0.00\pm0.12$ & $0.08\pm0.14$ \\
NGC~2712     & $ 3.2\pm0.8$ & $-0.05\pm0.06$ & $0.16\pm0.07$ \\
NGC~2955     & $ 1.7\pm0.2$ & $-0.04\pm0.07$ & $0.24\pm0.13$ \\
UGC~4000     & $ >15.0    $ & $-0.19\pm0.12$ & $0.33\pm0.11$ \\
UGC~4341     & $13.8\pm4.8$ & $ 0.12\pm0.11$ & $0.22\pm0.06$ \\
UGC~5026     & $ 1.9\pm0.4$ & $ 0.26\pm0.06$ & $0.16\pm0.05$ \\
UGC~5184     & $ 7.7\pm2.5$ & $-0.29\pm0.07$ & $0.26\pm0.11$ \\
\noalign{\smallskip}
\hline
\noalign{\medskip}
\end{tabular}
\end{small}
\label{tab:agemetalfa}
\end{center}
\end{table}
%
We list the central values of age, metallicity, and $\alpha$/Fe
enhancement and their corresponding errors in
Table~\ref{tab:agemetalfa} and show the histograms of their number
distribution in Fig.~\ref{fig:idb_histo_stpop}.

We found that the ages of the sample bulges have a bimodal
distribution (Fig.~\ref{fig:idb_histo_stpop}, left-hand panel) with
about half of them being old (14-15 Gyr) and the remaining ones
characterised by a young-to-intermediate age (1-5 Gyr). Such a large
range of ages for the bulges of isolated galaxies is consistent with
the results obtained by \citet{katkov2015}. We measured prominent
emission lines, which are indicative of on-going star formation, only
in the centre of sample galaxies with younger bulges. 
The metallicity of the sample bulges spans a large range of values
(Fig.~\ref{fig:idb_histo_stpop}, middle panel) from high
(\ZH$\,=\,0.3$ dex) to sub-solar metallicity (\ZH$\,=-0.7$ dex).  
On the contrary, the $\alpha$/Fe enhancement of the sample bulges is
characterised by a narrow distribution of super-solar values
(\aFe$\,=\,0.1-0.3$ dex) peaked at \aFe$\,=\,0.2$ dex
(Fig.~\ref{fig:idb_histo_stpop}, right-hand panel).

In Fig.~\ref{fig:idb_histo_stpop}, we compare the number distributions
of the age, metallicity, and $\alpha$/Fe enhancement of the bulges of
our sample of isolated galaxies to those found by \citet{moreetal08},
who carried out a similar analysis on the bulge stellar populations of
galaxies in groups and clusters.
The isolated galaxies show a large fraction of very old bulges, which are
not observed in group and cluster galaxies.
There is no difference in the bulge metallicity distribution for most
of the isolated galaxies with respect to the group and cluster
galaxies, except for a couple of bulges with a very low metallicity.
The most significant difference between the bulges of isolated
galaxies and those of group and cluster galaxies is represented by the
very different distributions of their \aFe\ ratios. The bulges of
isolated galaxies have systematically higher values of $\alpha$/Fe
enhancement. This implies a difference in the star-formation timescale
with the inner regions of isolated bulges being formed more rapidly
with respect to bulges in high density environments.

The metallicity and $\alpha$/Fe enhancement are well correlated with
the central velocity dispersion in early-type galaxies
\citep{mehletal03, spoletal10} and in bulges of high
\citep{gandetal07, moreetal08} and low surface-brightness galaxies
\citep{morelli2012}. Cosmological hydrodynamic simulations
\citep{deluetal04, tassetal08} and chemodinamycal models
\citep{matteucci1994, kawgib03, kobayashi04} demonstrated that these
relations are the result of a mass-dependent star formation
efficiency. High mass galaxies have a higher efficiency in converting
gas-phase metals into new stars, giving rise to less prolonged star
formation events and higher $\alpha/$Fe enhancements. Our
  findings suggest that these results can be extended also to our
  isolated galaxies. Our conclusion is
that the most massive bulges of our sample are more metal rich and
characterised by a shorter star-formation timescale.

\begin{figure}
\centering
\includegraphics[angle=0,width=0.5\textwidth]{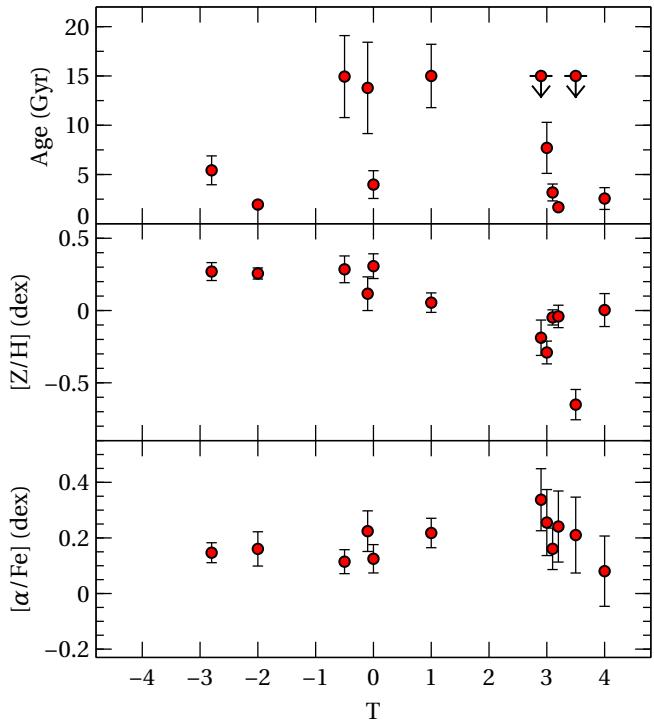}\\
\caption[]{Mean age (upper panel), total metallicity (middle panel),
  and total $\alpha/$Fe enhancement (lower panel) of the stellar
  populations of the bulges of the sample galaxies as a function of
  the galaxy morphological type.
\label{fig:T_agemetalfa}}
\end{figure}

Finally, we looked for a possible correlation between the stellar
population properties of the sample bulges and the morphological type
of their host galaxies. Indeed, very shallow correlations were found
by \citet{gandetal07} and \citet{moreetal08} whereas \citet{thda06}
and \citet{morelli2012} did not observe any trend. We did not find any
correlation between the galaxy morphological type and age,
metallicity, or $\alpha$/Fe enhancement of bulges in isolated galaxies
(Fig.~\ref{fig:T_agemetalfa}).
%
The absence of these correlations could be an indication
that the stellar populations of the bulges and discs in isolated
galaxies had an independent evolution.

Three out of 14 galaxies in our sample turned out to have a bar
  (Fig. \ref{fig:decomposition}). We found no correlation between the
  stellar population properties of the sample bulges and the presence
  of the bar. This could be a consequence that the bulge contribution
  is always dominating the light distribution inside \rbd . Similar
  results were found in \citet{seidel2016} and are consistent with the
  theoretical predictions by \citet{wozniak2007} and observational
  findings by \citet{perezbazquez2011} and
  \citet{jamespercival2016}. 
\begin{figure}
  \centering
  \includegraphics[width=0.51\textwidth]{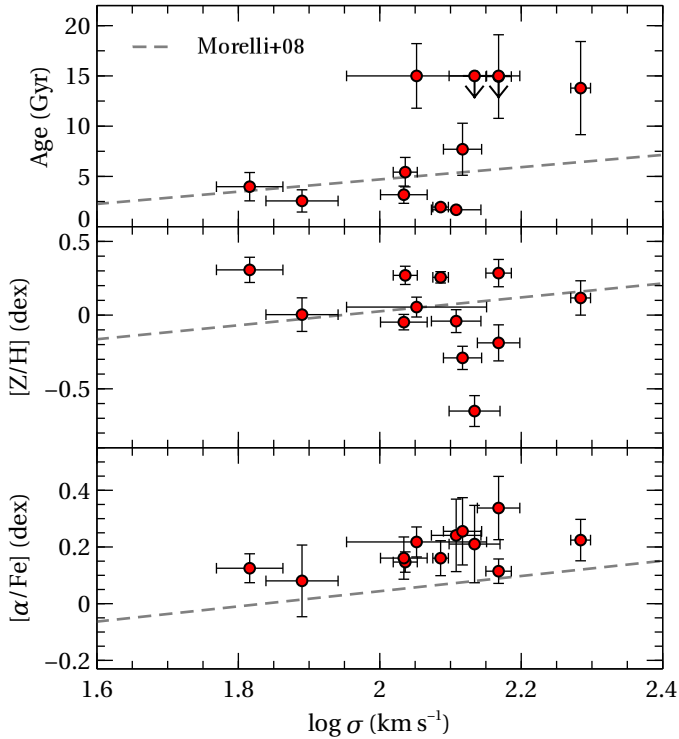}
  \caption{Mean age (upper panel), total metallicity (middle panel),
    and total $\alpha$/Fe enhancement (lower panel) of the stellar
    populations of the bulges of the sample galaxies as a function of
    the central velocity dispersion. In each panel the dashed line
    represents the correlation found by \citet{moreetal08} for the
    bulges of group and cluster galaxies. }
  \label{fig:idb_cstpop_sig}
\end{figure}

\subsection{Radial gradients of age, metallicity, and $\alpha/$Fe enhancement}
\label{sec:gradient}

Theoretical models predict different radial gradients of age,
metallicity, and $\alpha$/Fe enhancement as consequence of different
galaxy formation mechanisms. Therefore, the radial trends of the
stellar population properties within bulges are one of the most
important diagnostics to understand the processes regulating their
assembly.

From the theoretical point of view a steep metallicity gradient is
expected in bulges forming through pure monolithic collapse scenarios
\citep{eglbsa62, lars74, aryo87} and their modern versions
\citep{kawaetal01, kobayashi04}. Gas dissipation with subsequent
occurrence of star formation and blowing of galactic winds produce the
metallicity gradient. A relation between the steepening of the
gradient and mass is also expected \citep{pipietal10}. A strong
gradient in the $\alpha$/Fe enhancement is expected too
\citep{fesi02}. However, \citet{pipietal08} suggested that together
with the outside-in scenario, other important processes need to be
considered to explain the origin of the abundance ratios. In addition,
they pointed out that the interplay between the star formation
timescale and the gas flows is very important, since it acts both in
flattening the $\alpha$/Fe gradient and in enabling the galaxy to
harbour a metallicity gradient.

On the contrary, the metallicity gradient is expected to be very
shallow or absent when it is investigated in merger-based models
\citep{besh99}. This is due to the fact that mergers mix up all the
stars in the galaxy removing the gradients. Secondary star formation,
eventually happening during a wet merger, only rarely steepens the
gradient \citep{kobayashi04}. If this happens, the age radial profile
  should maintain a clear signature for several Gyr
  \citep{hopketal09} whereas dry mergers flatten all the pre-existing
gradients \citep{pipietal10}.

Secular evolution predicts a variety of opposite results for the
resulting gradients of the stellar populations as consequence of the
relative importance of different mechanisms acting in this scenario as
discussed in \citet{mooretal06}.

An issue in measuring the gradients of the age, metallicity, and
$\alpha/$Fe enhancement in bulges could be the contamination of their
stellar population by the light coming from the underlying disc
stellar component. To account for the light contamination from the
disc component, we mapped the radial gradients up to the radius
\rbd\ where the surface brightness contribution of the bulge component
is equal to the contribution from the other components. Although it
was not possible to completely remove the contribution from the disc,
limiting the analysis to the radius \rbd\ ensured us that the degree
of contamination is the same across all the sample of galaxies.

\begin{figure}
  \centering
  \includegraphics[width=0.51\textwidth]{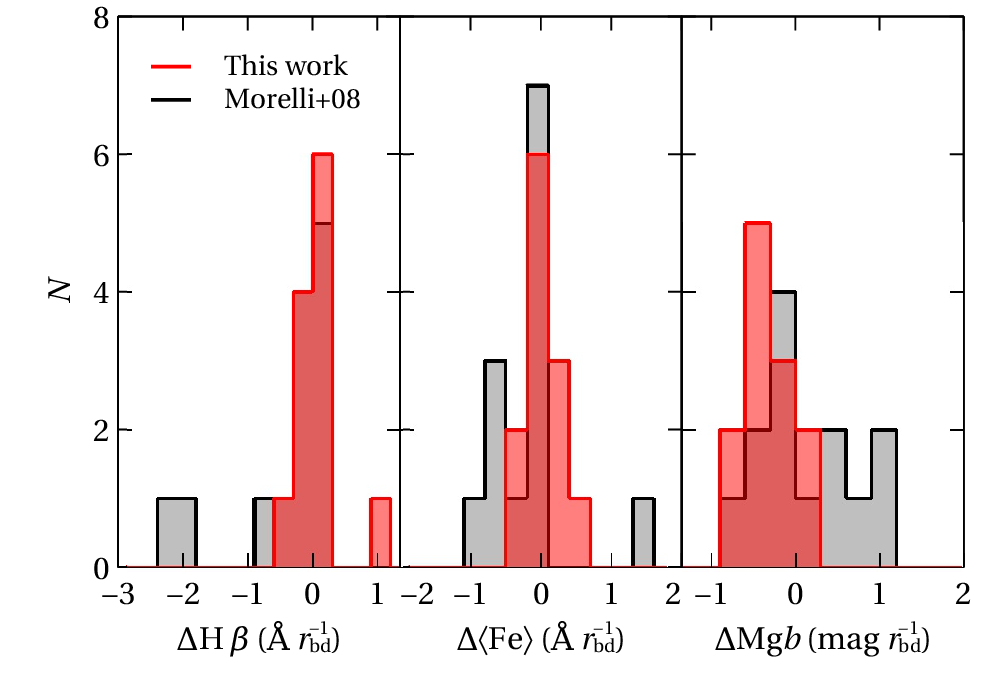}
  \caption{ Distribution of the gradients of \Hb\ (left-hand
    panel), \Fe\ (central panel), and \Mgb\ (right-hand panel) for the
    sample bulges (red histograms).  The distribution of the same
    quantities for the bulges of group and cluster galaxies studied by
    \citet{moreetal08} is plotted for a comparison (grey
    histograms). }
  \label{fig:indices_histo_grad}
\end{figure}

Bulges can have a very complicate and peculiar structure
\citep{peletier2007,morelli2010,corsini2012}. We have carefully
inspected the 2D photometric decomposition and stellar kinematics of
the sample galaxies to look for the presence of distinct structures in
our bulges, but, we did not find any evidence of them. Furthermore to
avoid possible misleading results due to the presence of spiral arms
or small star-forming regions, we adopted the following process to
derive the gradients of the line-strength indices along the major axis
of the sample galaxies.  Following \citet{mehletal03}, we derived the
gradient of each line-strength index by fitting the data inside \rbd .
The gradient was derived as the difference between the best-fitting
value at \rbd\ and in the central region within
0.1\reff\ \citep{moreetal08, morelli2012, morelli2015a}. The resulting
values of gradients for the \Hb, \Fe, and \Mgb\ line-strength indices
are reported in Table \ref{tab:ind_grad}
\renewcommand{\tabcolsep}{2pt}
\begin{table}
\caption{Gradients of the \Hb, \Fe, and \Mgb\/ line-strength
  indices of the sample bulges derived from the values in the centre
  and at \rbd , where the surface-brightness contributions of the
  bulge and remaining components are equal.}
\begin{center}
\begin{small}
\begin{tabular}{lrrrr}
\hline
\noalign{\smallskip}
\multicolumn{1}{c}{Galaxy} &
\multicolumn{1}{c}{\rbd} &
\multicolumn{1}{c}{$\Delta$ \Hb} &
\multicolumn{1}{c}{$\Delta$ \Fe} &
\multicolumn{1}{c}{$\Delta$ \Mgb\ } \\
\noalign{\smallskip}
\multicolumn{1}{c}{} &
\multicolumn{1}{c}{(kpc)} &
\multicolumn{1}{c}{(\AA\ r$^{-1}_{\rm bd}$)} &
\multicolumn{1}{c}{(\AA\ r$^{-1}_{\rm bd}$)} &
\multicolumn{1}{c}{(mag r$^{-1}_{\rm bd}$)} \\
\noalign{\smallskip}
\multicolumn{1}{c}{(1)} &
\multicolumn{1}{c}{(2)} &
\multicolumn{1}{c}{(3)} &
\multicolumn{1}{c}{(4)} &
\multicolumn{1}{c}{(5)} \\
\noalign{\smallskip}
\hline
\noalign{\smallskip}  
CGCG 034-050 & $1.40$ & $ 0.19 \pm 0.16$ & $-0.17 \pm 0.19$ & $-0.47 \pm 0.21$ \\
CGCG 088-060 & $0.73$ & $-0.24 \pm 0.19$ & $-0.34 \pm 0.25$ & $-0.82 \pm 0.24$ \\
CGCG 152-078 & $1.52$ & $ 0.02 \pm 0.14$ & $-0.10 \pm 0.23$ & $-0.37 \pm 0.18$ \\
CGCG 206-038 & $1.10$ & $ 0.03 \pm 0.17$ & $-0.23 \pm 0.21$ & $-0.64 \pm 0.21$ \\
IC 2473      & $1.99$ & $ 0.24 \pm 0.31$ & $ 0.53 \pm 0.46$ & $ 0.21 \pm 0.37$ \\
NGC 2503     & $0.70$ & $ 0.04 \pm 0.33$ & $ 0.25 \pm 0.43$ & $-0.22 \pm 0.35$ \\
NGC 2712     & $0.50$ & $-0.35 \pm 0.19$ & $ 0.03 \pm 0.26$ & $-0.37 \pm 0.26$ \\
NGC 2955     & $1.92$ & $-0.00 \pm 0.24$ & $-0.01 \pm 0.37$ & $ 0.02 \pm 0.28$ \\
UGC 4000     & $2.57$ & $-0.06 \pm 0.40$ & $ 0.14 \pm 0.56$ & $-0.07 \pm 0.53$ \\
UGC 4341     & $0.90$ & $-0.24 \pm 0.25$ & $-0.16 \pm 0.26$ & $-0.31 \pm 0.30$ \\
UGC 5026     & $1.76$ & $ 0.05 \pm 0.17$ & $-0.10 \pm 0.24$ & $-0.21 \pm 0.21$ \\
UGC 5184     & $0.53$ & $ 0.91 \pm 0.35$ & $ 0.17 \pm 0.32$ & $-0.34 \pm 0.31$ \\
\noalign{\smallskip}
\hline
\noalign{\medskip}
\end{tabular}
\end{small}
\label{tab:ind_grad}
\end{center}
\end{table}
and their distributions are shown in
Fig.~\ref{fig:indices_histo_grad}.  With few exceptions, the sample
galaxies display flat radial profiles of \Hb\/ and \Fe\ and a negative
gradient of \Mgb\/ (Table \ref{tab:ind_grad}).

\begin{figure}
  \centering
  \includegraphics[width=0.51\textwidth]{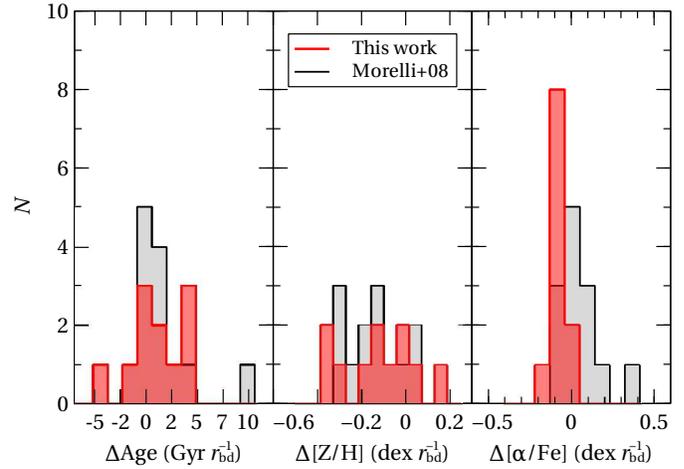}
  \caption{Distribution of the gradients of mean age (left-hand
    panel), total metallicity (central panel), and total $\alpha$/Fe
    enhancement (right-hand panel) for the sample bulges (red
    histograms).  The distribution of the same quantities for the
    bulges of group and cluster galaxies studied by \citet{moreetal08}
    is plotted for a comparison (grey histograms). }
  \label{fig:idb_histo_grad}
\end{figure}

We then converted the derived values of line-strength indices to the
corresponding age, metallicity, and $\alpha/$Fe enhancement of the
stellar population using the models by \citet{thmabe03}, as previously
done for the central values. The gradient of each stellar population
property was derived as the difference between the value derived at
\rbd\ and the central value obtained within 0.1\reff. The
  uncertainties on the resulting gradients were derived as done in
  \citet{morelli2012}.  We report the gradients of age, metallicity,
and $\alpha/$Fe enhancement of the sample bulges are listed in
Table~\ref{tab:agemetalfa_grad} and plot their number distributions in
Fig.~\ref{fig:idb_histo_grad}.

\renewcommand{\tabcolsep}{2pt}
\begin{table}
\caption{Gradients of mean age, total metallicity, and total
  $\alpha/$Fe enhancement of the stellar populations of the sample
  bulges derived from the central values and values at the radius
  \rbd\ where the surface-brightness contributions of the bulge and
  remaining components are equal.}
\begin{center}
\begin{small}
\begin{tabular}{lrrrr}
\hline
\noalign{\smallskip}
\multicolumn{1}{c}{Galaxy} &
\multicolumn{1}{c}{$\Delta$(Age)} &
\multicolumn{1}{c}{$\Delta$(\ZH)} &
\multicolumn{1}{c}{$\Delta$(\aFe)} \\
\noalign{\smallskip}
\multicolumn{1}{c}{} &
\multicolumn{1}{c}{(Gyr r$^{-1}_{\rm bd}$)} &
\multicolumn{1}{c}{(dex r$^{-1}_{\rm bd}$)} &
\multicolumn{1}{c}{(dex r$^{-1}_{\rm bd}$)} \\
\noalign{\smallskip}
\multicolumn{1}{c}{(1)} &
\multicolumn{1}{c}{(2)} &
\multicolumn{1}{c}{(3)} &
\multicolumn{1}{c}{(4)} \\
\noalign{\smallskip}
\hline
\noalign{\smallskip}  
CGCG 034-050 & $-5.11 \pm 4.7$ & $-0.10 \pm 0.07$ & $-0.04 \pm 0.10$ \\
CGCG 088-060 & $ 4.60 \pm 5.7$ & $-0.37 \pm 0.11$ & $-0.12 \pm 0.11$ \\
CGCG 152-078 & $ 0.99 \pm 2.1$ & $-0.21 \pm 0.10$ & $-0.10 \pm 0.09$ \\
CGCG 206-038 & $ 2.25 \pm 4.2$ & $-0.29 \pm 0.08$ & $-0.08 \pm 0.07$ \\
IC 2473      & $           - $ & $  -           $ & $           -  $ \\
NGC 2503     & $-0.12 \pm 4.2$ & $ 0.02 \pm 0.18$ & $-0.10 \pm 0.17$ \\
NGC 2712     & $ 1.00 \pm 2.3$ & $-0.11 \pm 0.08$ & $-0.09 \pm 0.14$ \\
NGC 2955     & $ 0.01 \pm 0.5$ & $-0.00 \pm 0.09$ & $ 0.02 \pm 0.17$ \\
UGC 4000     & $     -       $ & $    -         $ & $              $ \\
UGC 4341     & $ 4.03 \pm 4.2$ & $-0.34 \pm 0.19$ & $-0.06 \pm 0.12$ \\
UGC 5026     & $ 0.14 \pm 0.9$ & $-0.14 \pm 0.07$ & $-0.04 \pm 0.10$ \\
UGC 5184     & $-1.81 \pm 2.1$ & $ 0.17 \pm 0.13$ & $-0.07 \pm 0.19$ \\
\noalign{\smallskip}
\hline
\noalign{\medskip}
\end{tabular}
\end{small}
\label{tab:agemetalfa_grad}
\end{center}
\end{table}

Almost all the sample bulges show a null or very shallow age gradients
with a distribution peaked at $\Delta$(age)$\,=\,0$. The only
exception is the bulge of CGCG~034-050 ($\Delta$(age)$\,=\,-5.1\pm4.7$
Gyr) because the bulges of both CGCG~088-060 and NGC~4341 are
consistent within the errors with $\Delta$(age)$\,=\,0$ in spite of
their large age gradients. Our findings are in agreement with previous
results for early-type galaxies \citep{mehletal03, sancetal06p,
  spoletal10} and late-type bulges \citep{jabletal07}. The number
distribution of the age gradients of the bulges in isolated galaxies
is remarkably similar to that of bulges in group and cluster galaxies
(Fig.~\ref{fig:idb_histo_grad}, left-hand panel) suggesting that the
age distribution inside bulges is almost insensitive to environment.

The metallicity gradients of all the sample bulges are negative or
null, with the exception of UGC~5184 that has a slightly positive
gradient ($\Delta$(\ZH)$=0.17$ dex). Negative metallicity gradients
were also measured in early-type galaxies \citep{procetal02,
  mehletal03, sancetal06s, rawletal10} and in bulges of spiral
galaxies \citep{jabletal07, morelli2015a}. They are expected for
bulges assembled through a process of dissipative collapse
\citep{kobayashi04}. As for the age gradients, the number
distributions of the metallicity gradients of the bulges hosted in
isolated or in group and cluster galaxies are very similar to each
other (Fig.~\ref{fig:idb_histo_grad}, central panel) suggesting the
same formation scenario for bulges in different environments.

The gradients of $\alpha/$Fe enhancement are negative or null for all
the sample bulges and display a number distribution with a remarkable
peak at $\Delta($\aFe$)=-0.1$ dex (Fig.~\ref{fig:idb_histo_grad},
right-hand panel). This is a particularly interesting finding since it
is in contrast with previous results obtained for early-type galaxies
\citep{mehletal03, sancetal06s, spoletal10} bulges of unbarred
\citep{jabletal07} and barred galaxies \citep{sancetal11, adri12}, and
bulges of group and cluster galaxies \citep{moreetal08}.
The \aFe\ ratio is commonly used as a proxy of the star formation
timescale in galaxies \citep{thometal05} because it is regulated by
the different contributions to the enrichment of the interstellar
medium caused by Type II and Type I supernovae \citep{matteucci1986}.
Therefore, we conclude that the star formation process was more
prolonged in the outer parts of the sample bulges than in their
central regions. Numerical simulations predict for dissipative
collapse a strong inside-out formation process for bulges which gives
rise to a negative gradient in the $\alpha$/Fe enhancement
\citep{fesi02}. Therefore, the negative gradients of $\alpha$/Fe
enhancement found in the bulges of isolated galaxies is consistent
with the predictions of a dissipative collapse formation scenario.

\begin{figure*}
\centering
\includegraphics[width=0.49\textwidth]{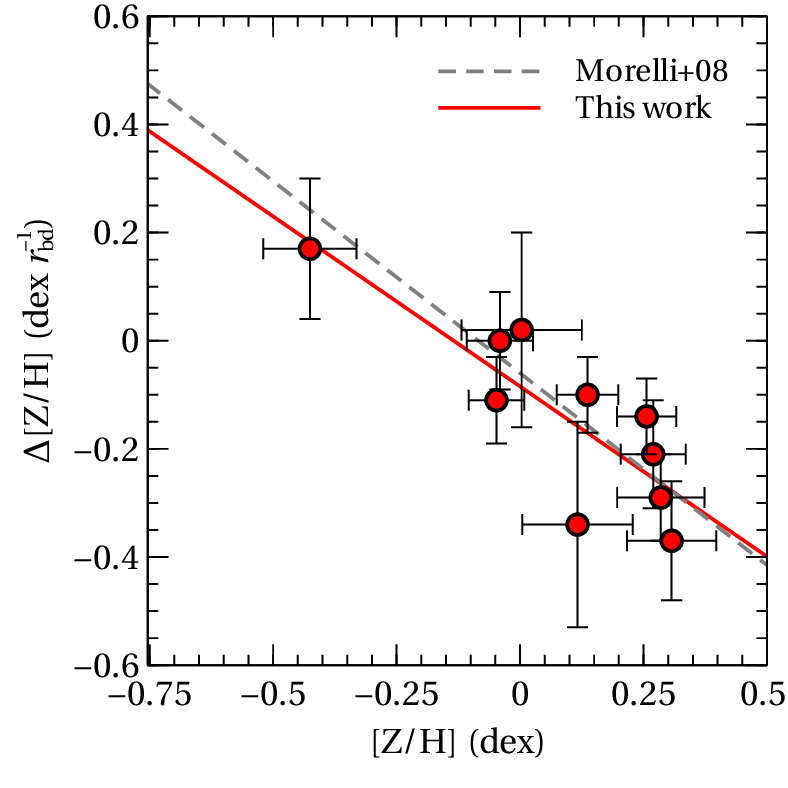}
\includegraphics[width=0.49\textwidth]{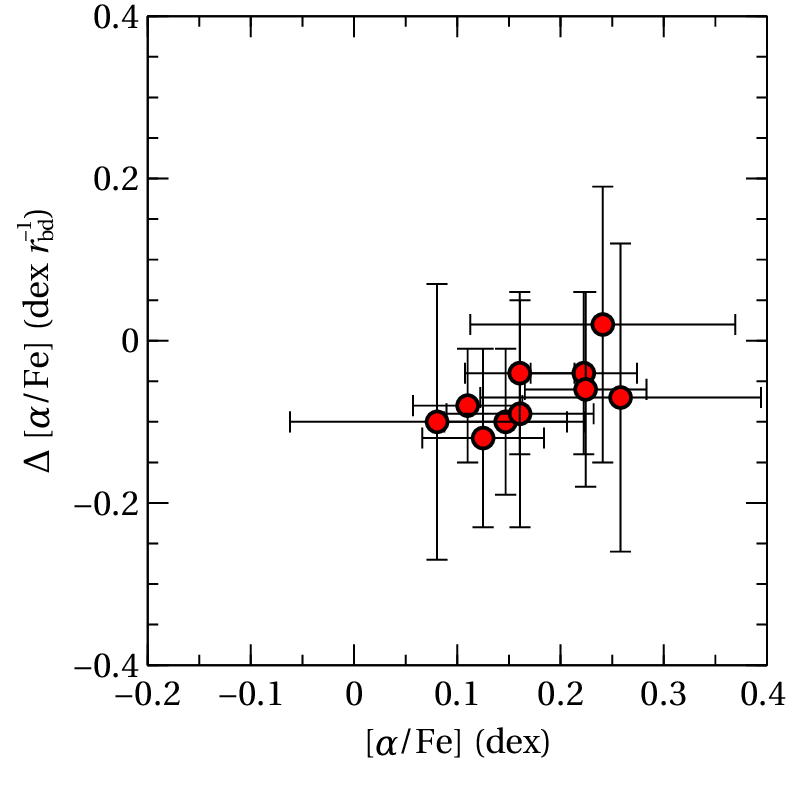}
\caption{Gradient and central value of metallicity (left-hand panel)
  and $\alpha$/Fe enhancement (right-hand panel) for the sample
  bulges. In the left-hand panel, the linear regression obtained for
  the bulges of the sample galaxies (red solid line) is compared to
  the correlation found by \citet{moreetal08} for the bulges of
  galaxies in groups and clusters. }
  \label{fig:idb_grads_cv}
\end{figure*}

We plot the central values and gradients of metallicity and
$\alpha/$Fe enhancement of the sample bulges in
Fig.~\ref{fig:idb_grads_cv}.
As \citet{moreetal08} and \citet{rawletal10}, we found a tight linear
correlation between the central values and gradients of metallicity
(Fig.~\ref{fig:idb_grads_cv}, left-hand panel) with a slope consistent
with that given by \citet{moreetal08}. If confirmed with a larger
sample of galaxies, this correlation is a further indication of the
importance of dissipative collapse in the assembly of bulges
\citep{aryo87,pipietal10}. We did not find any correlation between the
central values and gradients of $\alpha$/Fe enhancement
(Fig.~\ref{fig:idb_grads_cv}, right-hand panel) as also pointed out by
\citet{moreetal08} and \citet{rawletal10}.

\section{Conclusions}
\label{sec:conclusions}

We analysed the surface-brightness distribution, stellar kinematics,
and stellar population properties of a sample of isolated galaxies
selected from the CIG to constrain the dominant mechanism of the
assembly of their bulges. To this aim the properties of stellar
populations of the sample bulges were compared with those of bulges in
galaxies residing in groups and clusters.

A photometric decomposition of the SDSS $i$-band images was performed
to obtain the structural parameters of the sample galaxies.  We used
the structural parameters to identify the bulge-dominated radial range
of the sample galaxies by measuring the radius \rbd , where the bulge
contribution to the galaxy surface brightness dominates over that of
the remaining components.

We measured the stellar kinematics and radial profiles of the \Mgb,
\Mgd, \Hb, and \Fe\ line-strength indices from the major-axis spectra
we obtained at TNG. The kinematics of all the sample galaxies is very
regular giving further support to the idea that these objects are not
suffering interactions with the neighbour galaxies. The correlations
between the central values of the \Mgd\ and \Fe\ line-strength indices
and velocity dispersion were found to be consistent with those for
bulges of group and cluster galaxies \citep{idiaetal96, prugetal01,
  procetal02, moreetal08}.

We obtained the stellar population properties of the bulges of the
sample galaxies by deriving their central values of mean age, total
metallicity, and total $\alpha/$Fe from stellar population models with
variable element abundance ratios. The sample bulges are characterised
by a bimodal age distribution with intermediate-age ($\sim3$ Gyr) and
old systems ($\sim15$ Gyr), a large spread in metallicities ranging
from sub- to super-solar values, and $\alpha$/Fe enhancements peaked
at \aFe$\,=\,0.2$ dex. The higher \aFe\ ratios found for the bulges of
isolated galaxies indicate a shorter star-formation timescale with
respect their counterparts in high density environment
\citep{thometal05}.  On the contrary, the metallicity distribution is
very similar for bulges residing in different environments. 

The absence of a correlation between the bulge stellar populations and
galaxy morphology excludes a strong interplay between bulges and discs
during their evolution. This conclusion is also supported by the
findings of \citet{silchenko2012} and \citet{katkov2015} who formulate
the hypothesis that the morphological type of a field galaxy is
determined by the outer-gas accretion.  Finally, we derived the
gradients of the stellar population properties within the sample
bulges. Most of them have a null age gradient and a negative
metallicity gradient. This is in agreement with earlier findings for
bulges in cluster \citep{jabletal07,moreetal08} and high
surface-brightness galaxies \citep{morelli2012}. All the sample bulges
show a negative gradient for the $\alpha/$Fe enhancement. This is a
prediction of the dissipative collapse model of bulge formation and it
was never been observed before. The stellar population gradients are
believed to be flattened or even erased by merging and acquisition
events. Therefore, we suggest that the gradients imprinted during the
inside-out formation process are preserved in the bulges of isolated
galaxies, which suffered a limited number of interactions and mergers,
whereas the gradients are cancelled in the bulges of group and cluster
galaxies as a consequence of phenomena driven by environment.

\section*{Acknowledgements}

We would like to thank Lodovico Coccato and Enrico V. Held for the useful
discussion and their suggestions.

This investigation was based on observations obtained at the ESO
Telescopes at the La Silla Paranal Observatory under programmes
76.B-0375, and 80.B-00754.
This work was partially supported by Padua University through grants
60A02-5857/13, 60A02-5833/14, 60A02-4434/15, and CPDA133894.
LM acknowledges financial support from Padua University grant
CPS0204. JMA acknowledges support from the European Research Council
Starting Grant SEDmorph (P.I. V. Wild).
Part of the data used in this research were acquired through the Sloan
Digital Sky Survey (SDSS) Archive (http://www.sdss.org/). 
This research also made use of the HyperLeda Database
(http://leda.univ-lyon1.fr/) and NASA/IPAC Extragalactic Database
(NED) which is operated by the Jet Propulsion Laboratory, California
Institute of Technology, under contract with the National Aeronautics
and Space Administration (http://ned.ipac.caltech.edu/).


\appendix

\section{Photometric decomposition, stellar kinematic and line strength indices.}

The photometric decomposition of all the sample galaxies are plotted
in Fig.~\ref{fig:decomposition}.  The stellar
kinematics of all the sample galaxies are given in
Table~\ref{tab:kinematics_indices} and plotted in
Fig.~\ref{fig:kinematics}, respectively. The line-strength indices of
all the sample galaxies are given in
Table~\ref{tab:kinematics_indices} too and plotted in
Fig.~\ref{fig:kinematics}.

\begin{figure*}
\centering
\includegraphics[angle=-90.0,width=0.80\textwidth]{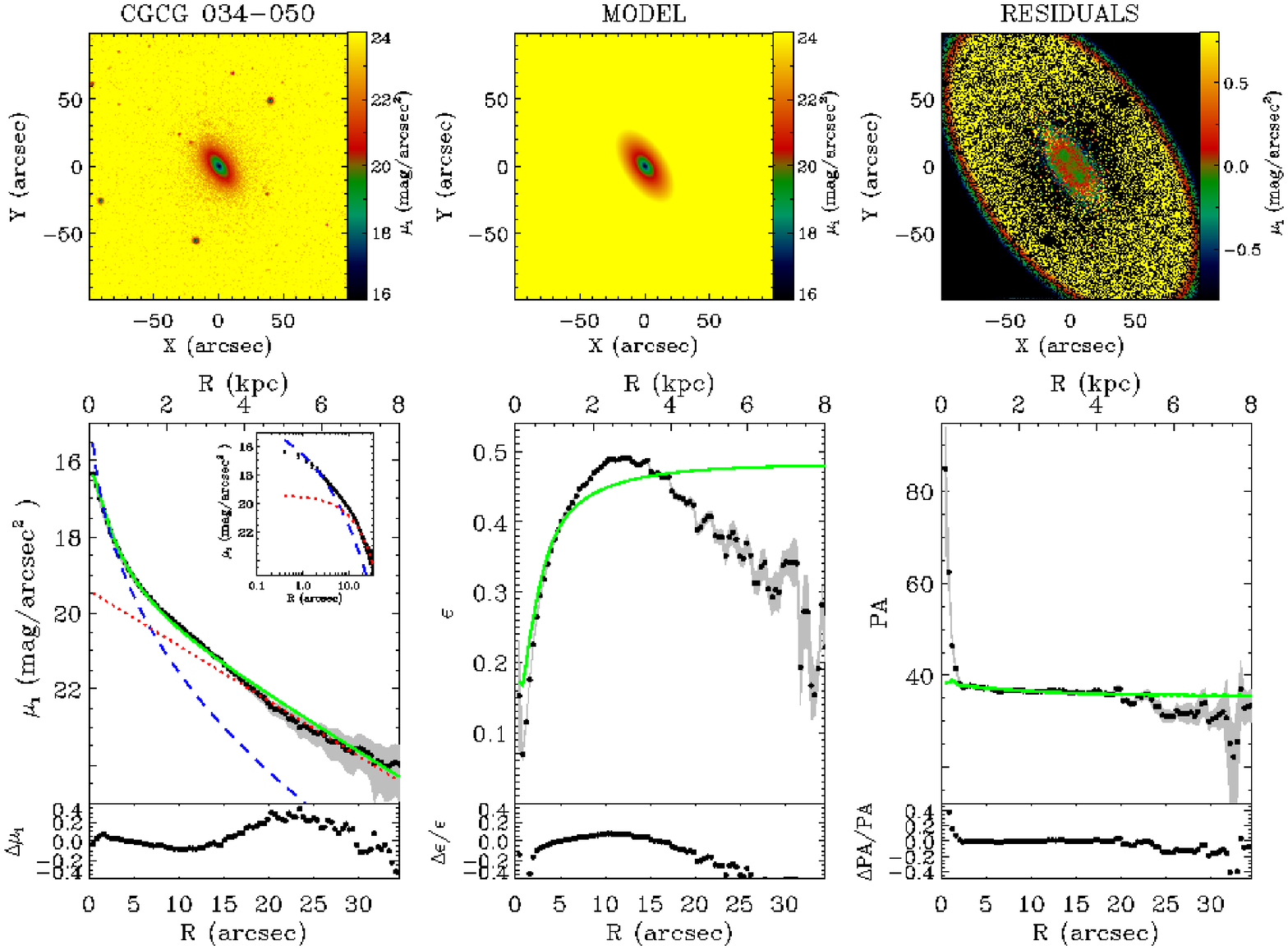}
\includegraphics[angle=-90.0,width=0.80\textwidth]{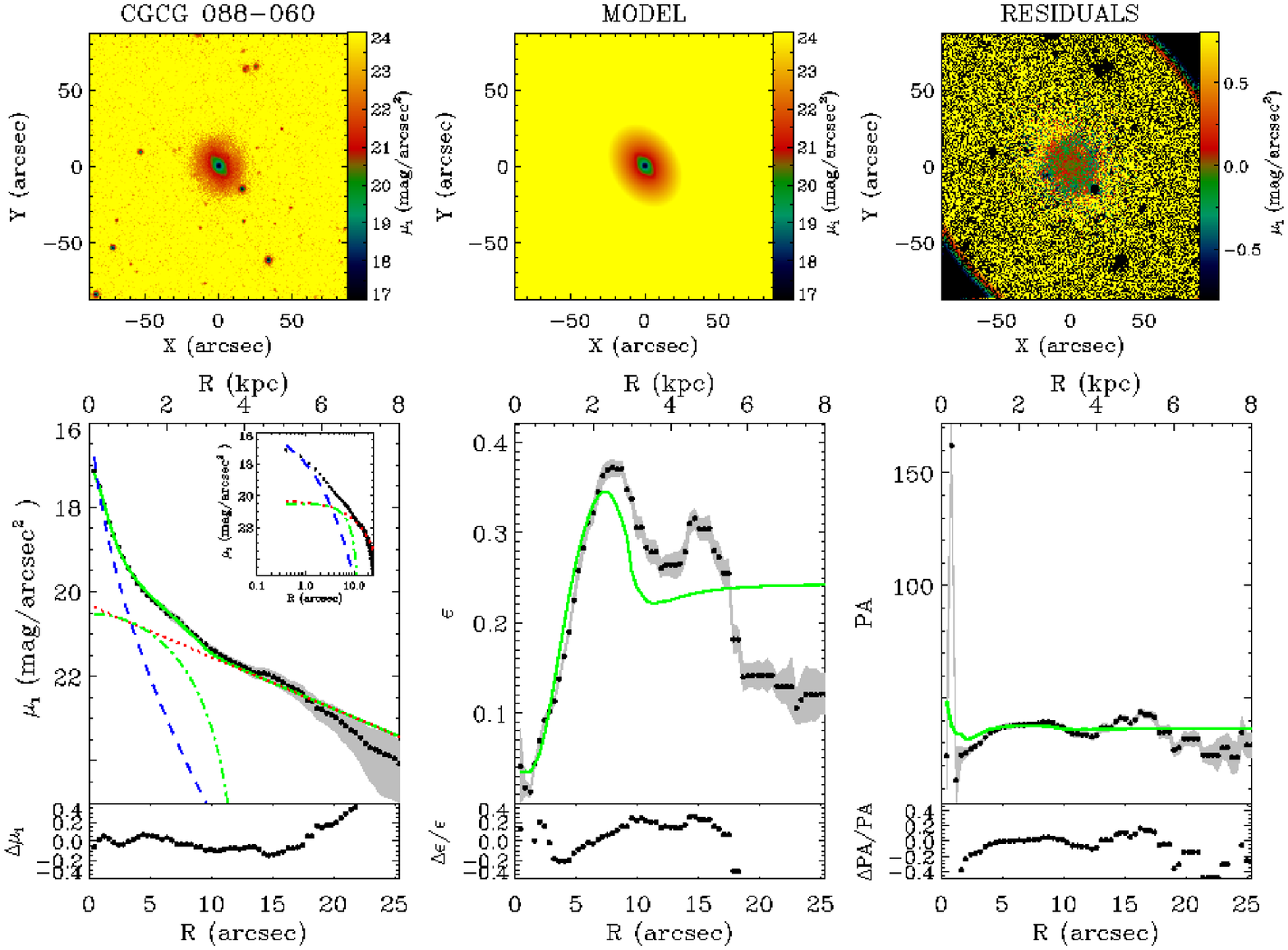}
\caption{Two-dimensional photometric decomposition of the SDSS
  $i$-band images of the sample galaxies. Upper panels (from left- to
  right-hand side): Map of the observed, modelled, and residual
  (observed-modelled) surface-brightness distribution of the
  galaxy. The surface brightness range of each image is indicated at
  the right-hand side of the panel. In each panel the spatial
  coordinates with respect to the galaxy centre are given in
  arcsec. Lower panels (from left- to right-hand side):
  Ellipse-averaged radial profile of surface-brightness, position
  angle, and ellipticity measured in the observed (black circles with
  error bars given as a grey area) and modeled image (green solid
  line). The radial profiles of the intrinsic surface-brightness
  contribution of the bulge (blue dashed line), disc (red dotted
  line), and bar (green dash and dotted line) are given in the lower
  left-hand panel with an inset showing the fit with a logarithmic
  scale for the distance to the galaxy centre. The difference between
  the ellipse-averaged radial profiles extracted from the observed and
  modeled images is also shown.}
\label{fig:decomposition}
\end{figure*}
\begin{figure*}
\centering
\includegraphics[angle=-90.0,width=0.80\textwidth]{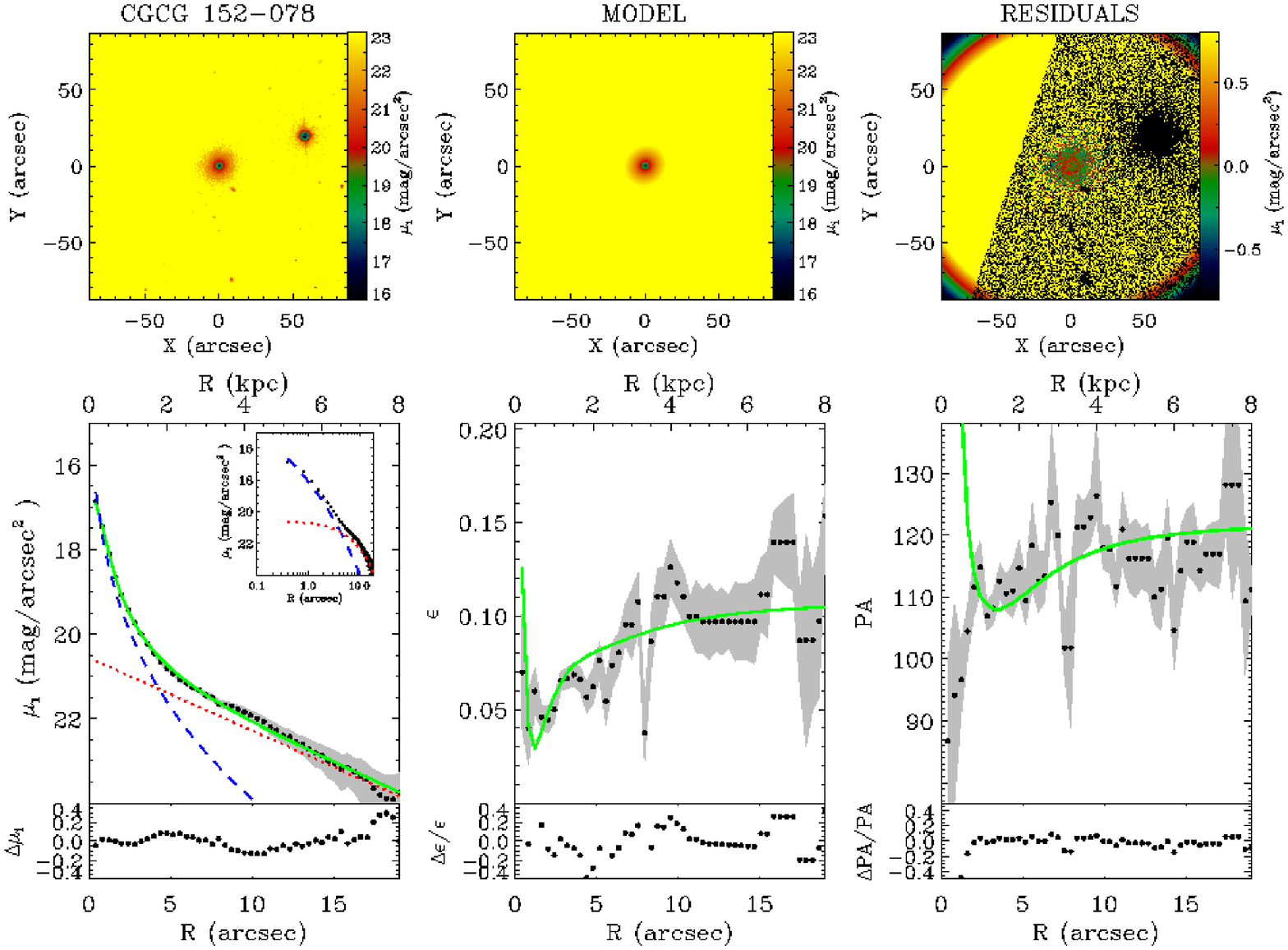}
\includegraphics[angle=-90.0,width=0.80\textwidth]{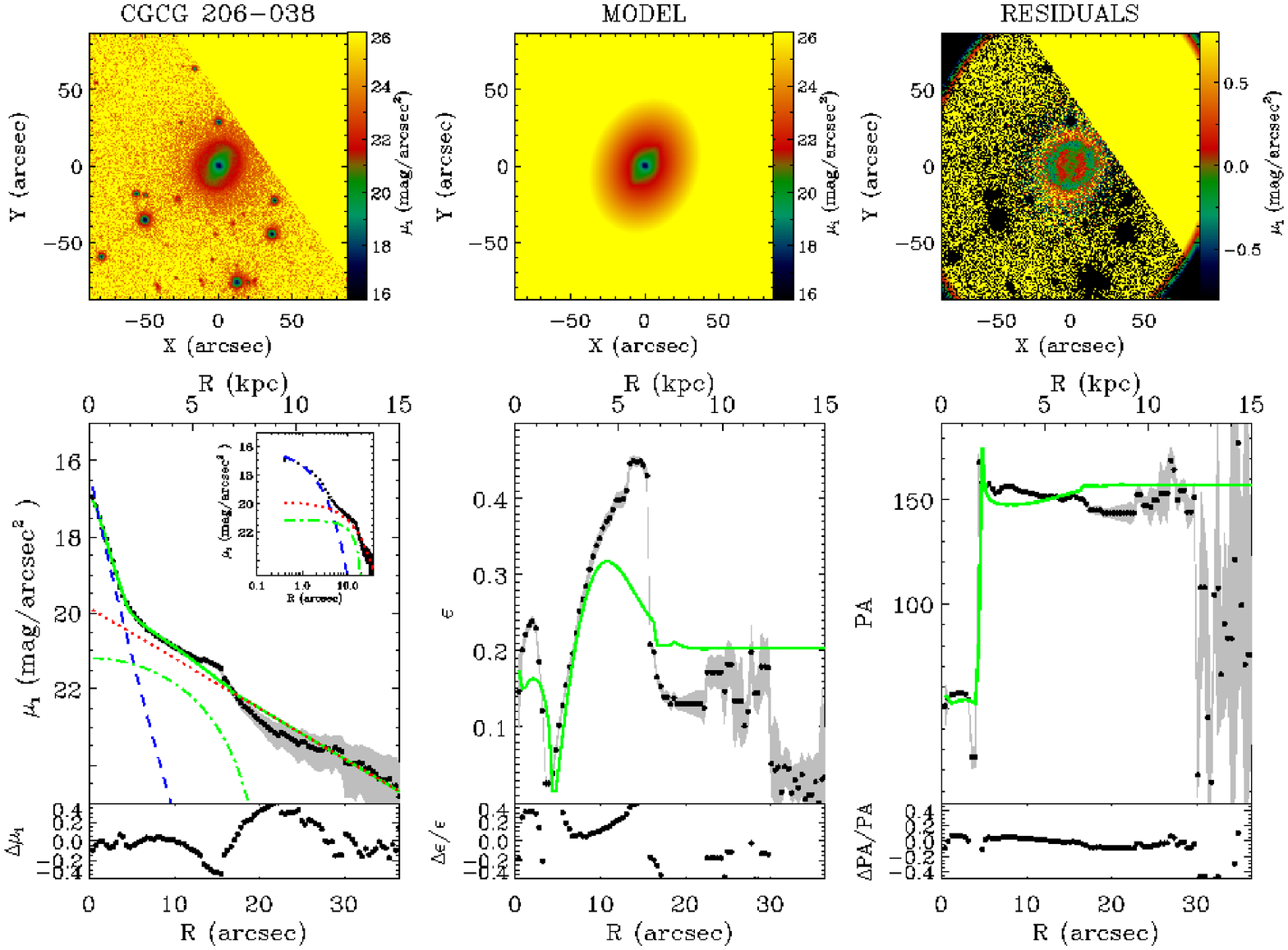}
\contcaption{}
\end{figure*}
\begin{figure*}
\centering
\includegraphics[angle=-90.0,width=0.80\textwidth]{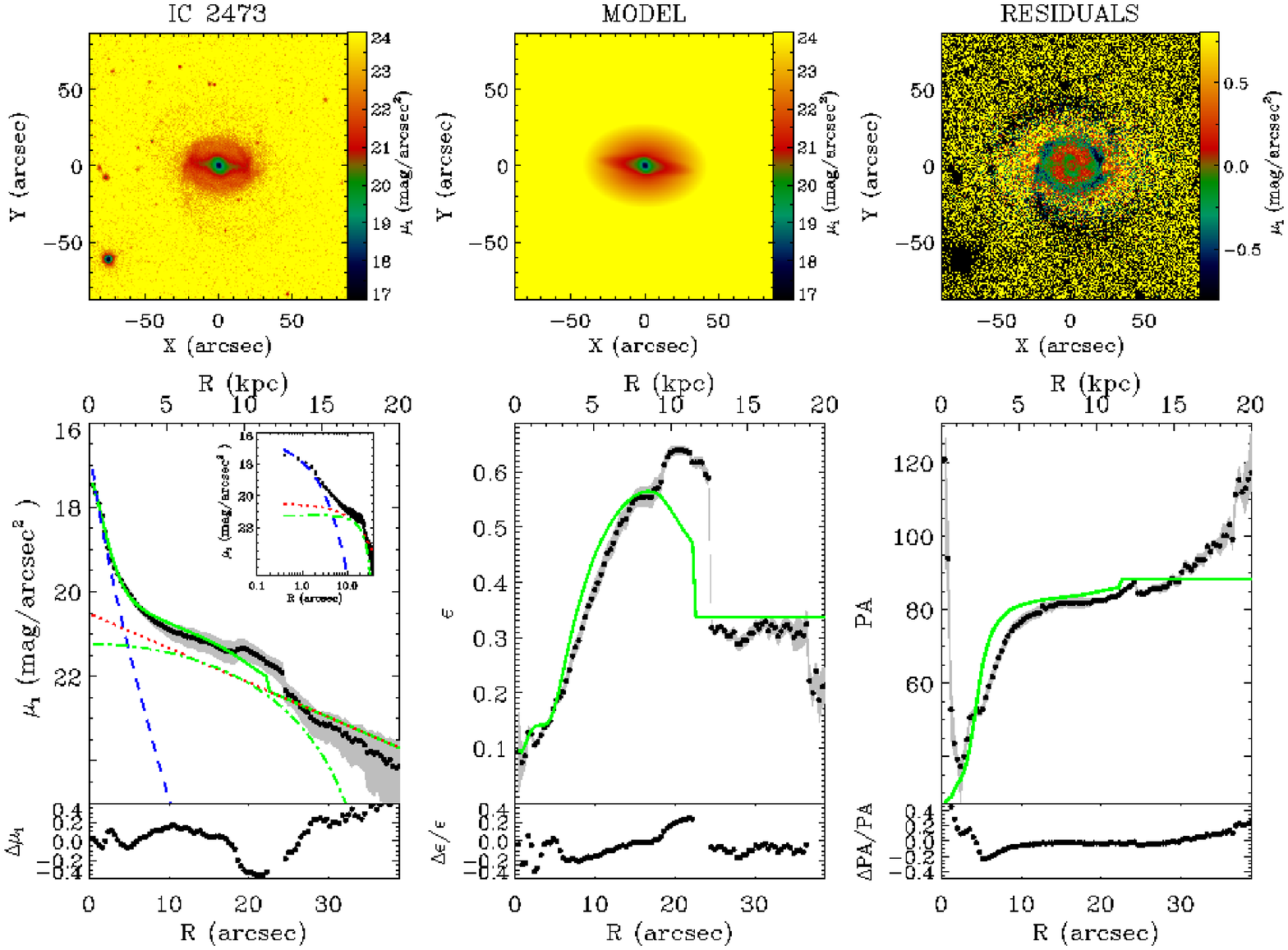}
\includegraphics[angle=-90.0,width=0.80\textwidth]{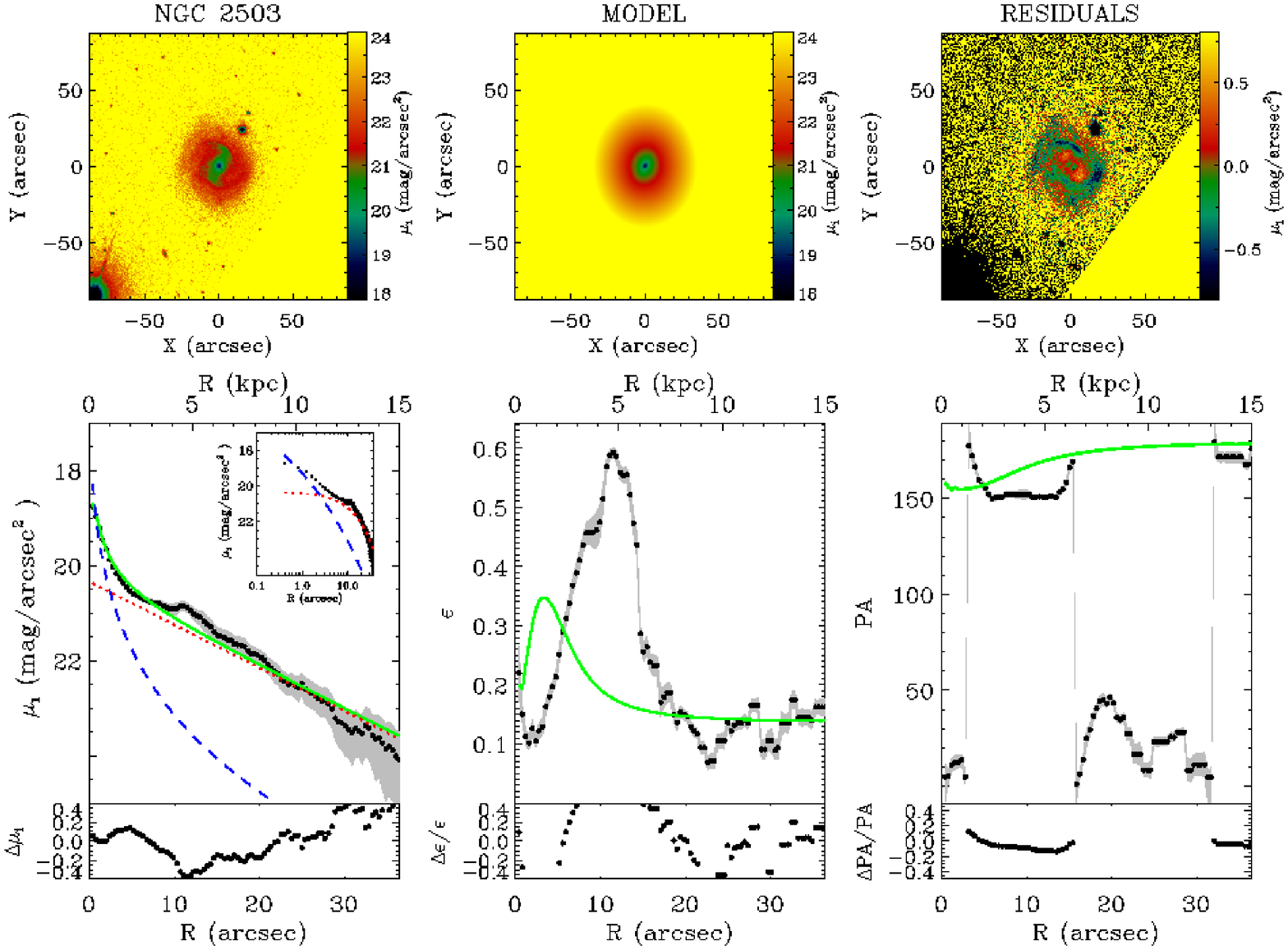}
\contcaption{}
\end{figure*}
\begin{figure*}
\centering
\includegraphics[angle=-90.0,width=0.80\textwidth]{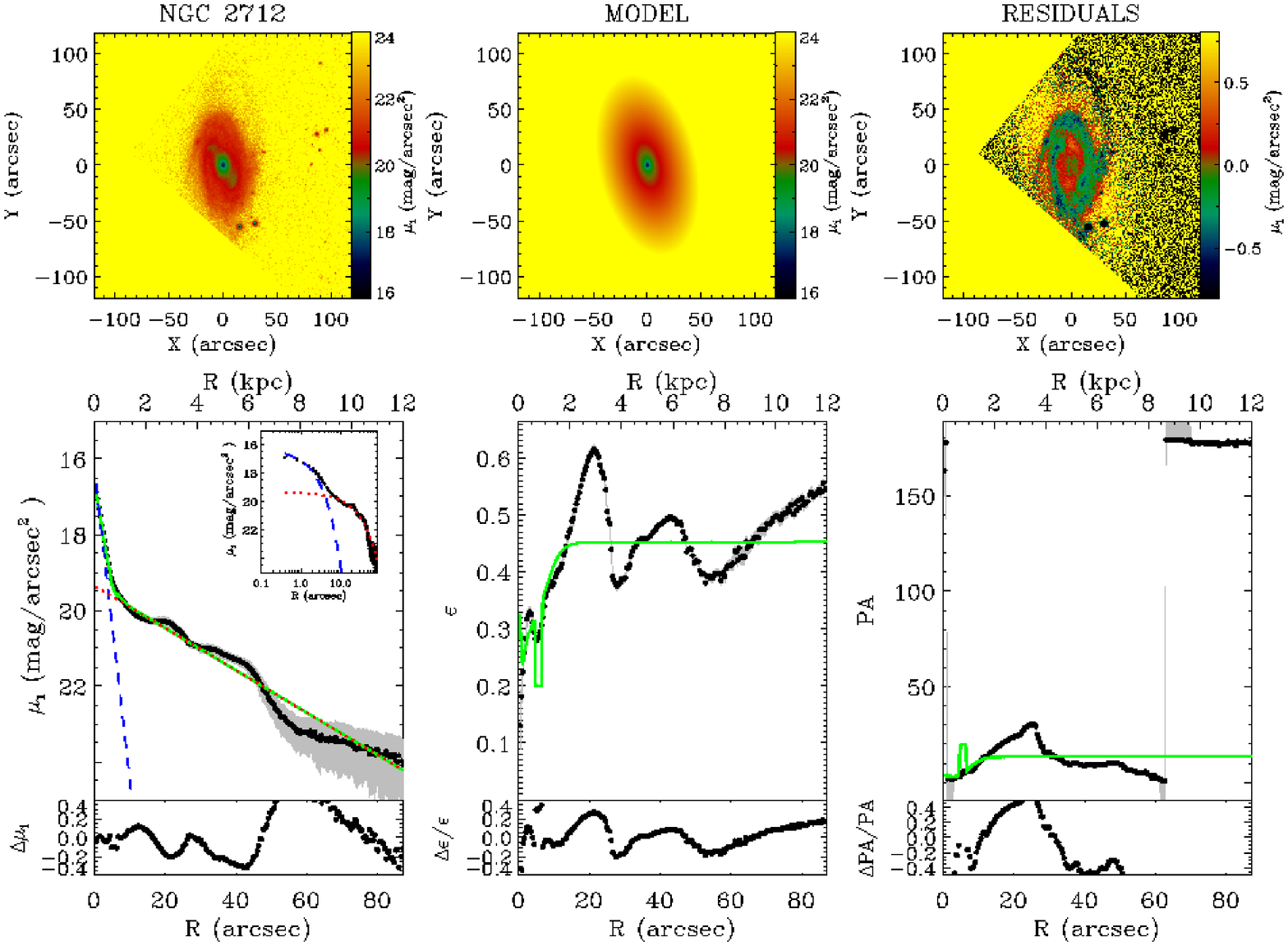}
\includegraphics[angle=-90.0,width=0.80\textwidth]{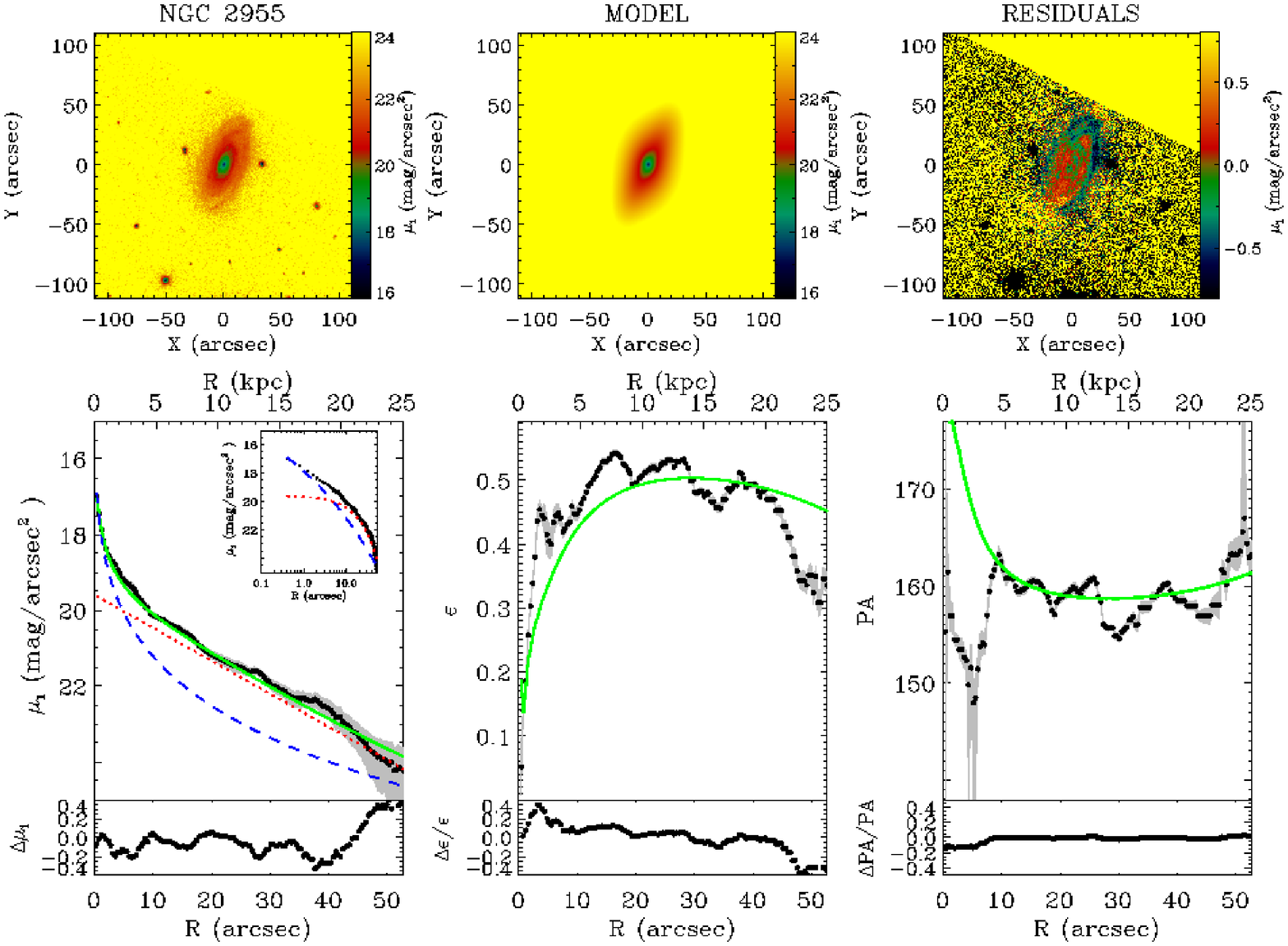}
\contcaption{}
\end{figure*}
\begin{figure*}
\centering
\includegraphics[angle=-90.0,width=0.80\textwidth]{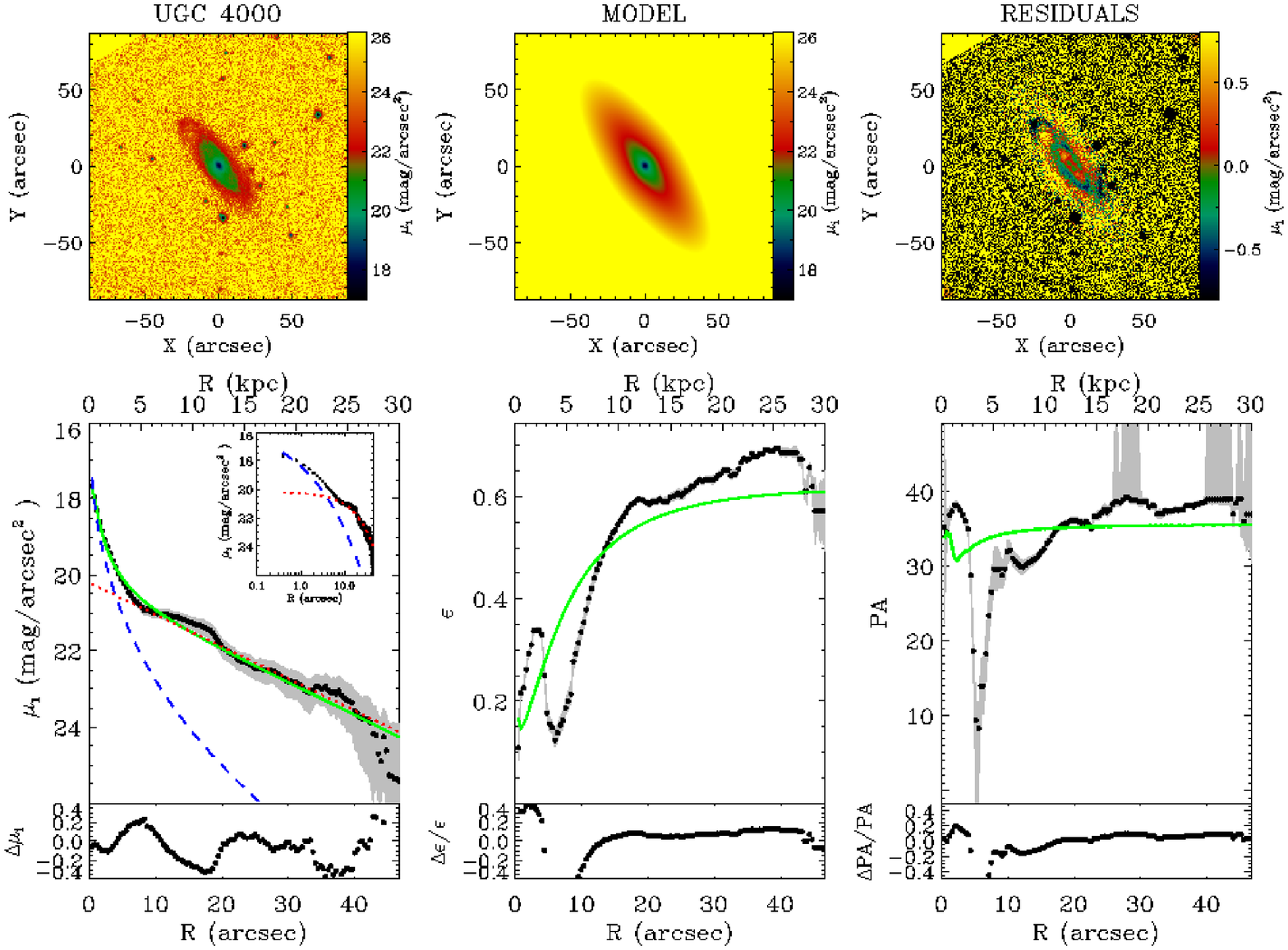}
\includegraphics[angle=-90.0,width=0.80\textwidth]{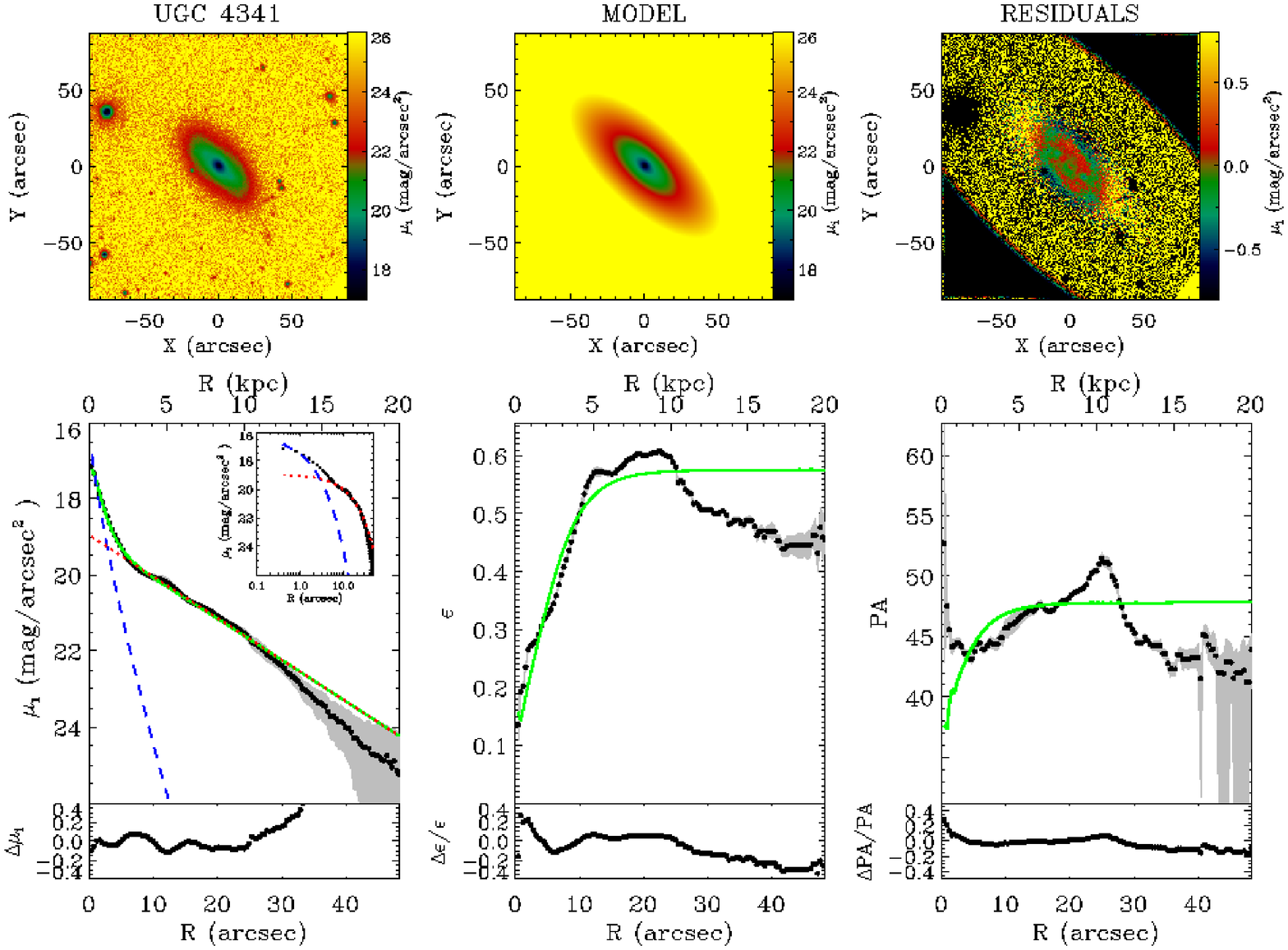}
\contcaption{}
\end{figure*}
\begin{figure*}
\centering
\includegraphics[angle=-90.0,width=0.80\textwidth]{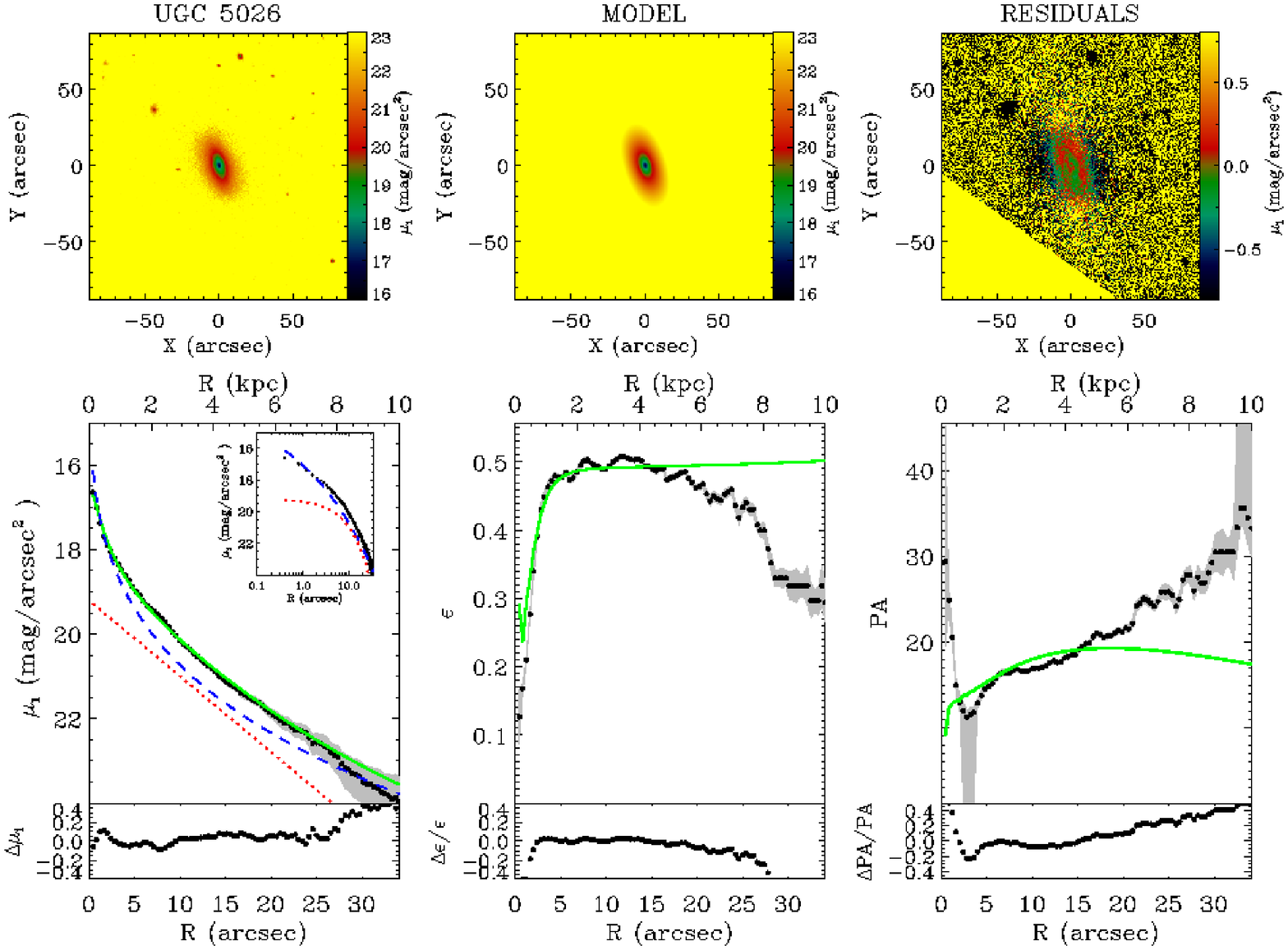}
\includegraphics[angle=-90.0,width=0.80\textwidth]{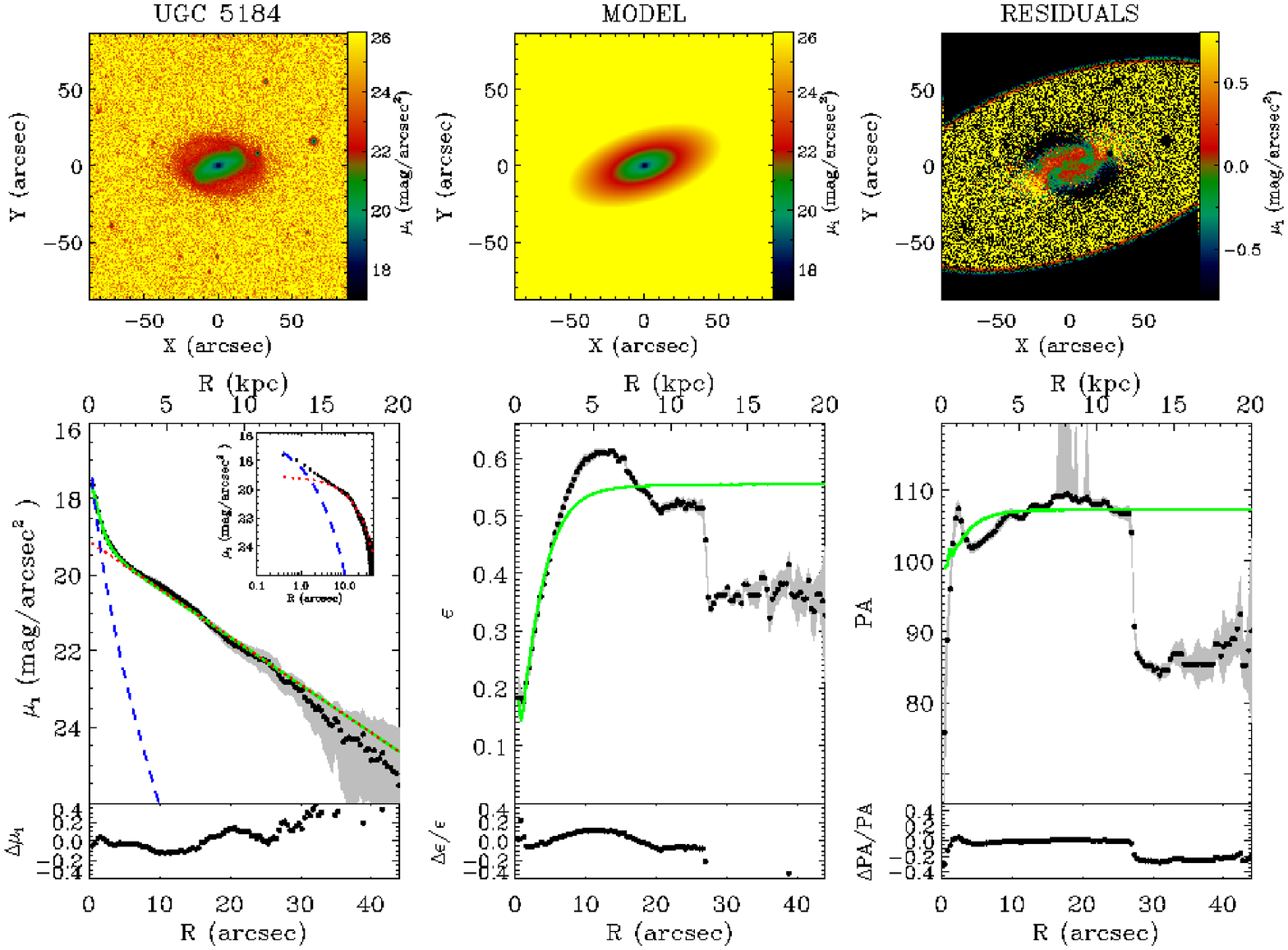}
\contcaption{}
\end{figure*}
\begin{scriptsize}
\begin{figure*} 
\caption{Stellar kinematics measured along the major axis of the
  sample galaxies. For each axis, the curves are folded around the
  nucleus. Blue asterisks and red circles refer to data measured along
  the approaching and residing sides of the galaxy, respectively. The
  radial profiles of the LOS velocity ($v$) after the
  subtraction of the systemic velocity, the velocity dispersion
  ($\sigma$), the third- and fourth-order coefficients of the
  Gauss-Hermite decomposition of the LOSVD ($h_3$ and $h_4$) are shown
  (panels from top to bottom). The vertical dashed line corresponds to
  the radius \rbd , where the surface-brightness contribution of the
  bulge is equal to that of the remaining components.}
\includegraphics[angle=90.0,width=0.498\textwidth]{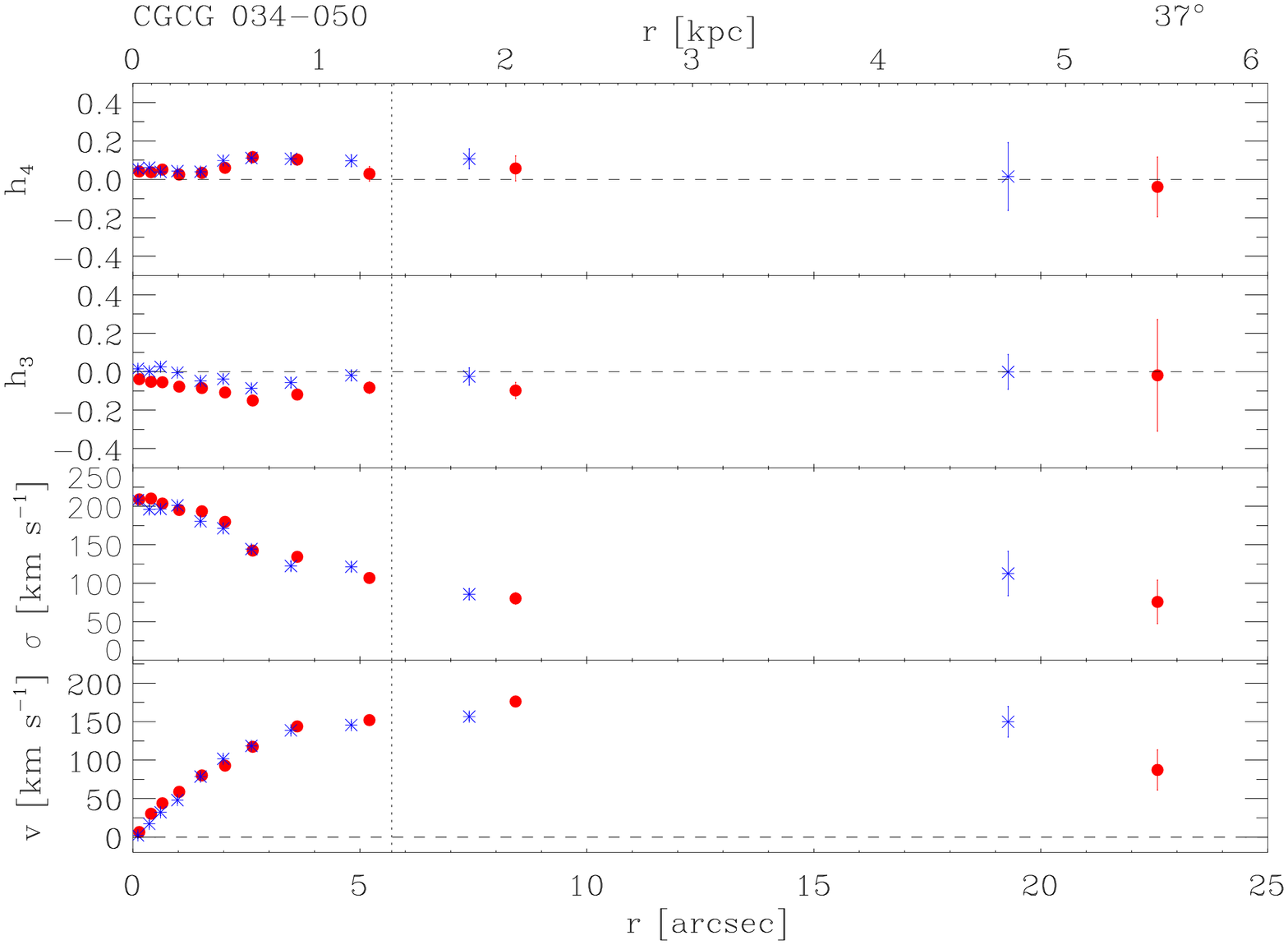}
\includegraphics[angle=90.0,width=0.498\textwidth]{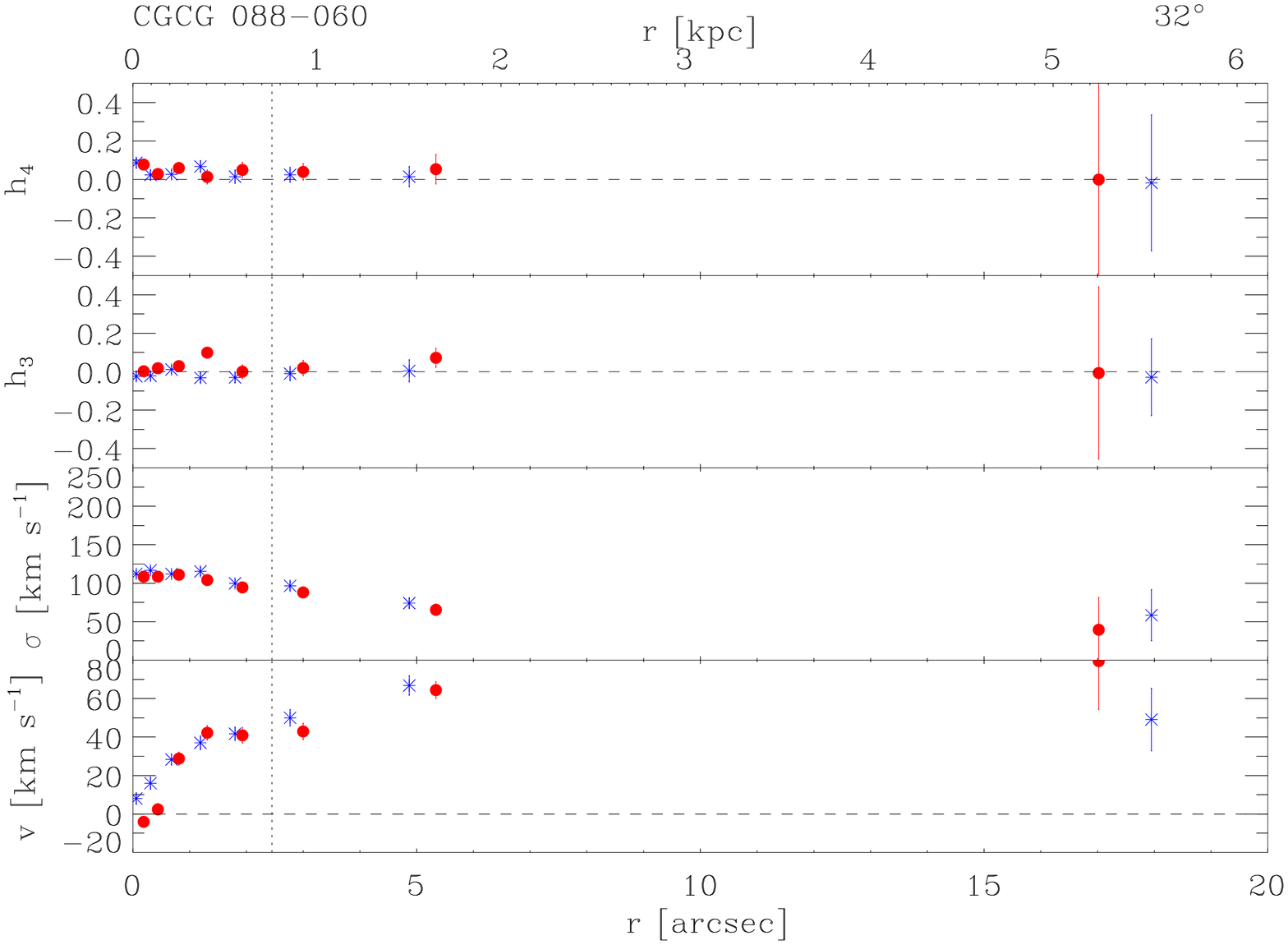}
\includegraphics[angle=90.0,width=0.498\textwidth]{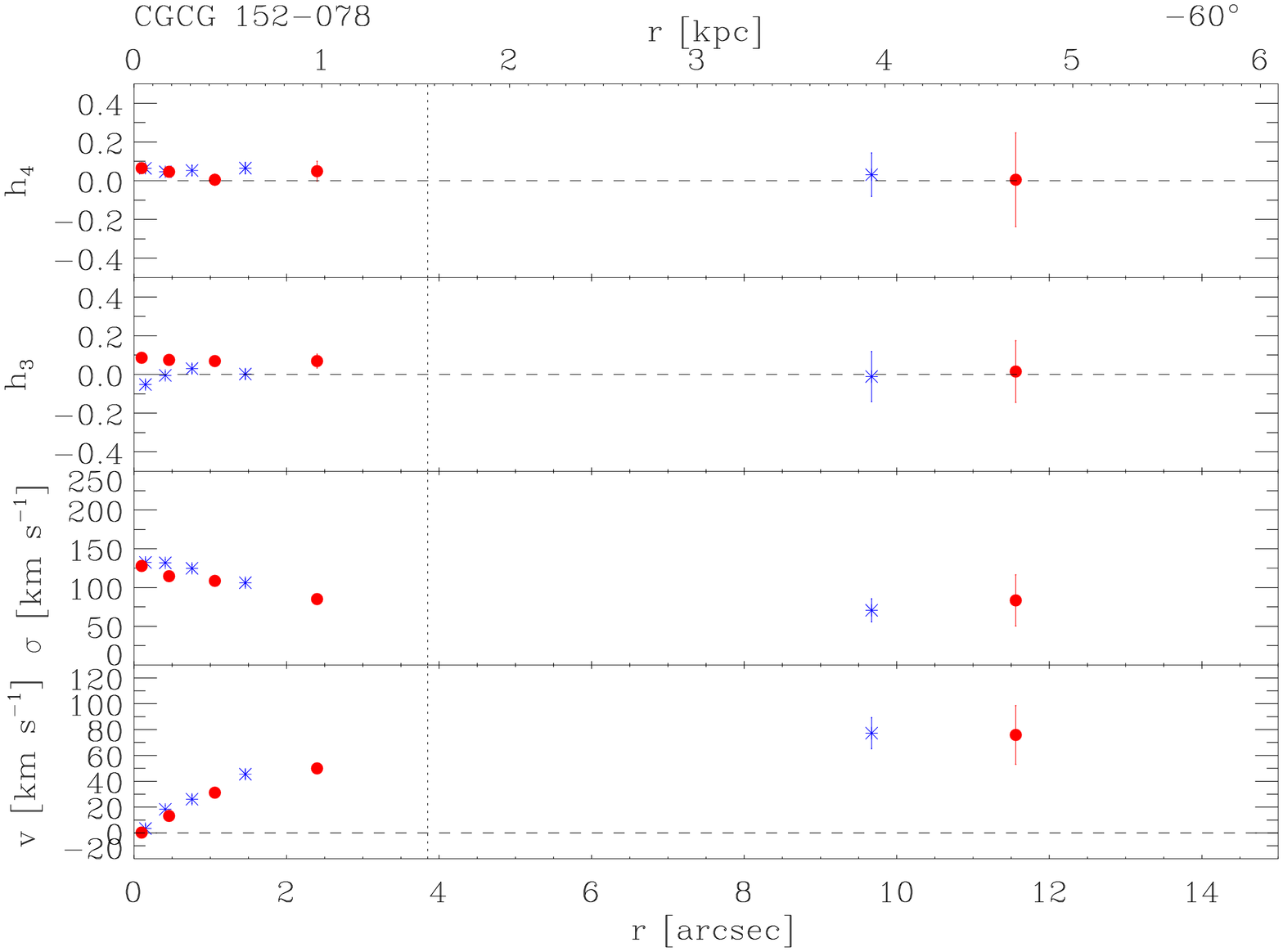}
\includegraphics[angle=90.0,width=0.498\textwidth]{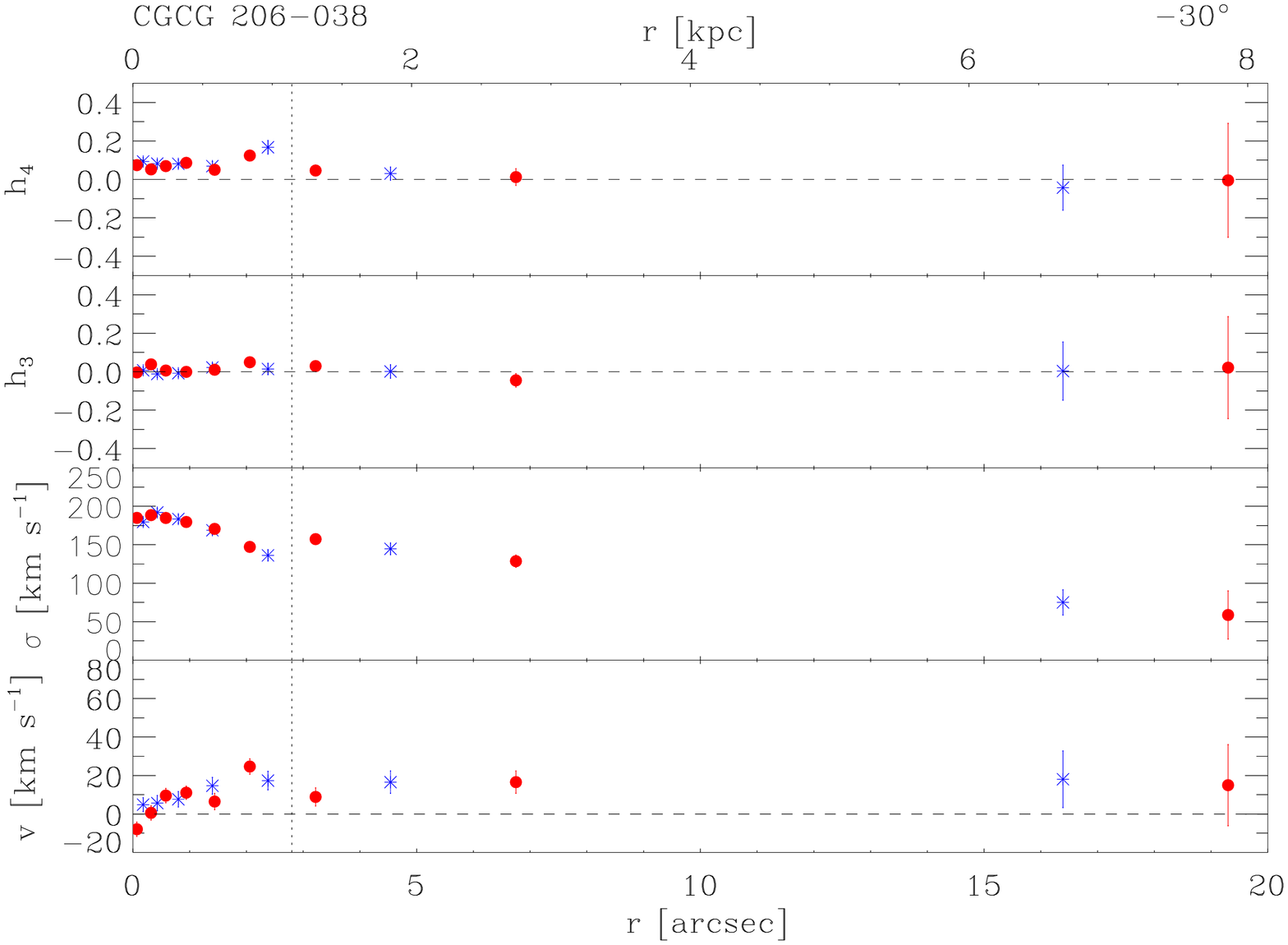}
\includegraphics[angle=90.0,width=0.498\textwidth]{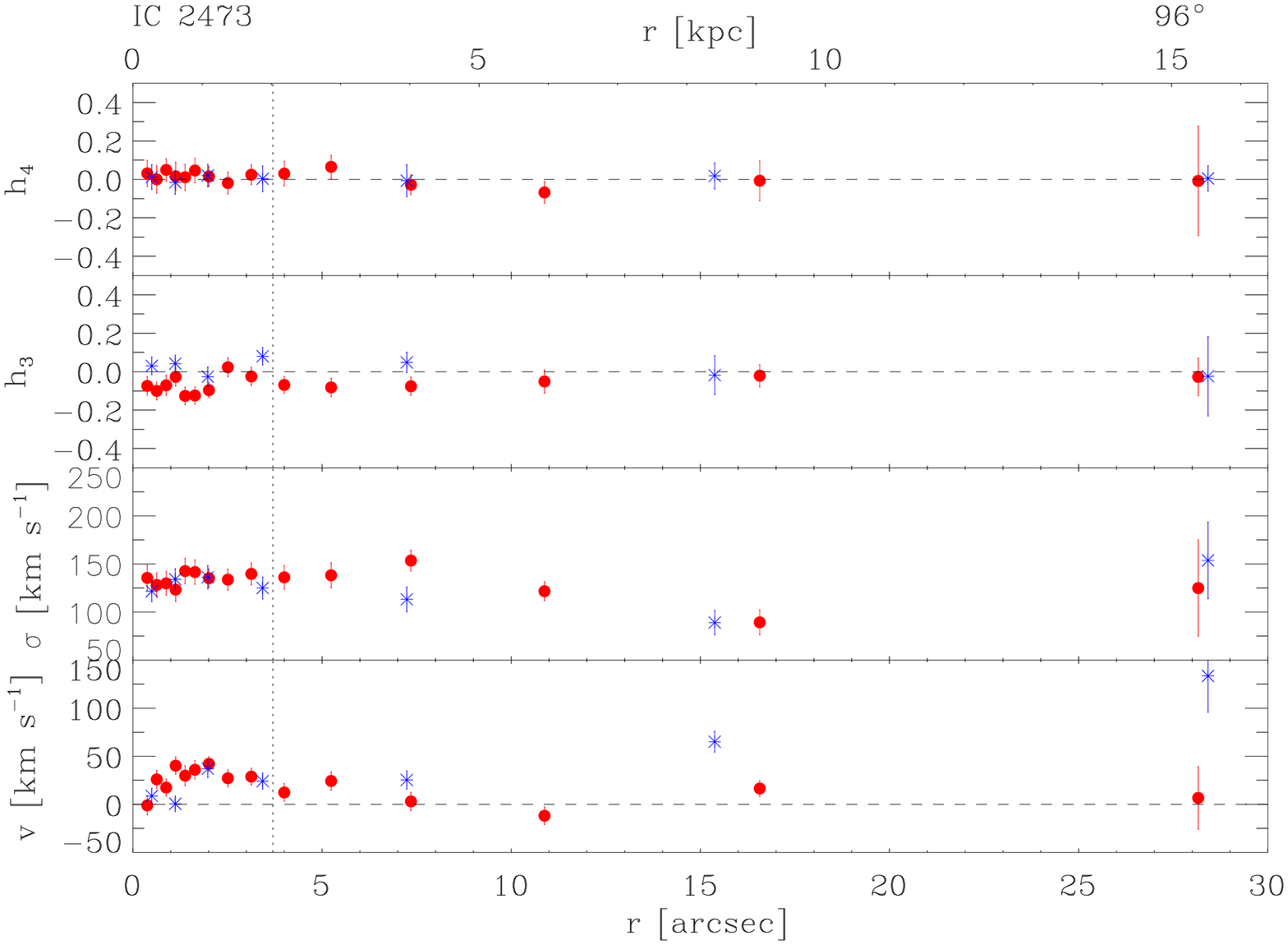}
\includegraphics[angle=90.0,width=0.498\textwidth]{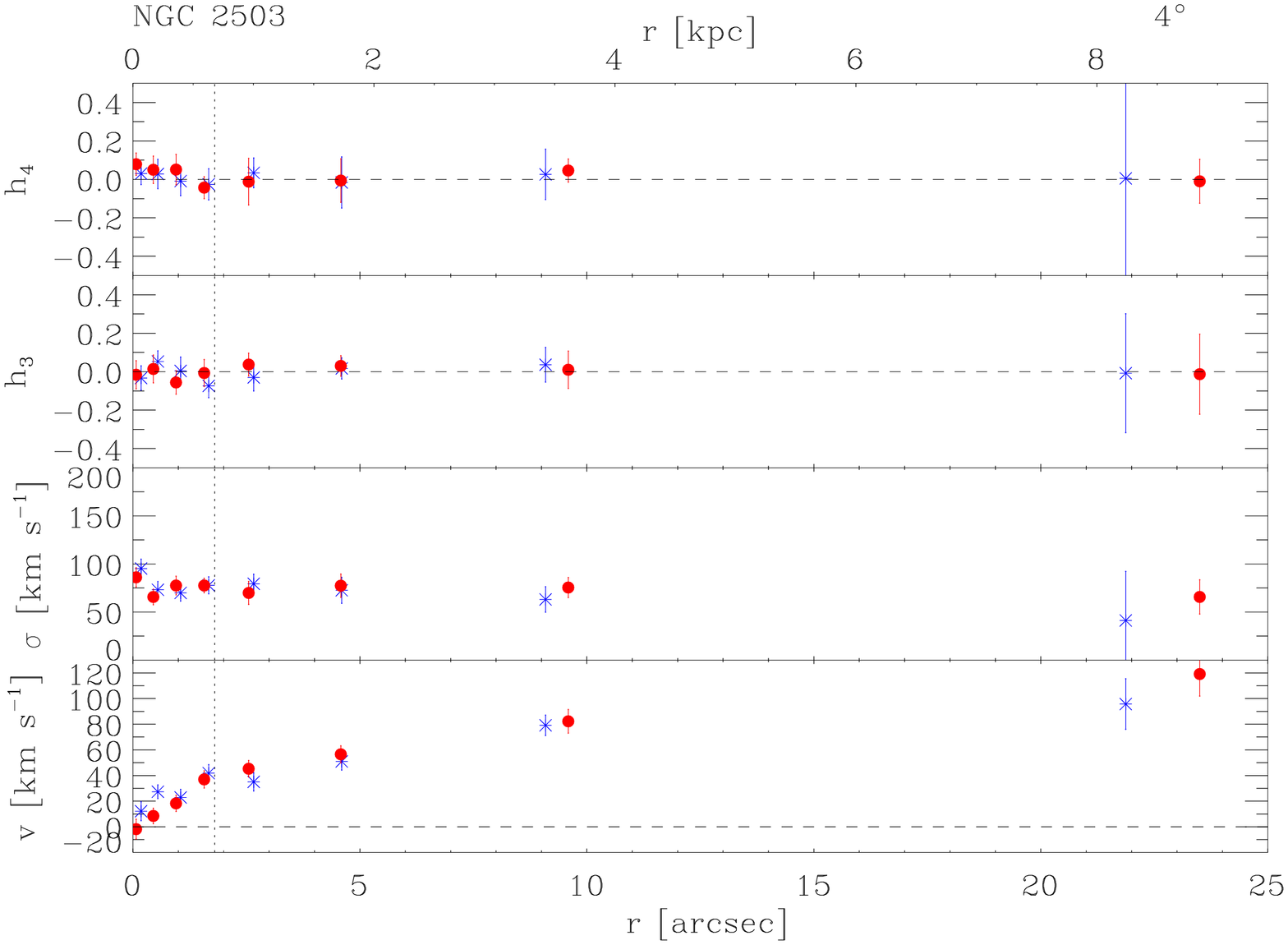}\\
\label{fig:kinematics}
\end{figure*}
\end{scriptsize}
\begin{figure*}
\centering
\includegraphics[angle=90.0,width=0.498\textwidth]{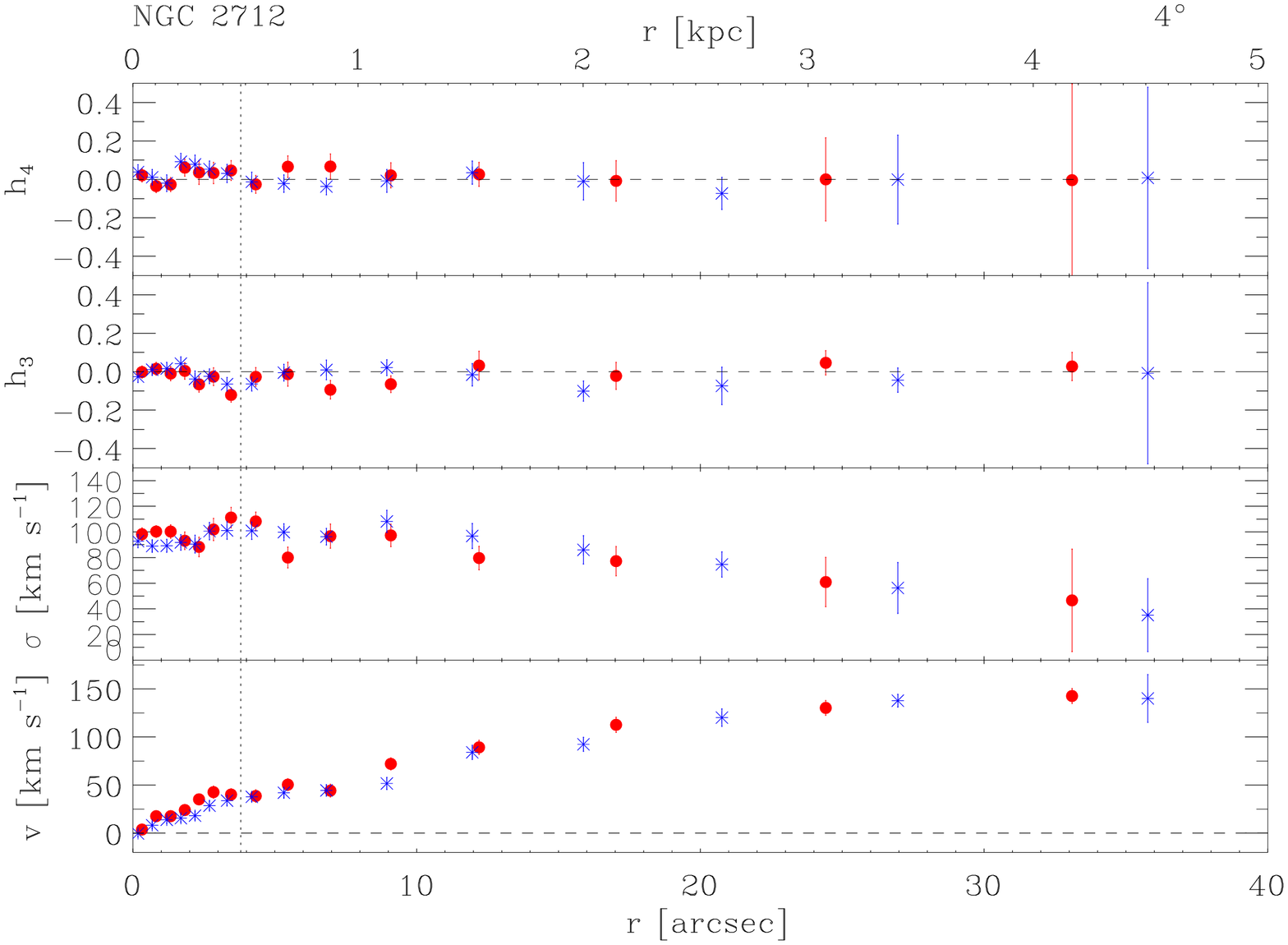}
\includegraphics[angle=90.0,width=0.498\textwidth]{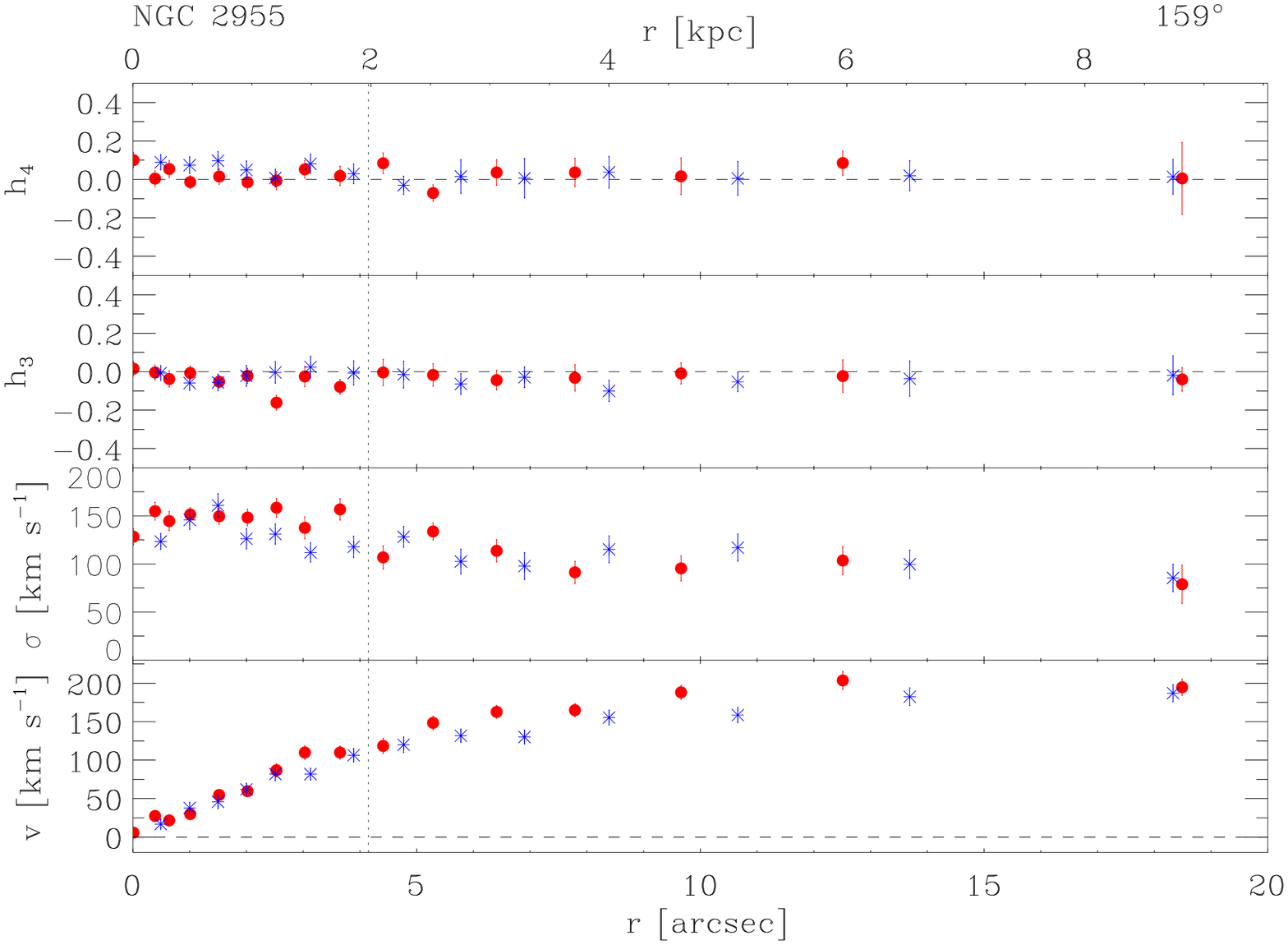}
\includegraphics[angle=90.0,width=0.498\textwidth]{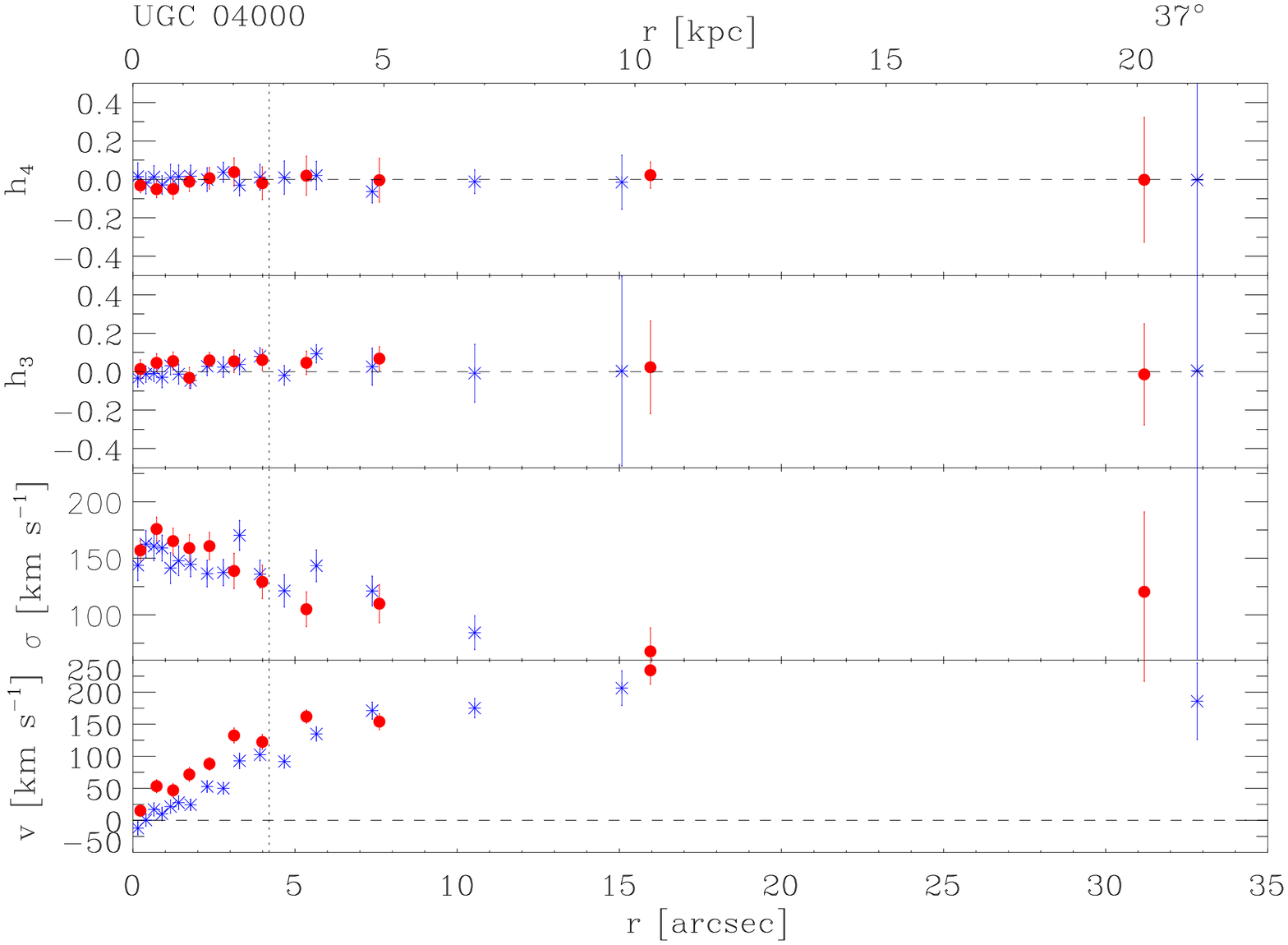}
\includegraphics[angle=90.0,width=0.498\textwidth]{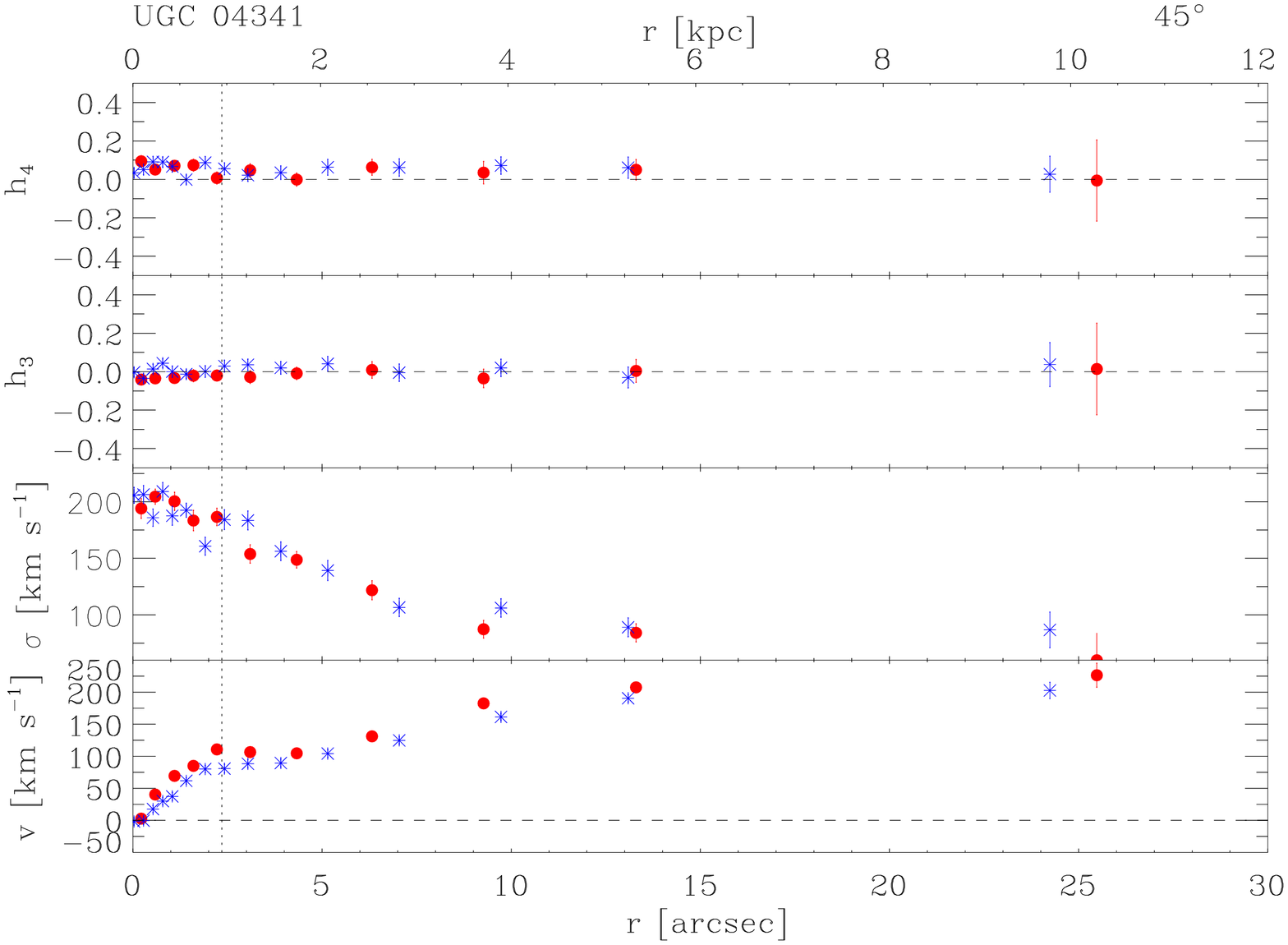}
\includegraphics[angle=90.0,width=0.498\textwidth]{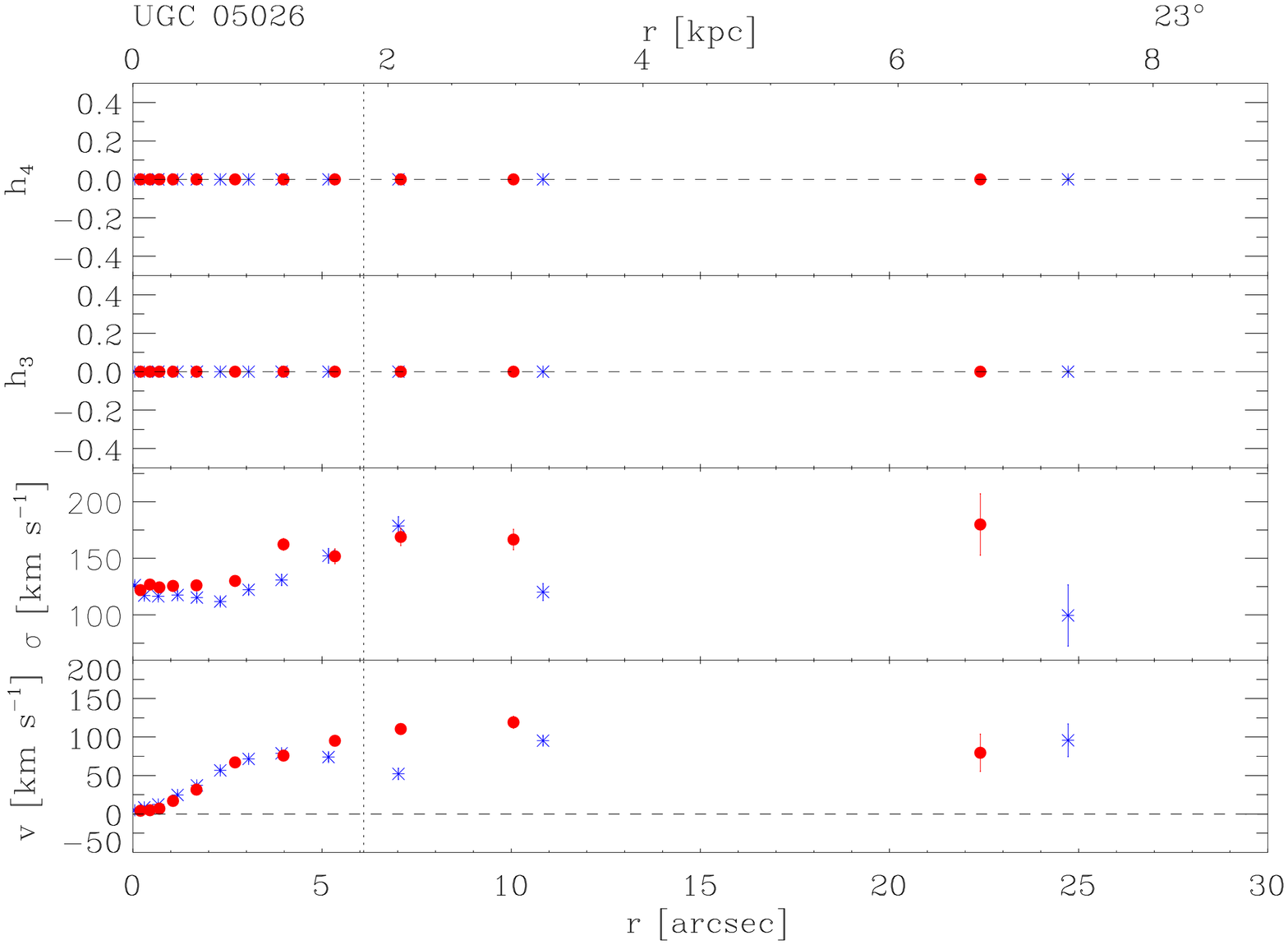}
\includegraphics[angle=90.0,width=0.498\textwidth]{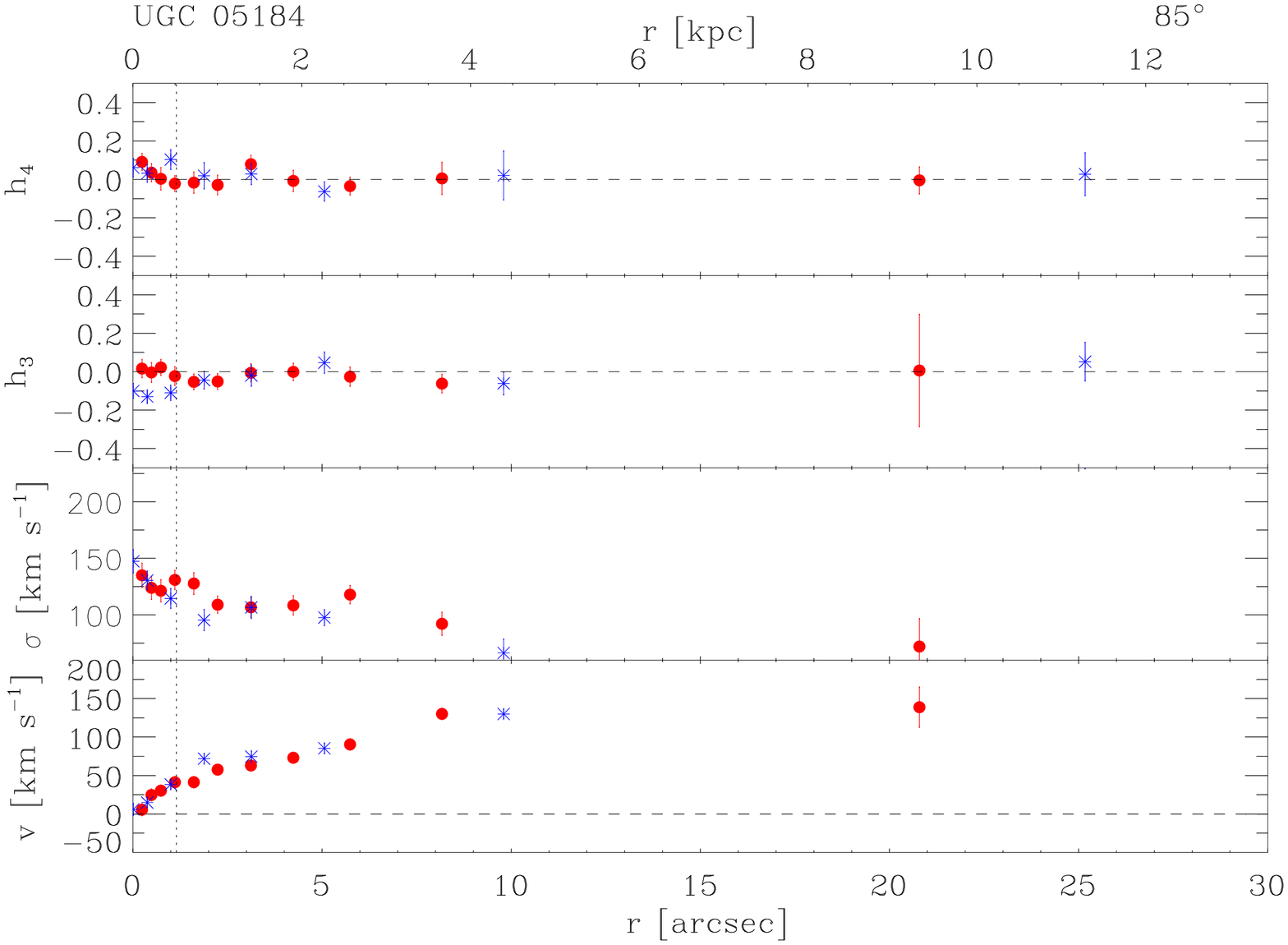}\\
\contcaption{}
\end{figure*}
\begin{figure*}
\centering
\includegraphics[angle=90.0,width=0.498\textwidth]{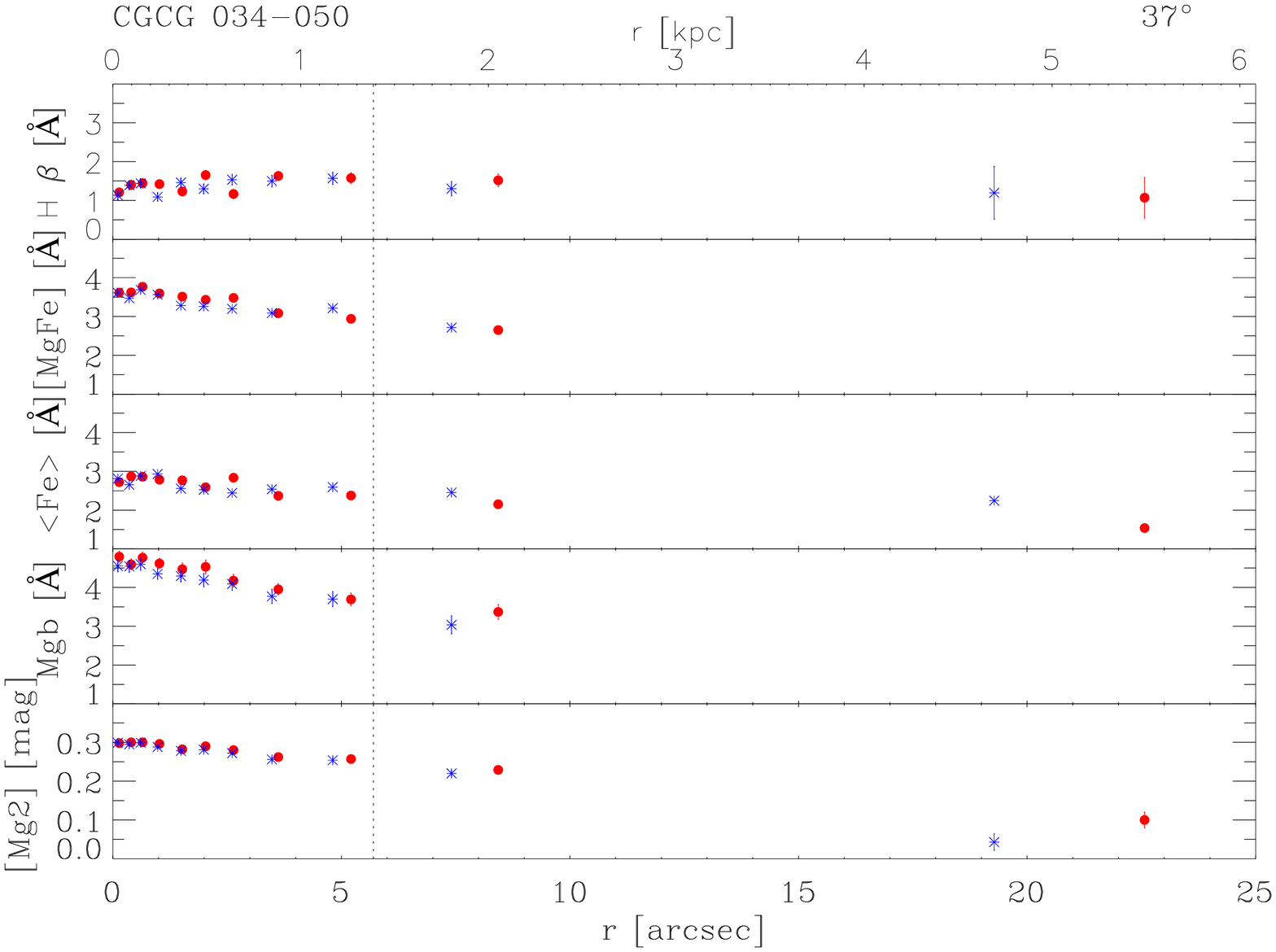}
\includegraphics[angle=90.0,width=0.498\textwidth]{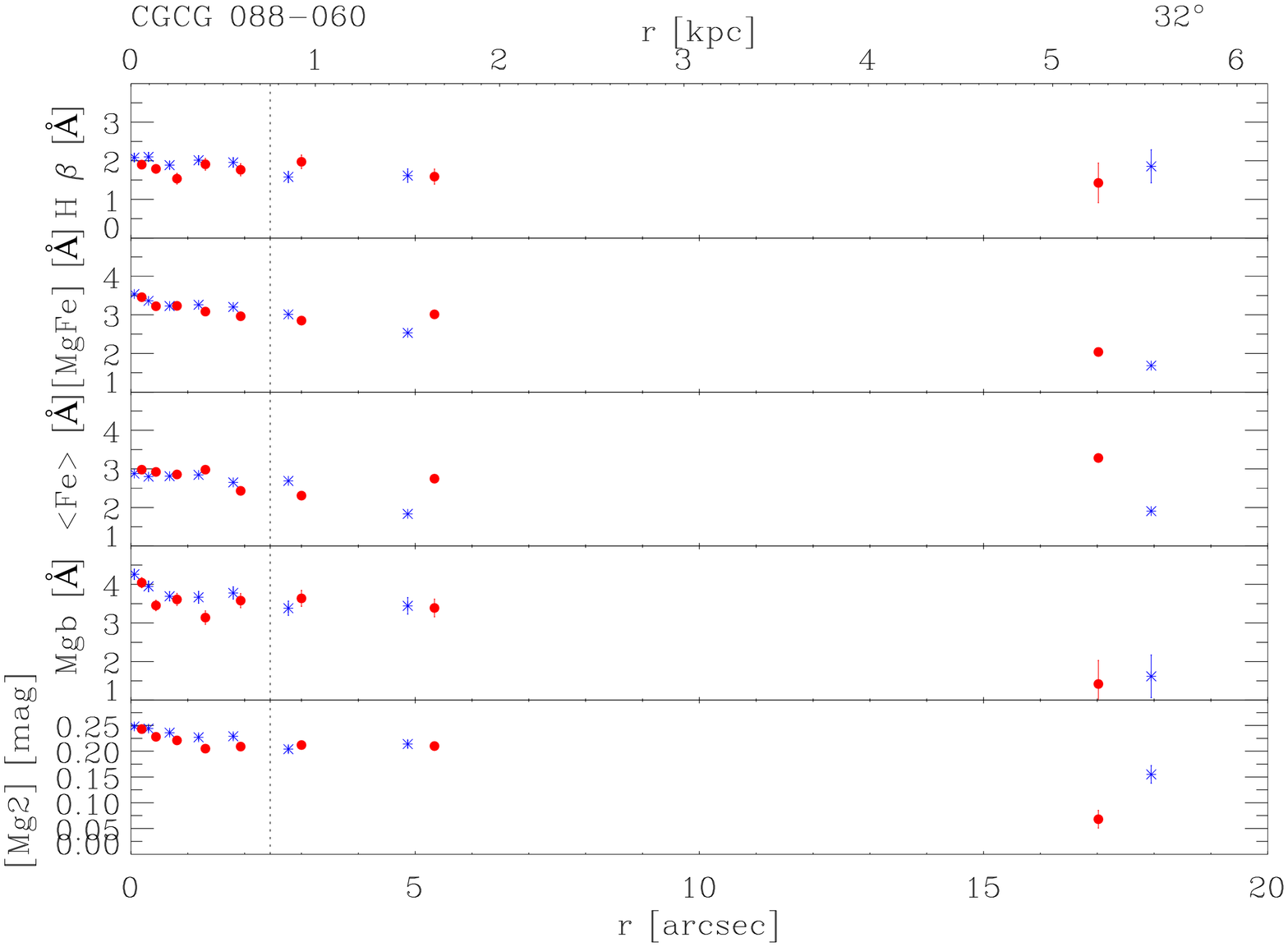}
\includegraphics[angle=90.0,width=0.498\textwidth]{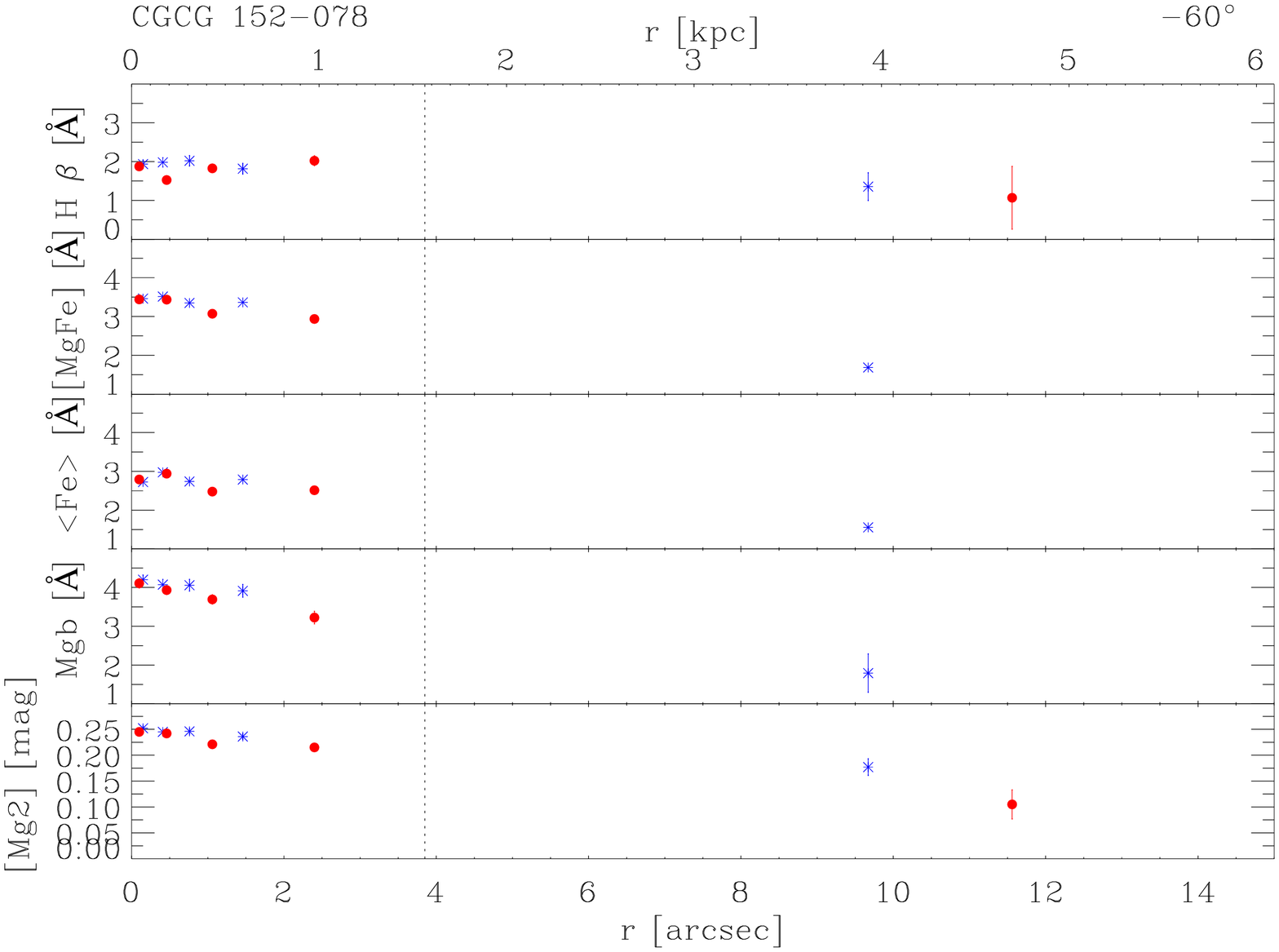}
\includegraphics[angle=90.0,width=0.498\textwidth]{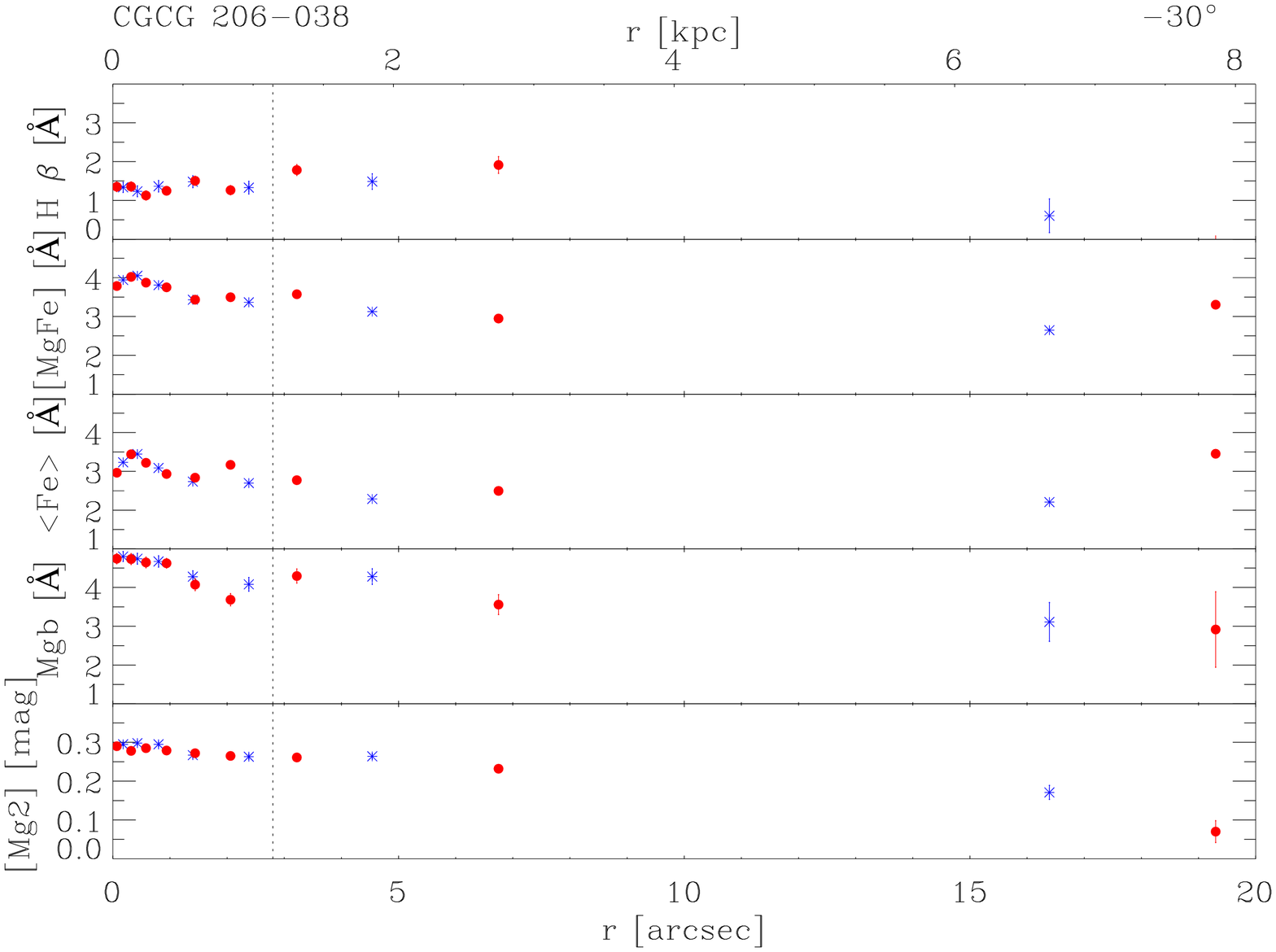}
\includegraphics[angle=90.0,width=0.498\textwidth]{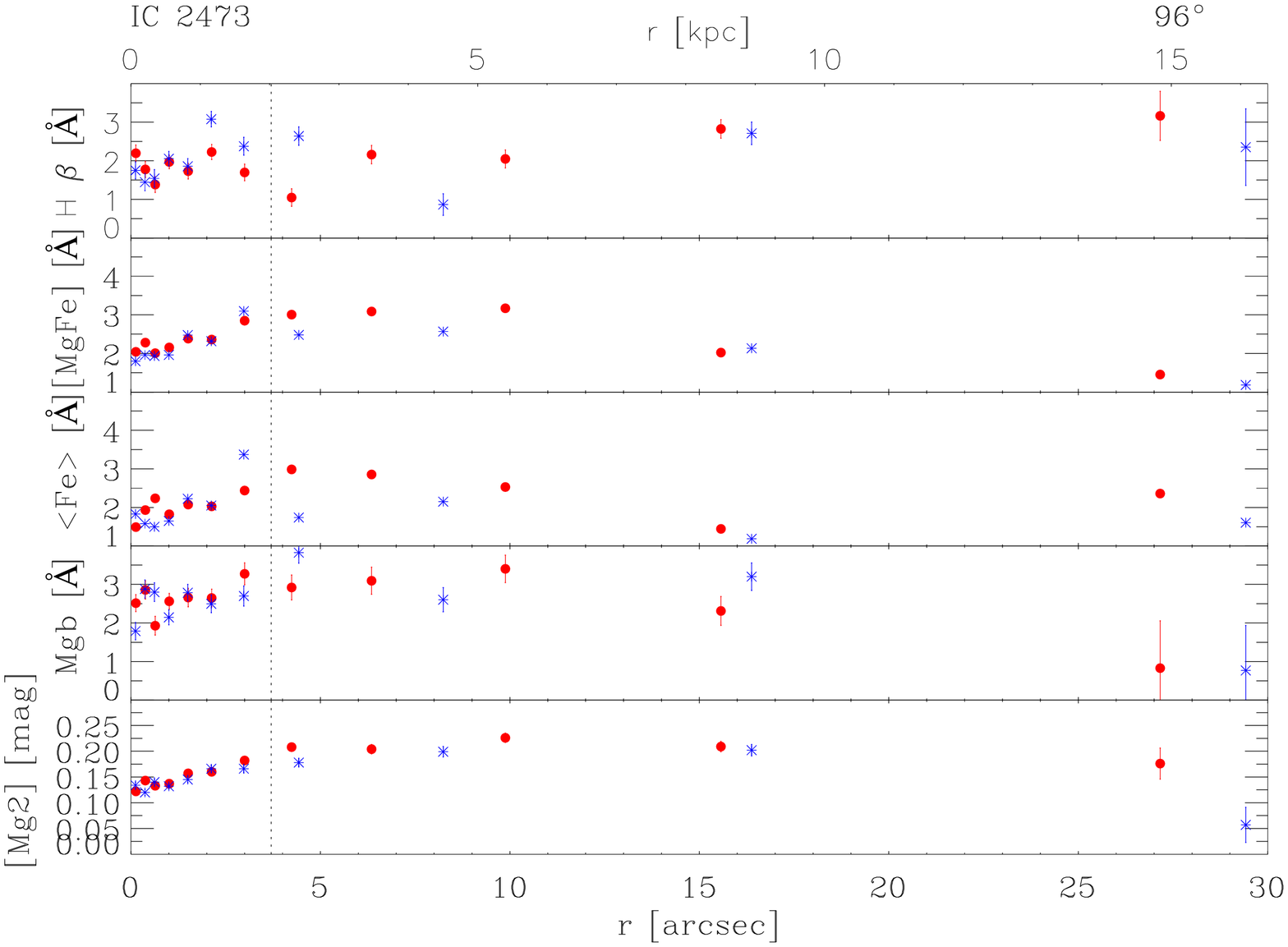}
\includegraphics[angle=90.0,width=0.498\textwidth]{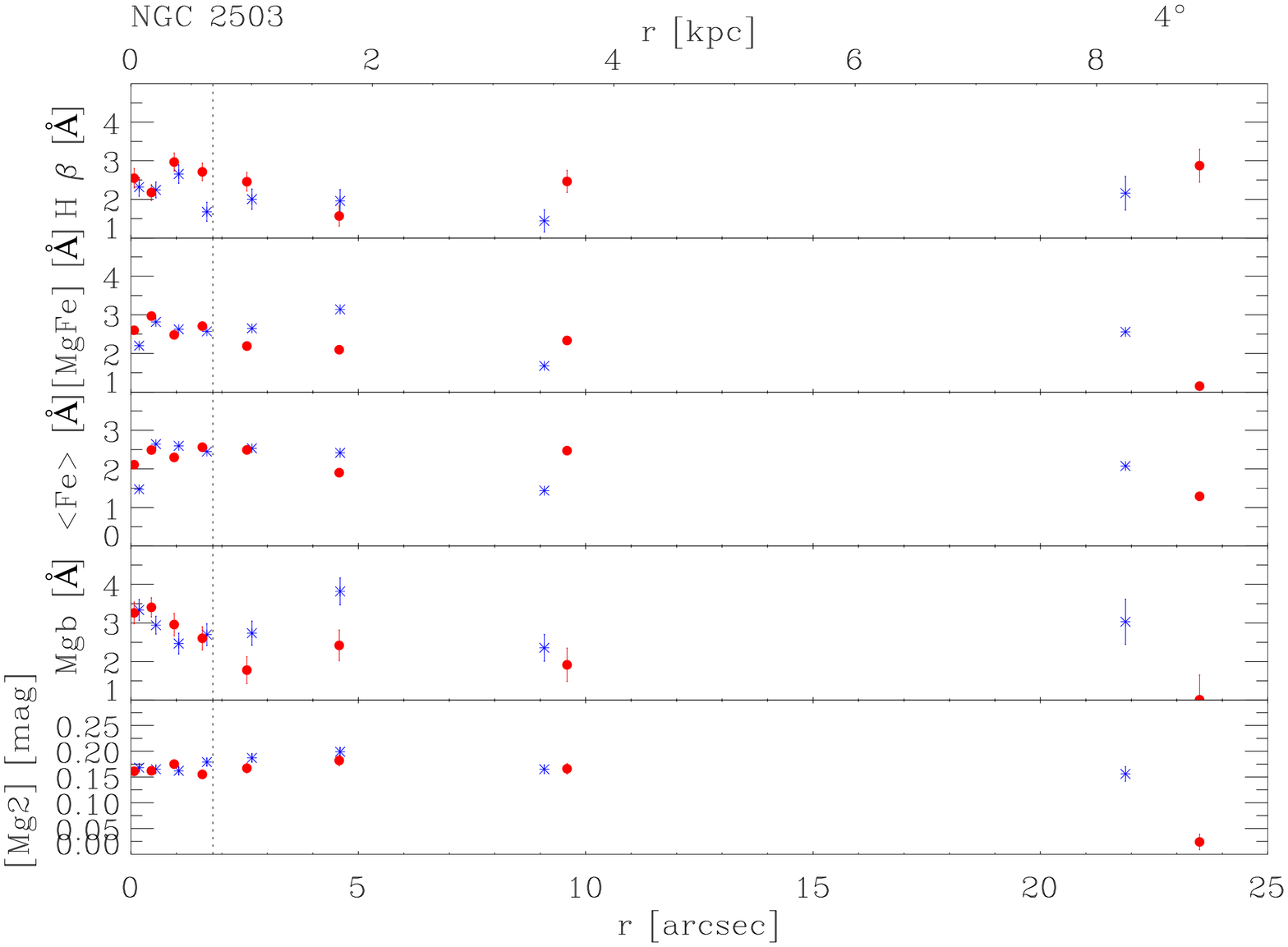}\\
\caption{Line-strength indices measured along the major axis of the
  sample galaxies. For each axis the curves are folded around the
  nucleus. Blue asterisks and red circles refer to data measured along
  the approaching and residing side of the galaxy, respectively.  The
  radial profiles of the line-strength indices \Hb, \MgFe, \Fe, \Mgb,
  and \Mgd\ are shown (from top to bottom panel).  The vertical dashed
  line corresponds to the radius \rbd , where the bulge contributes
  half of galaxy surface brightness. The name of the galaxy and
  position angle of the slit are given for each data set.}
\label{fig:indices}
\end{figure*}
\begin{figure*}
\centering
\includegraphics[angle=90.0,width=0.498\textwidth]{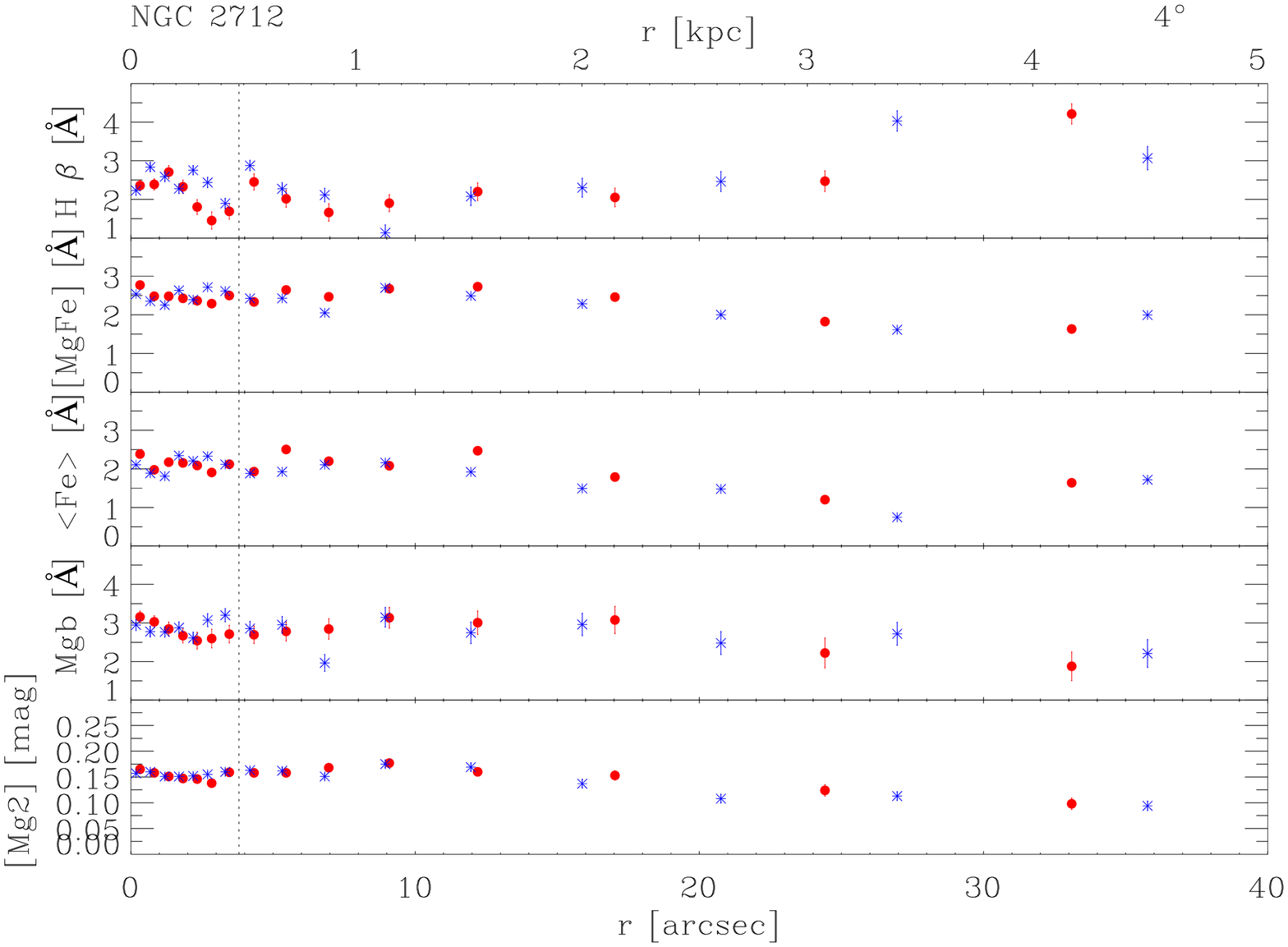}
\includegraphics[angle=90.0,width=0.498\textwidth]{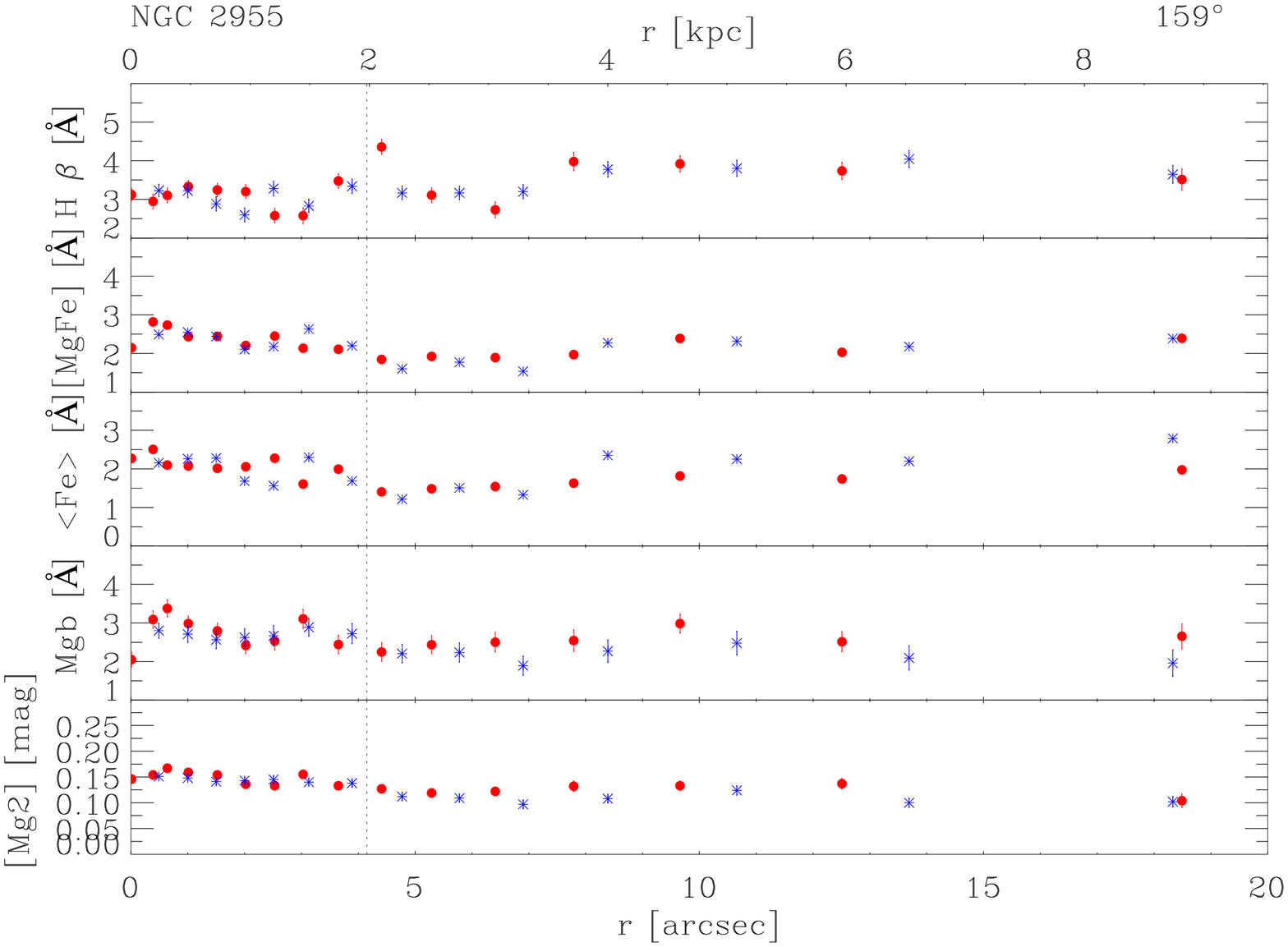}
\includegraphics[angle=90.0,width=0.498\textwidth]{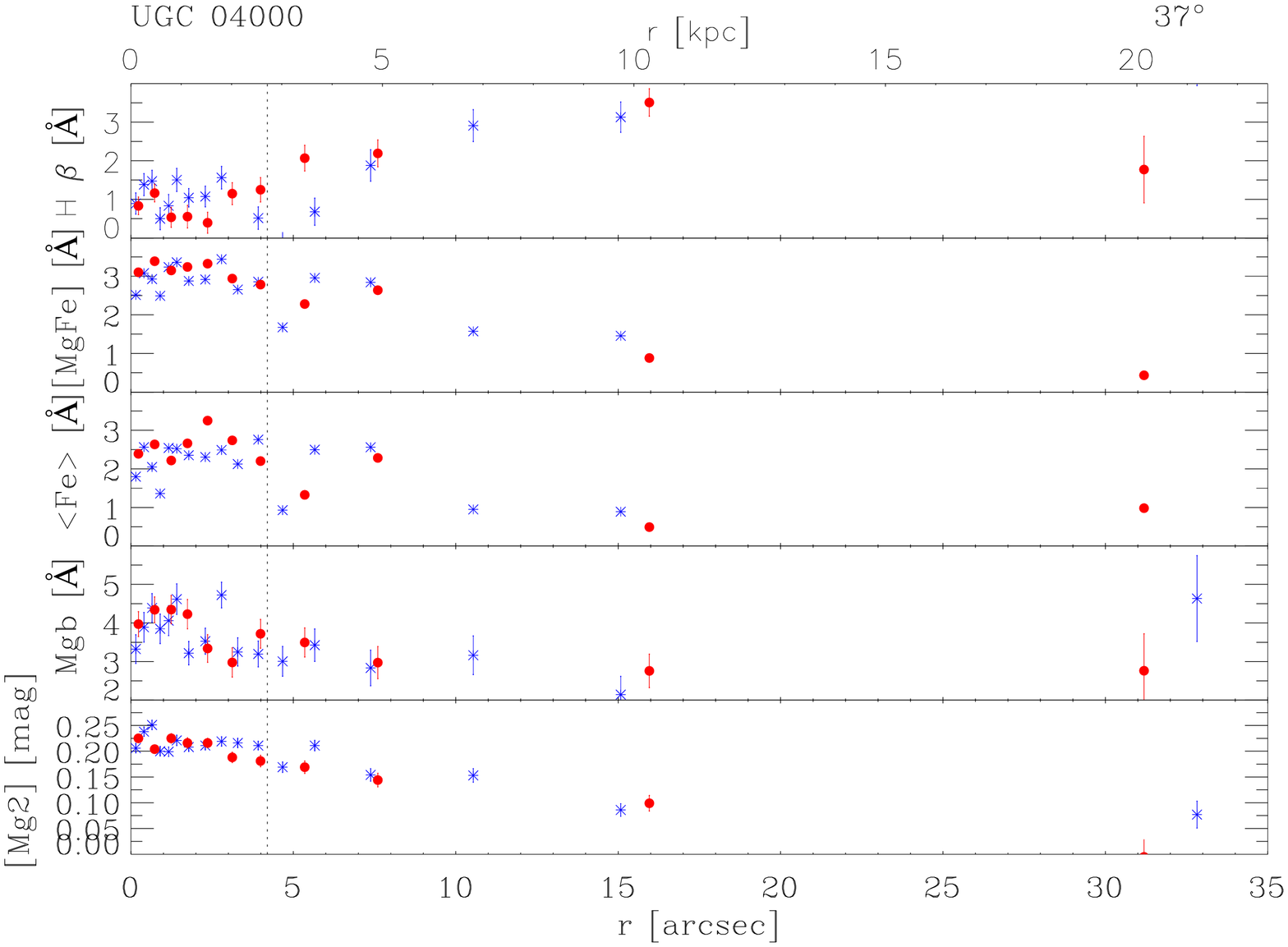}
\includegraphics[angle=90.0,width=0.498\textwidth]{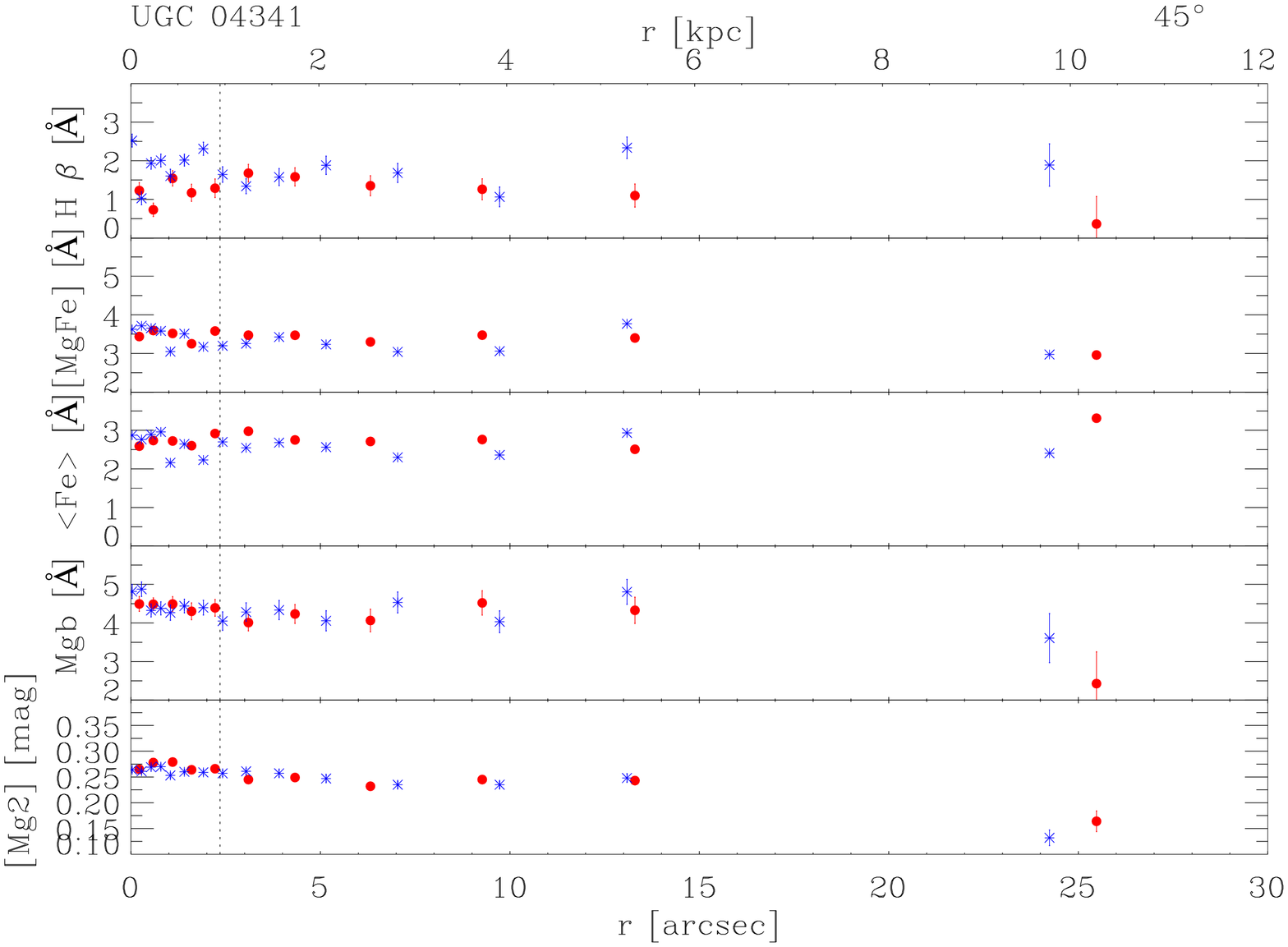}
\includegraphics[angle=90.0,width=0.498\textwidth]{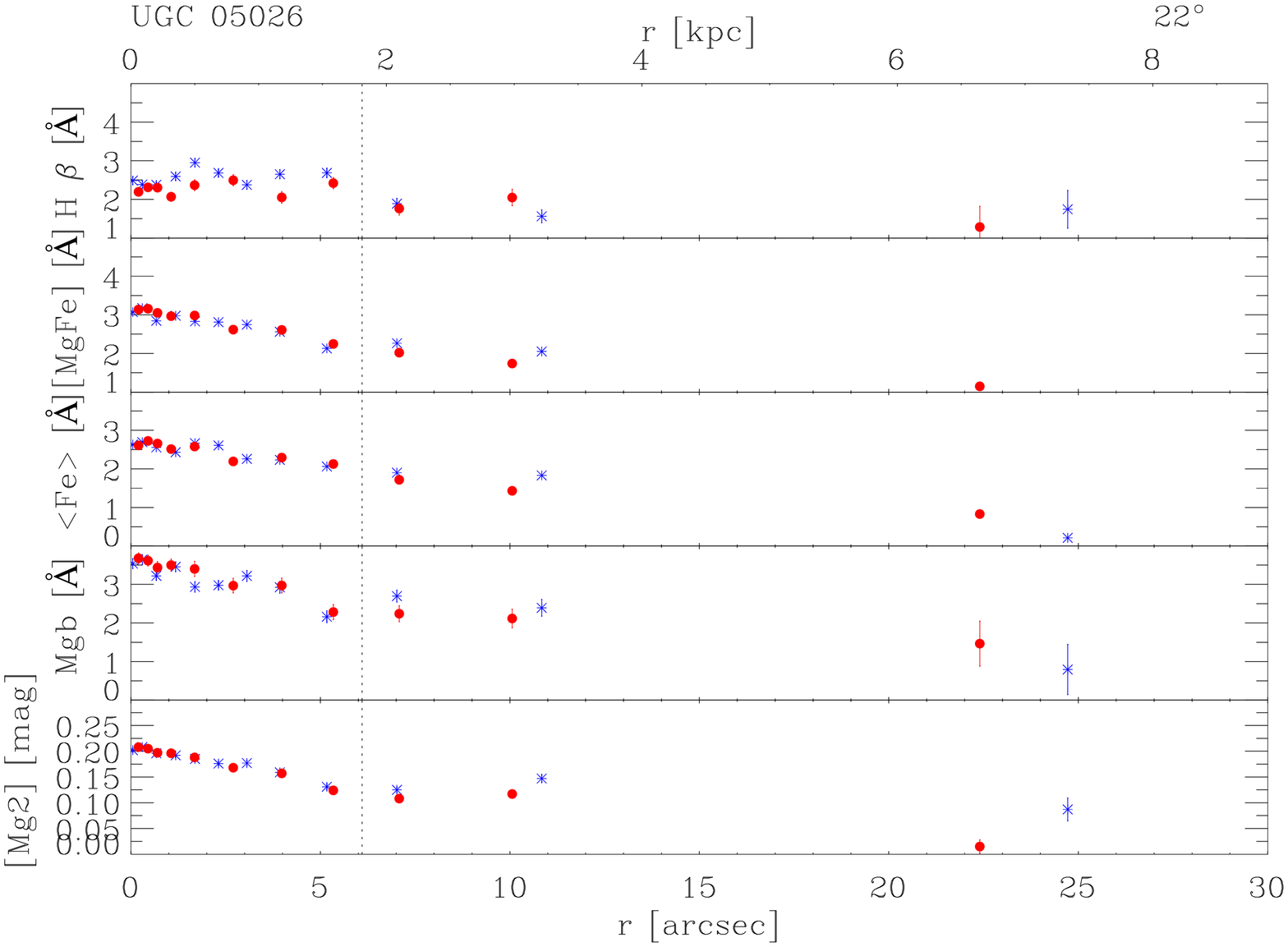}
\includegraphics[angle=90.0,width=0.498\textwidth]{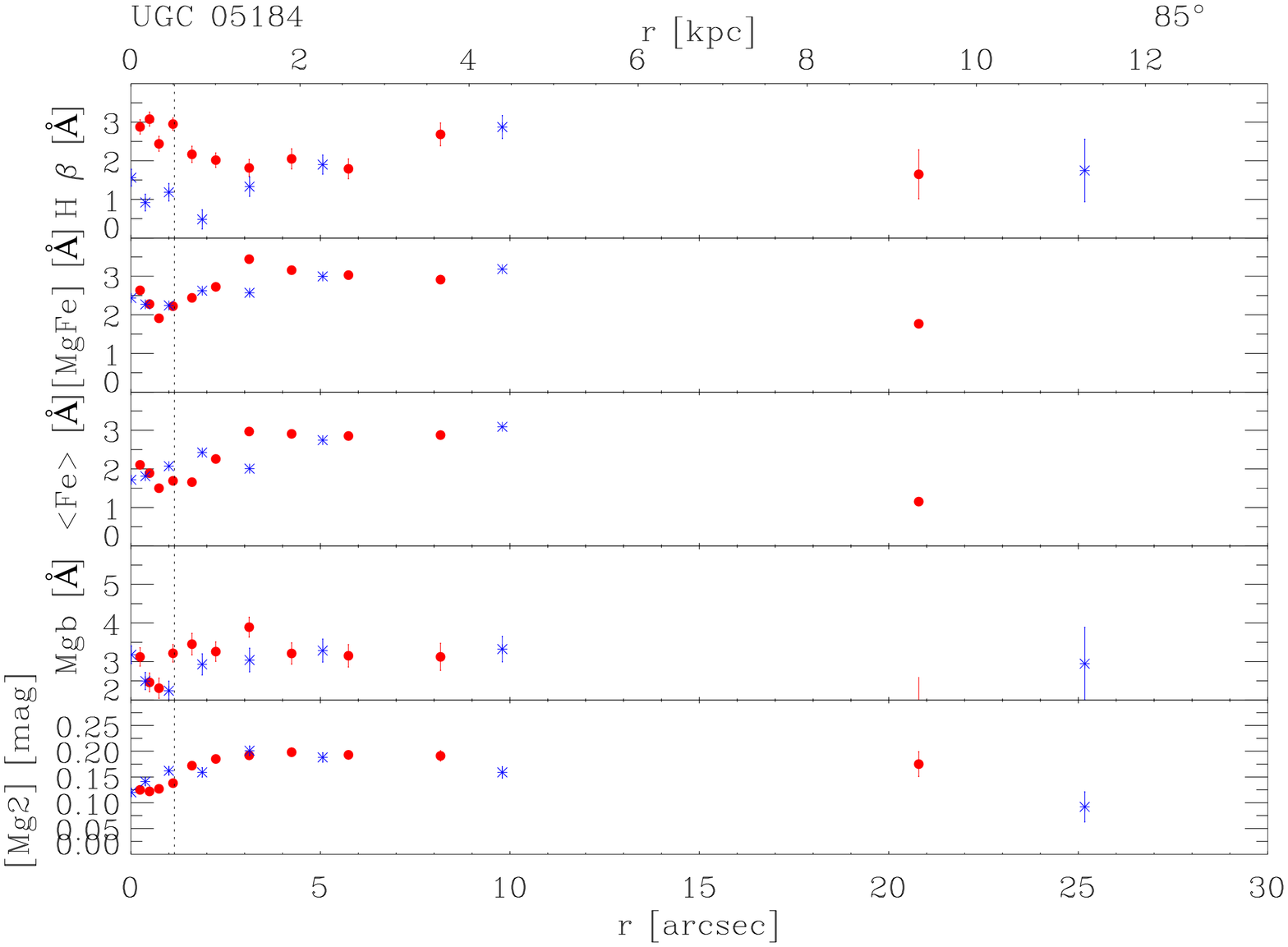}\\
\contcaption{}
\end{figure*}
\renewcommand{\tabcolsep}{2pt}
\begin{table*}
\caption{Stellar kinematics and line-strength indices measured along
  the major axis of the sample galaxies.  The columns show the
  following. (1): radius; (2): LOS velocity after subtraction of
  systemic velocity; (3): LOS velocity dispersion; (4): third-order
  Gauss-Hermite coefficient; (5): fourth-order Gauss-Hermite
  coefficient; (6)-(10): equivalent width of the line-strength
  indices.}
\label{tab:kinematics_indices}.
\begin{scriptsize}
\begin{tabular}{rrrrrrrrrrrr}
\hline
\noalign{\smallskip}
\multicolumn{1}{r}{r~~~~~~~} &
\multicolumn{1}{c}{$v$} &
\multicolumn{1}{c}{$\sigma$} &
\multicolumn{1}{c}{{$h_3$}} &
\multicolumn{1}{c}{{$h_4$}} &
\multicolumn{1}{c}{\Hb}&
\multicolumn{1}{c}{\Mgd} &
\multicolumn{1}{c}{\Mgb} &
\multicolumn{1}{c}{{\rm Fe5270}} &
\multicolumn{1}{c}{{\rm Fe5335}}\\
\multicolumn{1}{r}{(arcsec)} &
\multicolumn{1}{c}{(\kms)} &
\multicolumn{1}{c}{(\kms)} &
\multicolumn{1}{c}{} &
\multicolumn{1}{c}{} &
\multicolumn{1}{c}{(\AA)} &
\multicolumn{1}{c}{(mag)} &
\multicolumn{1}{c}{(\AA)} &
\multicolumn{1}{c}{(\AA)} &
\multicolumn{1}{c}{(\AA)}\\
\multicolumn{1}{c}{(1)} &
\multicolumn{1}{c}{(2)} &
\multicolumn{1}{c}{(3)} &
\multicolumn{1}{c}{(4)} &
\multicolumn{1}{c}{(5)} &
\multicolumn{1}{c}{(6)} &
\multicolumn{1}{c}{(7)} &
\multicolumn{1}{c}{(8)} &
\multicolumn{1}{c}{(9)} &
\multicolumn{1}{c}{(10)} \\
\noalign{\smallskip}
\hline
\noalign{\smallskip}
{\bf CGCG 034-050 } \\
\noalign{\smallskip}
\hline
\noalign{\smallskip}
$   -19.28 $ &    $ -149.9 \pm  19.8  $ & $ 112.6 \pm  28.8  $ & $  0.001 \pm 0.090  $ & $  0.015 \pm 0.176  $ & $ 1.196  \pm   0.592 $ & $    0.043 \pm 0.043  $ & $   -1.802  \pm   0.650 $ & $    1.220  \pm   0.691 $ & $    3.270  \pm   0.704 $ \\
$    -7.41 $ &    $ -156.6 \pm   4.7  $ & $  85.7 \pm   7.2  $ & $  0.025 \pm 0.045  $ & $  0.107 \pm 0.052  $ & $ 1.303  \pm   0.154 $ & $    0.220 \pm 0.220  $ & $    3.036  \pm   0.198 $ & $    2.389  \pm   0.183 $ & $    2.523  \pm   0.174 $ \\
$    -4.81 $ &    $ -145.6 \pm   3.9  $ & $ 121.4 \pm   5.9  $ & $  0.019 \pm 0.027  $ & $  0.097 \pm 0.031  $ & $ 1.569  \pm   0.126 $ & $    0.254 \pm 0.254  $ & $    3.699  \pm   0.169 $ & $    3.045  \pm   0.154 $ & $    2.146  \pm   0.151 $ \\
$    -3.48 $ &    $ -138.7 \pm   3.8  $ & $ 122.3 \pm   5.8  $ & $  0.057 \pm 0.025  $ & $  0.107 \pm 0.031  $ & $ 1.499  \pm   0.111 $ & $    0.256 \pm 0.256  $ & $    3.769  \pm   0.158 $ & $    2.521  \pm   0.144 $ & $    2.557  \pm   0.133 $ \\
\hline
\end{tabular}
\end{scriptsize}
\end{table*}
\begin{table*}
\begin{scriptsize}
\contcaption{}
\begin{tabular}{rrrrrrrrrrrr}
\hline
\noalign{\smallskip}
\multicolumn{1}{r}{r~~~~~~~} &
\multicolumn{1}{c}{$v$} &
\multicolumn{1}{c}{$\sigma$} &
\multicolumn{1}{c}{{$h_3$}} &
\multicolumn{1}{c}{{$h_4$}} &
\multicolumn{1}{c}{\Hb}&
\multicolumn{1}{c}{\Mgd} &
\multicolumn{1}{c}{\Mgb} &
\multicolumn{1}{c}{{\rm Fe5270}} &
\multicolumn{1}{c}{{\rm Fe5335}}\\
\multicolumn{1}{r}{(arcsec)} &
\multicolumn{1}{c}{(\kms)} &
\multicolumn{1}{c}{(\kms)} &
\multicolumn{1}{c}{} &
\multicolumn{1}{c}{} &
\multicolumn{1}{c}{(\AA)} &
\multicolumn{1}{c}{(mag)} &
\multicolumn{1}{c}{(\AA)} &
\multicolumn{1}{c}{(\AA)} &
\multicolumn{1}{c}{(\AA)}\\
\multicolumn{1}{c}{(1)} &
\multicolumn{1}{c}{(2)} &
\multicolumn{1}{c}{(3)} &
\multicolumn{1}{c}{(4)} &
\multicolumn{1}{c}{(5)} &
\multicolumn{1}{c}{(6)} &
\multicolumn{1}{c}{(7)} &
\multicolumn{1}{c}{(8)} &
\multicolumn{1}{c}{(9)} &
\multicolumn{1}{c}{(10)} \\
\noalign{\smallskip}
\hline
\noalign{\smallskip}
$      9.37$ & $ -77.3 \pm  12.0  $ & $ 70.6  \pm  14.8 $  & $ -0.011   \pm  0.129 $ & $  0.031  \pm   0.112 $ & $    1.355 \pm    0.364 $ & $    0.177 \pm  0.016 $ & $    1.793  \pm   0.427  $ & $   1.621 \pm    0.636 $ & $    1.493 \pm    0.584$\\ 
\noalign{\smallskip}                                                                                                         
\noalign{\smallskip}
\hline
{\bf CGCG 206-038 } \\
\noalign{\smallskip}
\hline
\noalign{\smallskip}
$    -19.30 $ & $  15.0\pm   21.1  $ & $   58.7 \pm    31.3 $ & $ -0.021 \pm  0.265  $ & $  -0.005 \pm  0.296   $ & $   -0.762 \pm    0.674 $ & $    0.070  \pm 0.028 $ & $    2.916 \pm    0.772 $ & $    4.110  \pm   0.932 $ & $    2.796 \pm    0.951 $ \\
$     -6.75 $ & $  16.6\pm    5.8  $ & $  128.6 \pm     7.9 $ & $  0.045 \pm  0.035  $ & $   0.012 \pm  0.043   $ & $    1.913 \pm    0.168 $ & $    0.232  \pm 0.008 $ & $    3.558 \pm    0.227 $ & $    2.370  \pm   0.243 $ & $    2.624 \pm    0.252 $ \\
\hline
\end{tabular}
\end{scriptsize}
\end{table*}
\begin{table*}
\begin{scriptsize}
\contcaption{}
\begin{tabular}{rrrrrrrrrrrr}
\hline
\noalign{\smallskip}
\multicolumn{1}{r}{r~~~~~~~} &
\multicolumn{1}{c}{$v$} &
\multicolumn{1}{c}{$\sigma$} &
\multicolumn{1}{c}{{$h_3$}} &
\multicolumn{1}{c}{{$h_4$}} &
\multicolumn{1}{c}{\Hb}&
\multicolumn{1}{c}{\Mgd} &
\multicolumn{1}{c}{\Mgb} &
\multicolumn{1}{c}{{\rm Fe5270}} &
\multicolumn{1}{c}{{\rm Fe5335}}\\
\multicolumn{1}{r}{(arcsec)} &
\multicolumn{1}{c}{(\kms)} &
\multicolumn{1}{c}{(\kms)} &
\multicolumn{1}{c}{} &
\multicolumn{1}{c}{} &
\multicolumn{1}{c}{(\AA)} &
\multicolumn{1}{c}{(mag)} &
\multicolumn{1}{c}{(\AA)} &
\multicolumn{1}{c}{(\AA)} &
\multicolumn{1}{c}{(\AA)}\\
\multicolumn{1}{c}{(1)} &
\multicolumn{1}{c}{(2)} &
\multicolumn{1}{c}{(3)} &
\multicolumn{1}{c}{(4)} &
\multicolumn{1}{c}{(5)} &
\multicolumn{1}{c}{(6)} &
\multicolumn{1}{c}{(7)} &
\multicolumn{1}{c}{(8)} &
\multicolumn{1}{c}{(9)} &
\multicolumn{1}{c}{(10)} \\
\noalign{\smallskip}
\hline
\noalign{\smallskip}
$     -0.95$ & $   18.3\pm    6.3   $ & $  77.6 \pm   9.7  $ & $  0.056 \pm  0.061  $ & $   0.051 \pm   0.079  $ & $    2.967  \pm   0.253  $ & $   0.175 \pm 0.008 $ & $   2.959  \pm   0.257 $ & $    1.803  \pm   0.321  $ & $   2.793 \pm    0.304  $ \\ 
$     -0.45$ & $    8.5\pm    6.0   $ & $  65.7 \pm   8.1  $ & $ -0.014 \pm  0.072  $ & $   0.050 \pm   0.071  $ & $    2.177  \pm   0.212  $ & $   0.162 \pm 0.007 $ & $   3.406  \pm   0.209 $ & $    2.709  \pm   0.261  $ & $   2.268 \pm    0.253  $ \\ 
$     -0.07$ & $   -1.8\pm    7.7   $ & $  86.1 \pm  10.0  $ & $  0.016 \pm  0.073  $ & $   0.079 \pm   0.058  $ & $    2.548  \pm   0.248  $ & $   0.161 \pm 0.008 $ & $   3.261  \pm   0.245 $ & $    2.029  \pm   0.309  $ & $   2.188 \pm    0.310  $\\  
$      0.18$ & $  -12.1\pm    7.4   $ & $  95.3 \pm   9.7  $ & $ -0.033 \pm  0.063  $ & $   0.030 \pm   0.057  $ & $    2.320  \pm   0.245  $ & $   0.168 \pm 0.007 $ & $   3.336  \pm   0.255 $ & $    1.426  \pm   0.306  $ & $   1.519 \pm    0.303  $\\
\hline
\end{tabular}
\end{scriptsize}
\end{table*}
\begin{table*}
\begin{scriptsize}
\contcaption{}
\begin{tabular}{rrrrrrrrrrrr}
\hline
\noalign{\smallskip}
\multicolumn{1}{r}{r~~~~~~~} &
\multicolumn{1}{c}{$v$} &
\multicolumn{1}{c}{$\sigma$} &
\multicolumn{1}{c}{{$h_3$}} &
\multicolumn{1}{c}{{$h_4$}} &
\multicolumn{1}{c}{\Hb}&
\multicolumn{1}{c}{\Mgd} &
\multicolumn{1}{c}{\Mgb} &
\multicolumn{1}{c}{{\rm Fe5270}} &
\multicolumn{1}{c}{{\rm Fe5335}}\\
\multicolumn{1}{r}{(arcsec)} &
\multicolumn{1}{c}{(\kms)} &
\multicolumn{1}{c}{(\kms)} &
\multicolumn{1}{c}{} &
\multicolumn{1}{c}{} &
\multicolumn{1}{c}{(\AA)} &
\multicolumn{1}{c}{(mag)} &
\multicolumn{1}{c}{(\AA)} &
\multicolumn{1}{c}{(\AA)} &
\multicolumn{1}{c}{(\AA)}\\
\multicolumn{1}{c}{(1)} &
\multicolumn{1}{c}{(2)} &
\multicolumn{1}{c}{(3)} &
\multicolumn{1}{c}{(4)} &
\multicolumn{1}{c}{(5)} &
\multicolumn{1}{c}{(6)} &
\multicolumn{1}{c}{(7)} &
\multicolumn{1}{c}{(8)} &
\multicolumn{1}{c}{(9)} &
\multicolumn{1}{c}{(10)} \\
\noalign{\smallskip}
\hline
\noalign{\smallskip}
$     -6.28$ & $  -131.7\pm    8.5   $ & $  102.6 \pm 12.8   $ & $   0.064 \pm  0.053  $ & $  0.015 \pm  0.015   $ & $    3.166  \pm    0.194  $ & $   0.109 \pm 0.007  $ & $    2.236  \pm    0.236 $ & $    1.269 \pm     0.266 $ & $    1.749 \pm     0.347 $ \\
$     -5.27$ & $  -120.0\pm   10.0   $ & $  128.2 \pm 10.7   $ & $   0.015 \pm  0.069  $ & $ -0.031 \pm -0.031   $ & $    3.168  \pm    0.182  $ & $   0.112 \pm 0.007  $ & $    2.206  \pm    0.228 $ & $    1.090 \pm     0.252 $ & $    1.346 \pm     0.333 $\\ 
$     -4.39$ & $  -106.3\pm    8.7   $ & $  117.8 \pm 11.0   $ & $   0.007 \pm  0.063  $ & $  0.029 \pm  0.029   $ & $    3.339  \pm    0.189  $ & $   0.138 \pm 0.007  $ & $    2.724  \pm    0.244 $ & $    1.884 \pm     0.265 $ & $    1.490 \pm     0.347 $\\ 
$     -3.63$ & $   -81.8\pm    7.6   $ & $  112.0 \pm 10.0   $ & $  -0.024 \pm  0.055  $ & $  0.081 \pm  0.081   $ & $    2.830  \pm    0.165  $ & $   0.140 \pm 0.006  $ & $    2.889  \pm    0.220 $ & $    2.531 \pm     0.237 $ & $    2.060 \pm     0.298 $\\
\hline
\end{tabular}
\end{scriptsize}
\end{table*}
\begin{table*}
\begin{scriptsize}
\contcaption{}
\begin{tabular}{rrrrrrrrrrrr}
\hline
\noalign{\smallskip}
\multicolumn{1}{r}{r~~~~~~~} &
\multicolumn{1}{c}{$v$} &
\multicolumn{1}{c}{$\sigma$} &
\multicolumn{1}{c}{{$h_3$}} &
\multicolumn{1}{c}{{$h_4$}} &
\multicolumn{1}{c}{\Hb}&
\multicolumn{1}{c}{\Mgd} &
\multicolumn{1}{c}{\Mgb} &
\multicolumn{1}{c}{{\rm Fe5270}} &
\multicolumn{1}{c}{{\rm Fe5335}}\\
\multicolumn{1}{r}{(arcsec)} &
\multicolumn{1}{c}{(\kms)} &
\multicolumn{1}{c}{(\kms)} &
\multicolumn{1}{c}{} &
\multicolumn{1}{c}{} &
\multicolumn{1}{c}{(\AA)} &
\multicolumn{1}{c}{(mag)} &
\multicolumn{1}{c}{(\AA)} &
\multicolumn{1}{c}{(\AA)} &
\multicolumn{1}{c}{(\AA)}\\
\multicolumn{1}{c}{(1)} &
\multicolumn{1}{c}{(2)} &
\multicolumn{1}{c}{(3)} &
\multicolumn{1}{c}{(4)} &
\multicolumn{1}{c}{(5)} &
\multicolumn{1}{c}{(6)} &
\multicolumn{1}{c}{(7)} &
\multicolumn{1}{c}{(8)} &
\multicolumn{1}{c}{(9)} &
\multicolumn{1}{c}{(10)} \\
\noalign{\smallskip}
\hline
\noalign{\smallskip}
$     -9.73 $ & $ -161.4\pm    5.8  $ & $  106.1 \pm    8.1 $ & $  -0.020\pm  0.045  $ & $   0.072  \pm  0.045   $ & $  1.064 \pm    0.206 $ & $    0.235  \pm 0.007   $ & $   4.034 \pm    0.258 $ & $    2.268 \pm    0.299 $ & $    2.453 \pm    0.223 $ \\    
$     -7.04 $ & $ -124.8\pm    5.9  $ & $  106.6 \pm    8.0 $ & $   0.005\pm  0.045  $ & $   0.062  \pm  0.045   $ & $  1.684 \pm    0.203 $ & $    0.235  \pm 0.007   $ & $   4.532 \pm    0.264 $ & $    1.713 \pm    0.278 $ & $    2.886 \pm    0.218 $ \\    
$     -5.15 $ & $ -104.2\pm    6.4  $ & $  139.2 \pm    8.8 $ & $  -0.041\pm  0.038  $ & $   0.063  \pm  0.041   $ & $  1.882 \pm    0.192 $ & $    0.247  \pm 0.007   $ & $   4.057 \pm    0.266 $ & $    2.599 \pm    0.267 $ & $    2.528 \pm    0.206 $ \\    
$     -3.91 $ & $  -89.5\pm    6.3  $ & $  156.3 \pm    8.2 $ & $  -0.020\pm  0.033  $ & $   0.035  \pm  0.035   $ & $  1.578 \pm    0.178 $ & $    0.257  \pm 0.007   $ & $   4.334 \pm    0.256 $ & $    2.743 \pm    0.252 $ & $    2.615 \pm    0.201 $ \\
\hline
\end{tabular}
\end{scriptsize}
\end{table*}
\begin{table*}
\begin{scriptsize}
\contcaption{}
\begin{tabular}{rrrrrrrrrrrr}
\hline
\noalign{\smallskip}
\multicolumn{1}{r}{r~~~~~~~} &
\multicolumn{1}{c}{$v$} &
\multicolumn{1}{c}{$\sigma$} &
\multicolumn{1}{c}{{$h_3$}} &
\multicolumn{1}{c}{{$h_4$}} &
\multicolumn{1}{c}{\Hb}&
\multicolumn{1}{c}{\Mgd} &
\multicolumn{1}{c}{\Mgb} &
\multicolumn{1}{c}{{\rm Fe5270}} &
\multicolumn{1}{c}{{\rm Fe5335}}\\
\multicolumn{1}{r}{(arcsec)} &
\multicolumn{1}{c}{(\kms)} &
\multicolumn{1}{c}{(\kms)} &
\multicolumn{1}{c}{} &
\multicolumn{1}{c}{} &
\multicolumn{1}{c}{(\AA)} &
\multicolumn{1}{c}{(mag)} &
\multicolumn{1}{c}{(\AA)} &
\multicolumn{1}{c}{(\AA)} &
\multicolumn{1}{c}{(\AA)}\\
\multicolumn{1}{c}{(1)} &
\multicolumn{1}{c}{(2)} &
\multicolumn{1}{c}{(3)} &
\multicolumn{1}{c}{(4)} &
\multicolumn{1}{c}{(5)} &
\multicolumn{1}{c}{(6)} &
\multicolumn{1}{c}{(7)} &
\multicolumn{1}{c}{(8)} &
\multicolumn{1}{c}{(9)} &
\multicolumn{1}{c}{(10)} \\
\noalign{\smallskip}
\hline
\noalign{\smallskip}
$   -1.88$ & $    -71.8\pm    6.2   $ & $  95.4 \pm    9.2  $ & $   0.044\pm    0.046 $ & $   0.019 \pm   0.068  $ & $    0.482  \pm   0.300 $ & $    0.159  \pm 0.009  $ & $    2.927 \pm    0.305 $ & $    2.267 \pm    0.320 $ & $    2.580  \pm   0.222 $ \\ 
$   -1.00$ & $    -38.4\pm    6.1   $ & $ 114.5 \pm    8.7  $ & $   0.110\pm    0.039 $ & $   0.103 \pm   0.051  $ & $    1.184  \pm   0.301 $ & $    0.162  \pm 0.008  $ & $    2.245 \pm    0.280 $ & $    2.460 \pm    0.285 $ & $    1.682  \pm   0.197 $ \\ 
$   -0.38$ & $    -14.8\pm    6.4   $ & $ 130.5 \pm    8.2  $ & $   0.130\pm    0.035 $ & $   0.032 \pm   0.046  $ & $    0.918  \pm   0.278 $ & $    0.141  \pm 0.007  $ & $    2.497 \pm    0.246 $ & $    2.385 \pm    0.256 $ & $    1.238  \pm   0.176 $ \\ 
$   -0.01$ & $     -6.0\pm    7.7   $ & $ 147.4 \pm   10.2  $ & $   0.100\pm    0.038 $ & $   0.061 \pm   0.048  $ & $    1.563  \pm   0.266 $ & $    0.120  \pm 0.007  $ & $    3.177 \pm    0.256 $ & $    2.053 \pm    0.274 $ & $    1.382  \pm   0.197 $ \\
\hline     
\end{tabular}
\label{tab:val_ind}
\end{scriptsize}
\end{table*}

\bsp

\label{lastpage}

\end{document}